\documentclass[iop]{emulateapj}

\newcommand{\CII}{\ion{C}{2}} \newcommand{\CIII}{\ion{C}{3}}
\newcommand{\CIV}{\ion{C}{4}} \newcommand{\HI}{\ion{H}{1}}
\newcommand{\HII}{\ion{H}{2}} \newcommand{\MgII}{\ion{Mg}{2}}
\newcommand{\NV}{\ion{N}{5}} \newcommand{\OI}{\ion{O}{1}}
\newcommand{\OVI}{\ion{O}{6}} \newcommand{\OVII}{\ion{O}{7}}
\newcommand{\SII}{\ion{S}{2}} \newcommand{\SiII}{\ion{Si}{2}}
\newcommand{\SiIII}{\ion{Si}{3}} \newcommand{\SiIV}{\ion{Si}{4}}

\newcommand{\eqw}{\ensuremath{W_{\lambda}}}
 \newcommand{\fuse}{{\sl
    FUSE}} 
\newcommand{\Ha}{\ensuremath{{\rm H}\alpha}} \newcommand{\hst}{{\sl
    HST}} \newcommand{\kms}{\ensuremath{{\rm km\,s}^{-1}}}
\newcommand{\lya}{\ensuremath{{\rm Ly}\alpha}}

\begin{document}

\title{Characterizing the Circumgalactic Medium of Nearby Galaxies
  with HST/COS and HST/STIS  Absorption-Line Spectroscopy\altaffilmark{1}}

\author{John T. Stocke, Brian A. Keeney, Charles W. Danforth,
  J. Michael Shull, Cynthia S. Froning, James C. Green, Steven
  V. Penton} 
\affil{Center for Astrophysics and Space Astronomy,
  Department of Astrophysical and Planetary Sciences, University of
  Colorado, 389 UCB, Boulder, CO 80309, USA; john.stocke@colorado.edu}
\and 
\author{Blair D. Savage} 
\affil{Department of Astronomy, University of Wisconsin, Madison, WI 53706}

\altaffiltext{1}{Based on observations with the NASA/ESA {\sl Hubble Space Telescope}, obtained at the Space Telescope Science Institute, which is operated by AURA, Inc., under NASA contract NAS 5-26555.}

\shorttitle{UV Probes of the Circumgalactic Medium}
\shortauthors{Stocke et~al.}
\submitted{Accepted by \apj\ on 21 Dec 2012}

\begin {abstract}


The Circumgalactic Medium (CGM) of late-type galaxies is characterized
using UV spectroscopy of 11 targeted QSO/galaxy pairs at $z \leq 0.02$
with the {\sl Hubble Space Telescope} Cosmic Origins Spectrograph and
$\sim60$ serendipitous absorber/galaxy pairs at $z \leq 0.2$ with the
Space Telescope Imaging Spectrograph.  CGM warm cloud properties are
derived, including volume filling factors of 3--5\%,  cloud sizes of
0.1--30 kpc, masses of 10--$10^{8}~M_{\odot}$ and metallicities of
$\sim0.1$--$1\,Z_{\Sun}$.  Almost all warm CGM clouds within
$0.5\,R_{\rm vir}$ are metal-bearing and many have velocities
consistent with being bound, ``galactic fountain'' clouds.  For
galaxies with $L \gtrsim 0.1\,L^*$, the total mass in these warm CGM
clouds  approaches $10^{10}~M_{\odot}$, $\sim10$--15\% of the total
baryons in massive spirals   and comparable to the baryons in their
parent galaxy disks.  This leaves $\gtrsim 50$\% of massive
spiral-galaxy baryons ``missing''.  Dwarfs ($<0.1\,L^*$) have smaller
area covering factors and warm CGM masses ($\leq 5$\% baryon
fraction), suggesting that many of their warm clouds escape.
Constant warm cloud internal pressures as a function of impact
parameter ($P/k \sim 10~{\rm cm^{-3}\,K}$) support the inference that
previous  COS detections of broad, shallow \OVI\ and \lya\ absorptions
are of an extensive ($\sim400$--600 kpc), hot ($T\approx10^6$~K)
intra-cloud gas which is very massive ($\geq10^{11}~M_{\odot}$).
While the warm CGM clouds cannot account for all the ``missing
baryons'' in spirals, the hot intra-group gas can, and could account
for $\sim 20$\% of the cosmic baryon census at $z \sim 0$ if this hot
gas is ubiquitous among spiral groups.
    
\end {abstract}

\section{Introduction}
\label{intro} 

The characterization of the Circumgalactic Medium (CGM) is necessary
for any detailed understanding of galaxy formation and evolution, but
its direct detection has been, so far, elusive. The theoretical case
for a massive CGM is demonstrated by the continuing high star
formation rate (SFR) in spiral galaxies \citep{binney87, chomiuk11} as
well as the detailed metallicity history in galaxies \citep*[e.g., the
  ``G dwarf problem'';][]{larson72, binney87, chiappini01}, requiring
that any successful model of galactic evolution is not a ``closed
box''.  Low-metallicity gas must be accreted by each star-forming
galaxy to explain these basic observables ($\sim 1~M_{\odot}\,{\rm
  yr}^{-1}$ for the Milky Way), but how much gas is present in the CGM
at any one time? And how much of this is accreted from outside the
system versus how much is recycled from the galaxy through the CGM?
Additionally, there exists a substantial deficiency of detected
baryons in spiral galaxies relative to the cosmic ratio of baryons to
dark matter \citep [e.g.,][]{mcgaugh00, klypin01} that seems to
require a CGM baryonic mass much  greater than the total amount in the
galaxy's disk. The direct measurement of the amount, extent,
ionization state (and thus total mass), metallicity and origin of the
multi-phase CGM \citep [a.k.a. the galactic ``halo'';][] {spitzer56}
remains largely uncharacterized. This is due both to its low density
(and thus low emission measure) and also to its range of temperatures
(thought to be $10^4$--$10^6$~K) which makes it impossible to detect
in emission  beyond a few kpc above galactic disks using current
instruments. 

For our own Galaxy, the detection of some small amount of CGM gas has
been made possible by observing the so-called high velocity clouds
(HVCs) using \HI\ 21-cm emission \citep*{wakker97, wakker01,
  putman12}, but the number of baryons in HVCs is not
substantial. Recently, various methods for determining, or at least
bracketing HVC distances, have found that these clouds are only a few
kpc away \citep{bland99, putman03, wakker07, lehner11}.  Many remain
without distance estimates, leading to suggestions that a subset of
HVCs are $\gg10$~kpc away and much more massive \citep{blitz99}. There
is little support for this  conjecture however \citep{putman12}.  At
present, the total infall rate of \HI\ 21-cm detected HVC mass is  an
order of magnitude short of that required to sustain the current level
of star formation in the Milky Way.  

However, it is possible to use background AGN (quasars, QSOs, BL Lacs
and Seyferts; we will use the abbreviation QSO to refer to these
various classes of AGN collectively), which have large far-UV (FUV)
fluxes ($\geq 10^{-15}~{\rm ergs\,cm^{-2}\,s^{-1}}\,$\AA$^{-1}$) to
probe the full extent of the CGM in both the Milky Way  and in other
galaxies. In our own Galaxy's halo, the discovery of highly-ionized
HVCs \citep*{sembach95, sembach03, collins04} using UV spectroscopy of
QSOs has revealed a much larger reservoir of infalling gas
\citep*[$\sim 1~M_{\odot}\,{\rm yr}^{-1}$;][]{shull09, collins09} than
the \HI\ 21-cm HVCs. But only in some cases \citep{lehner11} can the
distance to these highly-ionized HVCs be determined, allowing their
total mass to be estimated. Nevertheless, the mass infall rate
estimated by \citet{shull09} is sufficient to fuel much of the
on-going Milky Way SFR \citep [2--$4~M_{\odot}\,{\rm yr}^{-1}$;] []
{diehl06, robitaille10, shull11}. 

But what is the origin of this infalling material? Observations of
similar clouds around other galaxies can generalize their presence to
other star-forming galaxies and provide an elementary understanding of
a galactic ecology. Using UV spectroscopy of QSOs to study our own
Galaxy's gaseous halo and that of other galaxies are nicely
complementary. For our own Galaxy, CGM clouds within a few kpc of the
disk are directly detected, and their infalling or outflowing
kinematics are readily measured, but their distances are often poorly
known. On the other hand, QSO probes of the CGM of other galaxies
detect ``warm'', photoionized and ``warm-hot'', collisionally-ionized
clouds at greater galactocentric distances, providing an
easily-measured lower bound on their  galactocentric
distances. However, in these cases, cloud kinematics are usually
uncertain. And, until recently, very few QSOs near galaxies have been
bright enough to provide sufficient targets for multiple probes of
single galaxies \citep{keeney12}. 

The advent of the Cosmic Origins Spectrograph \citep[COS;][]{green12,
  osterman11} on the {\sl Hubble Space Telescope} (\hst) has allowed
much fainter background target QSOs to be observed. This has
facilitated detailed studies of the CGM by targeting fainter QSOs
which are  projected within the virial radius of a foreground
galaxy. Before COS, the Faint Object Spectrograph (FOS) was used to
conduct a substantial ``Key Project'' survey \citep{bahcall93,
  weymann95, jannuzi98} that detected only the strongest
\lya\ absorbers ($\eqw \geq 250$~m\AA). While many of these do appear
associated with bright galaxies \citep{lanzetta95}, these absorbers
are mostly well outside the virial radius of the nearby bright
galaxies and are at distances too great to determine if somewhat
fainter galaxies are much closer to the sight line. Later the Goddard
High Resolution Spectrograph (GHRS) and the Space Telescope Imaging
Spectrograph (STIS) found many weaker absorbers \citep*{morris91,
  tripp98, penton00a, penton00b, penton04, danforth05, danforth08}
which were shown to have a much looser association with galaxies
\citep*{morris93, bowen97, impey99, penton02, penton04, wakker09}.  In
the GHRS and STIS eras there was little choice of QSO targets for high
resolution and signal-to-noise UV spectroscopy and any probes of
foreground CGM gas were almost entirely ``serendipitous''. However,
given the very small covering factor of galaxies and their CGM on the
sky, there are few detected serendipitous CGM absorbers, and, when
detected, these absorbers are most often in the outermost parts of the
CGM at close to or just beyond the virial radius. The ten-fold
increase in FUV sensitivity of COS now allows a substantial list of
possible targets, including numerous QSOs close enough on the sky to
foreground galaxies to probe the inner and outer CGM of nearby
galaxies. 

The tactic taken by several other \hst/COS observers,
as first reported in \citet{tumlinson11}, is to use the vast database
of the Sloan Digital Sky Survey (SDSS) to locate foreground galaxies
near UV-bright targets. Owing to the flux-limited nature of the SDSS
photometry and spectroscopy, this approach allows a study of the CGM
of luminous galaxies out to $z\approx0.2$--0.3. These redshifts
maximize COS detectability of both higher order \HI\ Lyman lines and
the critical \OVI\ absorption doublet, which probes so-called
``warm-hot'' gas at $T \approx 10^5$--$10^{6.3}$~K. However, the
ionization mechanism for this transition remains controversial, as
both photo-ionization \citep{dave07, tripp08} and collisional
ionization due to shocks \citep*{cen99, shull12} have been
proposed. This ambiguity makes the interpretation of CGM
\OVI\ detections more uncertain, but the use of \OVI\ absorption is
essential to fully characterize the CGM gas. 

The tactic taken by the COS Science Team (hereafter called Guaranteed
Time Observers or GTOs) is complementary to the Tumlinson
et~al. approach. We have searched for bright QSOs near on the sky to
very nearby ($z\leq0.02$) galaxies with a range of luminosities
($<0.01\,L^*$ to $L^*$) and morphologies (massive spirals to dwarf
irregulars, including star bursting systems, and low surface
brightness galaxies). In this way we have probed the CGM of a variety
of late-type galaxies. Although the sample has limited size and is
somewhat biased in its target selection, we have nevertheless
constructed it with the goal of characterizing the CGM of late-type
galaxies of various luminosities and morphologies for input into
models of galactic evolution. At low redshift the diagnostic
absorption lines of low ions \SiII, \CII\ and \SiIII\ and the high
ions \SiIV\ and \CIV\ as well as the \HI\ \lya\ line are available
within the COS FUV bandpass (1150--1800\AA) for detection and
study. While \OVI\ absorption is more controversial, these FUV low-
and high-ionization metal lines have strengths that are well-modeled
for the most part assuming photo-ionization provided by the
extragalactic ionizing radiation field. This extragalactic background
has been well-characterized locally \citep{shull12, haardt12} and so
can be used to model warm ($T\sim10^4$~K) CGM clouds and the CGM in
general. Throughout this paper we will refer to these clouds as ``warm
CGM'' absorbers, an observational definition based originally on the
popular three-phase interstellar medium model. A confusion in
terminology has arisen recently when theoretical modelers refer to any
gas at less than the virial temperature of a system ($T\sim10^6$~K for
a massive galaxy) as ``cold''.  It is quite plausible that ``cold
accretion'' \citep*{keres09a, keres09b} could consist largely of what
we term here ``warm'' CGM clouds. 

Using a sample of very nearby galaxies also offers the possibility of
in-depth scrutiny of the host galaxy. For example, deep \Ha\ imaging
for star formation rates, long-slit emission-line spectroscopy for
galaxy rotation curves, and \HI\ 21-cm imaging spectroscopy for
rotation curves and to search for extra-planar \HI\ emission can all
be brought to bear when the targeted galaxy is at $z\leq0.02$. Galaxy
metallicity measurements provided by \HII\ region spectroscopy and/or
Lick absorption line indices help constrain absorber metallicity and
provide important limits for successful physical models of CGM clouds
created within the galaxy either as ``fountains'' or unbound
``winds'';  i.e., unless there are very nearby galaxies with higher
luminosity (and potentially higher metallicity), $Z_{\rm abs} \leq
Z_{\rm gal}$ can be assumed as a useful constraint on models of CGM
clouds. It is even possible in some cases at low-$z$ to infer whether
the gas is infalling or outflowing by using internal extinction
patterns across the galaxy disk and assuming that any outflow or
infall is largely vertical to the galaxy's disk
\citep*{stocke10}. These complementary studies of the host galaxy to
the absorbers are not possible in such detail even at $z\sim0.3$.

For the purposes of this paper, we define the CGM as the gas within
the virial radius of the galaxy without reference to its dynamical
state; i.e., infalling from outside the galaxy \citep[extragalactic
  ``cold accretion'';][]{keres09a, keres09b, kacprzak10}, infalling
after first being outflowing \citep[i.e., a ``galactic fountain'';][]
{shull09, lehner11},  outflowing but bound \citep [also a galactic
  fountain;][] {keeney05}, or outflowing and unbound \citep [a
  ``galactic wind'' as advocated for many systems by][among others]
{lehnert96, heckman00, shapley03, weiner09, martin12}. While
absorption-line spectroscopy against the continuum source produced by
a starburst galaxy nucleus can determine infall versus outflow
unambiguously, the location of the absorbing material relative to the
galaxy potential is not well-determined. Thus, whether outflowing gas
escapes into the IGM is not well-determined either. In the case of a
QSO sight line passing close to a foreground galaxy, the impact
parameter provides a lower limit on the physical distance of the
absorbing gas and the radial velocity difference allows a reasonable,
but not fully unambiguous,  determination of  whether this gas escapes
from the galaxy into the IGM \citep [see, e.g.,][]{tumlinson11}.
Thus, both techniques, observing absorption against  starburst
continua and observing QSO/galaxy pairs, have significant limitations
as well as unique advantages.

In this paper we present a two-pronged approach to addressing the
nature of the CGM using UV spectroscopy of background QSOs near on the
sky to foreground galaxies. Two samples of absorber/galaxy pairs are
investigated:  targeted detections made using COS and serendipitous
detections from the STIS/\fuse\ archives.

In Section 2 we present a modest-sized sample of 11 QSO/galaxy pairs
(11 QSO targets probing 10 foreground galaxies) targeted for
observation with COS because the QSO sight line passes within the
virial radius of the foreground galaxy. The targeted galaxies are all
late-type and at very low redshift ($z\leq 0.02$). At these redshifts
the COS FUV spectra are very sensitive to \lya, as well as covering
the wavelengths of low and high ions plausibly photoionized, ranging
from \CII\ 1335 \AA\ and \SiIII\ 1206 \AA\ to \SiIV\ 1393, 1403
\AA\ and \CIV\ 1548, 1552 \AA. While \NV\ is present within the
spectral coverage, this doublet is usually very weak, and the
wavelengths of \OVI\ are not covered by these spectra so that little
or no information is available for the highest ions likely to be
present in these clouds. The detection of the low and high ions listed
above allows an estimate of the basic physical structure of the clouds
(e.g., density, size, ionized fraction and total mass) from standard
photo-ionization modeling while leaving the amount of hotter gas
poorly constrained. In addition, with only the first transition of the
Lyman series detectable at the redshifts of the foreground galaxies,
the column density of \HI\ is sometimes not well-constrained either
since CGM \lya\ is usually saturated,  creating some uncertainty in
absorber metallicity.  This small sample was chosen for observation by
the GTOs in the first three years of COS operation. The observational
details of the individual sight lines, their absorber detections and
photo-ionization modeling of the absorbing gas can be found in a
companion paper \citep [][Paper 2 hereafter] {keeney13}. Section 2
summarizes these targets and observations.

In Section 3 we present the analysis of a serendipitous sample of
QSO/galaxy pairs, again focusing on those sight lines which pass
within the virial radius of a foreground galaxy. In this case we have
used the sample of $\sim500$ \lya\ absorbers with ${N_{\rm H\,I}} \geq
10^{13.0}~{\rm cm}^{-2}$ (hereafter all column densities are quoted in
cm$^{-2}$) found in the high-resolution STIS FUV spectra of QSOs
\citep[DS08 hereafter] {danforth08}.  Since none of these QSOs was
chosen for observation due to the presence of a foreground galaxy, all
QSO/galaxy pairs found in this sample are serendipitous. Galaxy
catalogues compiled from large-angle spectroscopic surveys of galaxies
(e.g., SDSS and 2dF) were cross-correlated with the STIS sight line
locations to find $\sim700$ galaxies $\leq1$~Mpc from these sight
lines and foreground to the QSO.  Because the serendipitous absorber
sample was required to have both STIS and {\sl Far Ultraviolet
  Spectroscopic Explorer} (\fuse) spectroscopy, information on
\OVI\ absorption in CGM clouds is available in most
cases. \fuse\ spectra also yield coverage of the higher-order Lyman
lines and thus to curve-of-growth $N_{\rm H\,I}$ values;
photo-ionization models have been constructed for some of these
absorbers in the literature (e.g., Tripp et al. 2002; Tumlinson
et~al. 2005). New modeling of the serendipitous absorbers with several
metal-line detections are presented in Paper 2. A summary of results
of the photo-ionization modeling  of the metal-line absorbers in both
the serendipitous and targeted samples is presented in Section 4.

\begin{deluxetable*}{lcllrcc}

\tablecolumns{7}
\tablewidth{0pt}

\tablecaption{Summary of {\sl HST}/COS Observations
\label{tab:COS}}

\tablehead{\colhead{Target} & \colhead{$z_{\rm em}$\tablenotemark{a}} & \colhead{Grating} & \colhead{Obs. Date} & \colhead{$t_{\rm exp}$} & \colhead{$F_{\lambda}$\tablenotemark{b}} & \colhead{$\langle {\rm S/N} \rangle$\tablenotemark{c}} \\ & & & & \colhead{(s)} & \colhead{(FEFU)} }

\startdata
1ES~1028+511      & 0.360 & G130M & 2011~May~01 & 14652 & 3.1 & 21 \\
                  &       & G160M & 2011~May~10 & 14607 & 2.3 & 13 \\
1SAX~J1032.3+5051 & 0.173 & G130M & 2011~Oct~15 & 11387 & 1.2 & 13 \\
                  &       & G160M & 2011~Oct~23 & 11342 & 0.8 & ~8 \\
FBQS~J1010+3003   & 0.256 & G130M & 2011~May~19 & 10797 & 3.1 & 18 \\
                  &       & G160M & 2011~May~21 & 10752 & 4.6 & 11 \\
HE~0435--5304     & 0.425 & G130M & 2010~Apr~13 &  8373 & 2.5 & 15 \\
                  &       & G160M & 2010~Apr~13 &  8936 & 2.0 & 11 \\
                  &       & G285M & 2010~Mar~31 &  4286 & 0.9 & ~2 \\
HE~0439--5254     & 1.053 & G130M & 2010~Jun~10 &  8403 & 4.6 & 17 \\
                  &       & G160M & 2010~Jun~10 &  8936 & 4.1 & 12 \\
                  &       & G285M & 2010~Mar~28 &  4316 & 2.2 & ~4 \\
PG~0832+251       & 0.330 & G130M & 2011~Apr~19 &  6135 & 4.2 & 16 \\
                  &       & G160M & 2011~Apr~19 &  6758 & 2.1 & 14 \\
PMN~J1103--2329   & 0.186 & G130M & 2011~Jul~05 & 13342 & 2.4 & 20 \\
                  &       & G160M & 2011~Jul~06 & 13297 & 1.9 & 12 \\
RX~J0439.6--5311  & 0.243 & G130M & 2010~Feb~07 &  8177 & 4.3 & 19 \\
                  &       & G160M & 2010~Feb~07 &  8934 & 3.1 & 11 \\
                  &       & G285M & 2010~May~26 &  4286 & 1.1 & ~2 \\
SBS~1108+560      & 0.767 & G130M & 2011~May~12 &  8388 & 0.2 & 16 \\
                  &       & G160M & 2011~May~12 &  8850 & 4.8 & 14 \\
SBS~1122+594      & 0.852 & G130M & 2010~Nov~07 &  9875 & 2.3 & 14 \\
                  &       & G160M & 2010~Nov~07 & 10462 & 2.9 & 13 \\
                  &       & G285M & 2010~Nov~08 & 10048 & 2.0 & ~6 \\
VII~Zw~244        & 0.131 & G130M & 2009~Sep~24 &  8866 & 8.4 & 31 \\
                  &       & G160M & 2009~Sep~24 &  6349 & 6.9 & 18 
\enddata

\tablenotetext{a}{The emission line redshift of the QSO as listed in the NASA Extragalactic Database (NED), except for HE~0435--5304, whose redshift ($z=0.425$) was measured from its coadded COS spectrum (NED lists $z=1.231$ for this QSO).}
\tablenotetext{b}{Continuum level as measured at 1250, 1550, and 2800~\AA\ in the coadded G130M, G160M, and G285M spectra, respectively.  Flux levels are listed in femto-erg flux units (FEFUs), where 1 FEFU = $10^{-15}~{\rm ergs\,s^{-1}\,cm^{-2}\,\mbox{\AA}^{-1}}$.}
\tablenotetext{c}{Median signal-to-noise ratio per resolution element in the grating passband, as measured by rms continuum deviations in the coadded spectra.} 

\end{deluxetable*}

Some new results on \OVI\ absorbers will also be presented in Sections
3 and 4.  Since \citet{stocke06} used these same absorber and galaxy
samples to investigate the galaxy environment of \OVI\ absorbers, this
paper will not add much new information to what has already been
published  previously. Recent work on \OVI\ absorbers by
\citet{prochaska11a} finds similar results to \citet{stocke06} in
closely associating the majority of \OVI\ absorbers with sub-$L^*$
galaxies, but both studies include only modest-sized absorber samples,
which largely overlap.  Due to its high ionization state and its large
$f$-value, \OVI\ 1032~\AA\ is a sensitive probe of very diffuse
photo-ionized gas \citep[$U\geq 10^{-1.5}$ or over-densities $\Delta_b
  \leq30$;][]{dave99, schaye01} or collisionally-ionized gas at $T\geq
10^5$~K. As such it provides our current best estimates for the spread
of metals away from galaxies: $\sim800$ kpc from $L^*$ galaxies and
$\sim450$ kpc from $0.1\,L^*$ galaxies \citep{stocke06}. However, all
of these \OVI\ results are based on quite small sample sizes which
will be enlarged soon using COS spectra. \OVI\ absorption shifts into
the COS band at $z\gtrsim 0.12$, which requires much deeper galaxy
survey work than what is used here  to further constrain the spread of
metals.
   
The GTO team is in the process of cataloguing all intervening
\lya\ and metal-line absorbers (especially \OVI) in COS GTO spectra
(C.W. Danforth et~al., in prep) as well as conducting a wide and deep
galaxy survey around each GTO sight line (B.A. Keeney et~al., in
prep).  Therefore, in Section 4 of this paper we combine the COS GTO
``targeted'' QSO/galaxy sample with a STIS-defined ``serendipitous''
QSO/galaxy sample to obtain  a  first look at CGM clouds, their
physical properties, masses and uncertain kinematics. 

In Section 5 we discuss the implications of these results for the
baryon census in spiral galaxies and galaxy groups and for galactic
chemical evolution. We also present the prospects for a better
understanding of the CGM of nearby galaxies which will be possible
when all currently available COS UV spectroscopy (GTO spectra as well
as those of \hst\ GOs [Guest Observers]) has been fully analyzed and
when detailed foreground galaxy spectroscopy near all COS sight lines
has been completed. The Summary Section lists our most important
results.

Throughout this paper we use the standard cosmological model with
$H_0 = 70.4~\kms\,{\rm Mpc}^{-1}$, $\Omega_{\Lambda}=0.727$,
$\Omega_{\rm m}=0.273$, and $\Omega_{\rm b} = 0.0455$
\citep{larson11}.

\section{The COS GTO QSO/Galaxy ``Targeted Survey''}

With a portion of the orbits allocated to the COS GTO Team, we have
conducted a modest-sized survey of the CGM of very nearby, late-type
galaxies. These observations were planned so as to obtain a peak
signal-to-noise ratio ${\rm (SNR)}\sim15$--20 per resolution element
of 18~\kms. For each target both a G130M and a G160M exposure were
obtained (see Table 1 for observing log).  We limited the total
exposures at the high end to avoid a re-pointing due to the South
Atlantic Anomaly. A few targets were observed with the G285M grating
for 1--2 orbits only to determine whether strong \MgII\ absorption was
present as might be expected for higher $N_{\rm H\,I}$ systems. None
were detected. Because of the very low-$z$ of the foreground galaxies
targeted, the expected location of \lya\ is close to the peak of the
COS detector + grating sensitivity. The low redshift of the target
galaxy also keeps the \CIV\ doublet in a G160M spectral region of
relatively high sensitivity. Thus, good measurements of line strengths
have been obtained in all cases for the low and high metal ions so
that viable photo-ionization models of CGM clouds can be
well-constrained. However, by targeting the CGM of very low-$z$
galaxies, only one transition of the Lyman series is present in the
COS bandpass which makes \HI\ column densities uncertain. Because the
important \OVI\ doublet is absent from the observed bandpass,
photo-ionization modeling of the absorbing clouds depends on either
\SiII, \SiIII\ and \SiIV, or \CII\ and \CIV.  A summary of the
\hst/COS observations in the GTO program on QSO/Galaxy Pairs is shown
in Table 1. A companion paper (Paper 2) will present more  details of
the observations, data analysis and photo-ionization modeling of these
absorbers.  For a specific example of the data handling and analysis
and the detailed procedure for the  photo-ionization modeling of these
CGM clouds, see the description of the three sightlines surrounding
the low-$z$ galaxy ESO~157--49 \citep{keeney12}. All observations for
this program except one  were successful in obtaining excellent
spectra near the planned SNR. Despite obtaining GALEX near- and far-UV
fluxes for SBS~1108+561, a previously undetected Lyman limit system
partially obscured \lya\ and \SiIII\ 1206~\AA\ at the redshift of the
foreground galaxy M~108 (i.e., SNR at \lya\ and \SiIII\ are much less
than the  value in Table 1). But various metal lines were detected at
M~108's redshift (see Table 2), so a detailed analysis of the two
absorption systems associated with M~108 was still possible.

\subsection{The Sample of ``Targeted'' QSO/Galaxy Pairs}

\begin{turnpage}
\begin{deluxetable*}{llccccccccl}

\tablecolumns{11}
\tablewidth{0pt}

\tablecaption{Targeted CGM Absorber Sample
\label{tab:GTOpairs}}

\tablehead{ \colhead{Target} & \colhead{Galaxy} & \colhead{$cz_{\rm abs}$} & \colhead{$cz_{\rm gal}$} & \colhead{$\rho$} & \colhead{$\phi$\tablenotemark{a}} & \colhead{$L_{\rm gal}$} & \colhead{$\log{N_{\rm H\,I}}$\tablenotemark{b}} & \colhead{$\rho/R_{\rm vir}$} & \colhead{$|\Delta v|/v_{\rm esc}$} & \colhead{Associated Metals} \\ & & \colhead{(\kms)} & \colhead{(\kms)} & \colhead{(kpc)} & \colhead{(deg)} & \colhead{($L*$)} }

\startdata
1ES~1028+511       & UGC~5740                  &   ~728  & ~649 & ~90 & \nodata & 0.007 & $13.50^{+0.18}_{-0.19}$ & 0.98--1.68 &   1.2--2.7   & none                                                  \\
1ES~1028+511       & SDSS~J103108.88+504708.7  &   ~961  & ~934 & ~25 &  $-57$  & 0.008 & $17.21^{+0.22}_{-3.20}$ & 0.26--0.46 &  0.24--0.50  & C\,{\sc iv}?                                          \\
1SAX~J1032.3+5051  & UGC~5740                  &   ~716  & ~649 & ~65 & \nodata & 0.007 & $13.07^{+0.33}_{-0.52}$ & 0.71--1.21 &  0.89--2.0   & none                                                  \\
FBQS~J1010+3003    & UGC~5478                  &   1384  & 1378 & ~48 &   $89$  & 0.011 & $17.79^{+0.11}_{-3.48}$ & 0.47--0.83 & 0.061--0.14  & none                                                  \\
HE~0435--5304      & ESO~157--49               &   1509  & 1673 & 172 &  $245$  & 0.16  & $13.76\pm0.12$          & 0.99--1.73 &   1.4--3.1   & none                                                  \\
                   &                           &   1635  &      &     &         &       & $13.91^{+0.09}_{-0.11}$ &            &  0.31--0.73  & none                                                  \\
                   &                           &   1710  &      &     &         &       & $13.58^{+0.16}_{-0.19}$ &            &  0.31--0.71  & none                                                  \\
HE~0439--5254      & ESO~157--49               &   1662  & 1673 & ~93 &   $-6$  & 0.16  & $14.38^{+0.13}_{-0.07}$ & 0.54--0.94 & 0.069--0.15  & C\,{\sc iv}, Si\,{\sc iii}/{\sc iv}                   \\
HE~0439--5254      & ESO~157--50               &   3849  & 3874 & ~88 &  $177$  & 0.53  & $14.04^{+0.08}_{-0.06}$ & 0.40--0.60 &  0.11--0.19  & C\,{\sc iv}                                           \\
PG~0832+251        & NGC~2611                  &   5227  & 5226 & ~53 &   $93$  & 0.63  & $18.45^{+0.14}_{-0.20}$ & 0.23--0.34 & 0.004--0.006 & many low + high ions                                  \\
                   &                           &   5425  &      &     &         &       & $15.01^{+2.29}_{-0.24}$ &            &  0.71--1.2   & C\,{\sc ii}/{\sc iv}, Si\,{\sc ii}/{\sc iii}/{\sc iv} \\
PMN~J1103--2329    & NGC~3511                  &   1194  & 1114 & 112 &   $97$  & 0.88  & $14.51^{+3.71}_{-0.10}$ & 0.46--0.65 &  0.34--0.54  & C\,{\sc iv}, Si\,{\sc iii}/{\sc iv}, N\,{\sc v}?      \\
RX~J0439.6--5311   & ESO~157--49               &   1671  & 1673 & ~74 &  $149$  & 0.16  & $14.41^{+0.12}_{-0.06}$ & 0.43--0.75 & 0.012--0.025 & C\,{\sc iv}, Si\,{\sc iii}/{\sc iv}                   \\
SBS~1108+560       & M~108                     &   ~665  & ~696 & ~20 &  $-81$  & 0.64  & $14.32^{+4.04}_{-0.22}$ & 0.09--0.13 & 0.087--0.14  & many low + high ions                                  \\
                   &                           &   ~778  &      &     &         &       & $14.20^{+3.99}_{-0.22}$ &            &  0.23--0.36  & C\,{\sc iv}, Si\,{\sc iii}/{\sc iv}                   \\
SBS~1122+594       & IC~691                    &   1204  & 1204 & ~32 &  $129$  & 0.091 & $17.71^{+0.35}_{-2.85}$ & 0.21--0.37 & 0.000--0.000 & C\,{\sc ii}/{\sc iv}, Si\,{\sc iii}/{\sc iv}          \\
VII~Zw~244         & UGC~4527                  &   ~712  & ~721 & ~~7 & \nodata & 0.003 & $17.75^{+0.19}_{-3.24}$ & 0.09--0.15 & 0.074--0.13  & C\,{\sc ii}/{\sc iv}, Si\,{\sc ii}/{\sc iii}/{\sc iv}
\enddata

\tablecomments{Column densities are given in units of ${\rm cm}^{-2}$.}

\tablenotetext{a}{The position angle of the QSO sight line with respect to the galaxy's major axis: $\phi \equiv {\rm PA(QSO) - PA(gal)}$.}
\tablenotetext{b}{H\,{\sc i} column density as determined from Voigt profile fits to the \lya\ line. Details of the fitting method can be found in \citet{keeney13}.}

\end{deluxetable*}
\end{turnpage}

The basic information on the COS GTO QSO/Galaxy Pairs sample is shown
in Table 2. These QSO targets were chosen to be bright enough to
provide excellent peak ${\rm SNR} \approx 15$--20 spectra to probe the
CGM of a variety of galaxy luminosities at $L<L^*$ and types within
the general category of star-forming galaxies. One $0.2\,L^*$ galaxy
(ESO~157--49) has three bright QSO targets around it
\citep{keeney12}. One of these three QSOs (HE~0439--5254) provides a
sight line past the major axis of a higher redshift, higher luminosity
($L=0.5\,L^*$) spiral, ESO~157--50. Both of these galaxies have only
weak \Ha\ emission indicating very modest star formation rates (${\rm
  SFRs} \lesssim 1~M_{\odot}\,{\rm yr}^{-1}$). Another low-SFR object,
a dwarf galaxy at $<0.01\,L^*$, SDSS~J103108.88+504708.7, has its halo
probed by two lines-of-sight at two different impact parameters.
These same two QSO sight lines (1ES~1028+511 and 1SAX~J1032.3+5051)
also probe the $0.01\,L^*$ dwarf Magellanic spiral UGC~5740 at
significantly larger impact parameters (0.6 and $0.9\,R_{\rm
  vir}$). The remaining galaxies are probed by single sight lines and
include a few late-type galaxies with much higher SFRs. Three modest
starburst galaxies (M108, NGC~3511 and NGC~2611) are probed along
their minor axes, while one star-bursting dwarf, IC~691, had a
previously-detected metal-line absorber  \citep{keeney06} close to its
minor axis, which we re-observed with COS. The very low surface
brightness (LSB) galaxy UGC~4527 with a very low SFR ($\leq
0.001~M_{\odot}\,{\rm yr}^{-1}$) rounds out the sample. Associated FUV
absorption has been found in every case and \HI\ + metal absorptions
definitely were detected in most (10 of 17) cases. One clear Lyman
limit system ($\log{N_{\rm H\,I}} = 18.39\pm0.06$) was found along the
minor axis of NGC~2611; otherwise the metal-bearing absorbers have
stronger detections of higher ionization lines like \SiIV\ and
\CIV. Where lower ionization metal lines were detected, \CIV\ is
generally stronger than \CII\ and/or \SiIII\ is stronger than
\SiII. Only one target (PG~0832+251) possesses a \fuse\ spectrum which
detects Ly$\beta$ and \OVI\ 1032, 1038~\AA. 

While this sample was chosen to investigate the CGM gas around
late-type  galaxies, including galaxies with a variety of
luminosities, morphologies  and SFRs, it is neither a complete nor an
unbiased sample. Several QSO  targets were selected for observation
due to  being projected close to the minor axis of a moderately
star-bursting, disk galaxy.  In these cases the disk galaxy  geometry
and the sign of the absorber/galaxy velocity difference allowed  the
determination of whether the absorber is infalling or outflowing gas
\citep{stocke10}.  Of the three Ly$\alpha$-only absorbers in the
HE~0435-5304 sightline, two are constrained to  be outflowing and one
infalling onto ESO~157-49 \citep{keeney12}. The absorber associated
with NGC~3511 in the PMN~J1103--2329 sightline is constrained to be
infalling, consistent with its low metallicity ($\sim15$\% solar; see
Section 4.3).  Both PG~0832+251/NGC~2611 and SBS~1108+560/M~108 have
absorbers whose radial velocities bracket the galaxy redshift, with
the higher redshift aborbers being infalling gas (Paper 2).   The
absorber in SBS~1122+594 is likely outflowing from the dwarf starburst
IC~691 but the extinction pattern in the galaxy is too patchy to be
certain of its orientation.

Table 2 contains the following information about this sample: (1) name
of the QSO target; (2) name of the nearby galaxy whose CGM is probed;
the heliocentric recession velocities of the absorber
($\pm10$--15~\kms; average of all species detected) in column (3) and
the galaxy ($\pm5$--10~\kms) in column (4); (5) the impact parameter
($\rho$) scaled to $h^{-1}_{70}$~kpc assuming a pure Hubble flow for
galaxy recession velocities; (6) the orientation of the QSO sight line
on the sky relative to the galaxy's major axis measured
counter-clockwise on the sky (i.e., $0\degr$ and $180\degr$ are along
the major axes while $90\degr$ and $270\degr$ are along the minor
axes). No entry in this column means that the nearby galaxy has no
well-defined major axis; (7) total $B$-band galaxy luminosity in $L^*$
units, from SDSS model magnitudes where available, otherwise from
galaxy magnitudes supplied in the NASA Extragalactic Database (NED);
(8) the logarithm of the absorber neutral hydrogen column  density in
cm$^{-2}$; (9) impact parameter in units of the virial radius ($R_{\rm
  vir}$ defined by two different scaling relations; see Section 3.1)
and (10) the absolute value of the absorber/galaxy velocity difference
($|\Delta v|$) in units of the escape velocity ($v_{\rm esc}$)
determined at the observed impact parameter using the galaxy mass
model of \citet[][see Section 3.1 for further
  discussion]{salucci07}. The range of values quoted in columns (9)
and (10)  refer to the two different definitions of $R_{\rm vir}$
described in Section 3.1 below.  As with the impact parameter, column
(10) reports the minimum of the 3D value of this quantity; it is
partially correlated with the value of the impact parameter through
the value of $v_{\rm esc}$. Column (11) lists detected metal line
absorption seen in conjunction with this absorber. As shown in Table
2, most of the galaxies probed are sub-$L^*$ but range from
$<0.01\,L^*$ to nearly $L^*$ with impact parameters ranging from
0.1--$1\,R_{\rm vir}$ in projection. 

While details of the spectral analysis, line identifications, and
photo-ionization modeling are presented elsewhere (Paper 2),  a
summary of these results are described in Sections 4.3 and
4.4. Examples of our detailed photo-ionization method and its results
are given in \citet{keeney12} for the three sightlines around
ESO~157-49. Nine of the 17 CGM absorbers found in these sight lines
contain metals with $\log{(Z/Z_{\Sun})} \approx -1$ to 0 where
photo-ionization modeling is possible.   We were surprised to find
similar, high-ionization absorbing gas associated with most of the
galaxies in our targeted sample.

\section{The STIS ``Serendipitous'' QSO/Galaxy Survey}

\subsection{The Absorber and Galaxy Samples} 

The 35 STIS sight lines used to define our ``serendipitous'' absorber
sample are as presented in DS08. The STIS sample includes very bright
FUV targets possessing both high resolution 7~\kms, moderate ${\rm
  SNR}\sim5$--15 STIS E140M spectra and also \fuse\ $\sim20$~\kms\ FUV
spectra. DS08 analyzed 650 \lya\ lines, and numerous associated
metal-lines spanning ionization states from \CII\ to \OVI; see DS08
for details concerning the line identifications of \HI, \SiIII, \CIII,
\SiIV, \CIV\ and \OVI, analyses of absorber systems, etc.  The lower
ionization detections and Lyman limit decrements associated with these
absorbers were added  after DS08 \citep [C.W. Danforth, private
  communication; see also] [] {tilton12}. Absorbers in the
serendipitous sample have higher Lyman series lines as well as
\OVI\ lines detectable; i.e., \OVI\ falls within the higher
sensitivity regions of the \fuse\ detector providing detections at
$\log{N_{\rm O\,VI}} \geq 13.2$ \citep{danforth05, stocke06} where
unobscured by Galactic absorption lines. In the current study we have
used only those $\sim500$ \lya\ absorbers with $\log{N_{\rm H\,I}}
\geq 13.0$  ($\eqw \geq 54$~m\AA), an absorber equivalent width
detectable in all 35 STIS spectra (DS08). It has been known for some
time \citep{lanzetta95, chen98} that a loose correlation exists
between \lya\ equivalent width and nearest galaxy distance \citep[see
  also][]{dave99} and so the higher column density absorbers are more
likely to be CGM clouds. Also \citet{penton02} found that at $\eqw
\leq 54$~m\AA\ an increasingly larger percentage of \lya\ absorbers
are found in galaxy ``voids'', $>3$ Mpc from the nearest known
galaxy. 

The sample of $\sim700$ galaxies at $\leq1$~Mpc from these 35 sight
lines is derived from a combined galaxy database with $>1$ million
entries last described in \citet{stocke06}, where it was used to
investigate the galaxy environments of nearby \OVI\ absorbers
discovered by \fuse. Two major changes have occurred since that last
use in 2006: SDSS DR4 has been replaced with SDSS DR8
\citep{eisenstein11} and numerous galaxies near several of these sight
lines have been catalogued in \citet{prochaska11b} and  analyzed in
\citet{prochaska11a}.  Details of the galaxy redshift database can be
found in \citet{stocke06} and \citet{penton04}. 

Even a pointed survey for galaxies as conducted by \citet {morris93},
\citet {mclin03}, or \citet {prochaska11b} has difficulties defining a
completeness limit given that both a limiting magnitude for galaxies
surveyed for redshifts as well as a maximum angular size scale for
target completeness must be defined. In this regard, both very nearby
and very distant absorbers present distinct challenges. For distant
absorbers faint limiting magnitudes (e.g., $r=19$--20 for the above
studies) require many galaxies to be surveyed with $>90$\%
completeness percentages at faint apparent magnitudes (percentage of
galaxies at each magnitude limit for which redshifts are
obtained). Further, faint galaxies are prone to misclassification. The
percentage of stars observed by mistake increases with magnitude at
least to $r\sim22$ and more galaxies are not targeted due to
misclassification at fainter magnitudes. However, even relatively
bright galaxies can be misclassified; e.g., a $17^{th}$ magnitude
galaxy  found to be the only one nearby to the strongest
\lya\ absorber in the FUV spectrum of 3C~273 was misclassified by
\citet{morris93} and only observed after a different galaxy classifier
was used to identify this object as a galaxy \citep [see][]{stocke04}.
Further, in a pointed survey care must be taken to ensure that no
galaxies are missed for observation either by being too close on the
sky to another targeted galaxy or because a galaxy falls in a sky area
in between multi-object mask setups. In this respect large-angle
galaxy surveys like SDSS and 2dF are excellent resources for this work
simply because they have fewer edges to the surveyed areas. Also, in
the absence of SDSS and/or 2dF some pointed surveys have too small a
field-of-view to obtain complete galaxy survey spectra out to a 1 Mpc
radius from a very low-$z$ absorber. Therefore, a combination of SDSS
and/or 2dF with a deeper, pointed survey is ideal, a process in which
we are now engaged for all COS GTO sight lines (Keeney et~al., in
prep.).

For this survey we have used only those regions complete to
well-defined apparent $r$-band magnitude limits in a circle around the
sight-line of radius at least 1 Mpc at the absorber's Hubble flow
distance.  For any absorber/galaxy association to be used
statistically, the galaxy's luminosity must be greater than the
completeness luminosity at its distance; i.e.,  a bright galaxy
identified as associated with an absorber is the closest at its
luminosity or higher.  In this way each potentially associated galaxy
with $L_{\rm gal}$ creates its own complete sample {\bf if} all
galaxies within 400~\kms\ and 1 Mpc of the absorber that have $L \geq
L_{\rm gal}$ have been observed and redshifts obtained. Only galaxy
samples defined by this procedure are used for statistical purposes.
To address some questions, an even more restrictive sample is used:
absorbers located in regions completely surveyed for galaxies to
$\lesssim 0.15\,L^*$ (see Section 4).

In order to investigate absorber/galaxy associations for galaxies of
differing luminosities, three luminosity bins are defined: (1)
luminous super-$L^*$ ($L> L^*$) galaxies; (2) sub-$L^*$ galaxies
($L=0.1$--$1\,L^*$)  and (3) dwarfs at $L< 0.1\,L^*$. Closest galaxies
to absorbers are determined for each luminosity bin given the
completeness constraints just described. We have used the same
luminosity bins as \citet{prochaska11a} to facilitate easy
comparisons. For the same reason, we have chosen the same ``retarded
velocity'' as \citet{prochaska11a} of $\pm400$~\kms, slightly greater
than the value we have used  previously
\citep[$\pm300$~\kms;][]{penton02}. This choice means that if $|\Delta
v| = |v_{\rm abs} - v_{\rm gal}| < 400$~\kms, the absorber and galaxy
are assumed to be at the same radial distance from us.  This value is
only slightly greater than the rotation speed of a massive galaxy. If
$|\Delta v| >$ 400~\kms, a radial distance defined by the Hubble flow
is assumed and the galaxy quickly attains a 3D space distance $>1$ Mpc
from the absorber by this formulation. As reported in
\citet{penton02}, \citet{stocke06} and \citet{prochaska11a}, the exact
choice of retarded velocity does not change the statistical results of
this work. However, it is important to note that the galaxy recession
velocities used in this study come from a variety of sources and,
therefore, have a variety of accuracies; e.g., $\pm10$~\kms\ if from
\HI\ 21-cm emission profiles; $\pm30$~\kms\ if from SDSS or \citet
           {prochaska11a}; and up to $\pm80$~\kms\ if from
           low-resolution spectroscopy obtained some time ago \citep
           [e.g.,][]{mclin03}.

Given the heterogeneous nature of the database, the magnitudes for
galaxies also have some variation in accuracy and precision. Comparing
magnitudes for the same galaxy from different sources, we find $\sigma
\approx 0.2$~ mag. In other cases magnitudes must be converted from a
different color using transformations (e.g., $B = (g+0.1) +
1.2\,(g-r)$ for SDSS model magnitudes)  as described in
\citet{penton02}. The CfA galaxy luminosity function of
\citet*{marzke94}  is adopted and sets $B^* = -19.57$. Given the
low-$z$ of our sample we  make no evolutionary or K-correction to
galaxy luminosities.

Lacking spatially-resolved spectroscopy for all these galaxies, we
infer total halo masses from galaxy luminosities only and then
calculate a virial radius as a physical quantity determined by
luminosity alone. However, when the theoretical definition of the
virial radius is folded through the Tully-Fisher relationship (i.e.,
mass-to-light ratio as a function of luminosity), $R_{\rm vir}$ is
expected to increase quite slowly with galaxy luminosity (i.e.,
$R_{\rm vir} \sim L^{0.3}$), so that any uncertainty in galaxy
luminosity does not create a large uncertainty in virial radius. Also,
the exact value of the virial radius is only {\bf indicative of} the
region over which the gravitation of a galaxy dominates the dynamics
of gas clouds in its vicinity. Theoretical models suggest that the CGM
extends to approximately the virial radius ($R_{\rm vir}$) and is
enriched with metals by supernova-driven galactic winds
\citep{stinson12, vandevoort12}, which may or may not escape the
galaxy's gravitational potential \citep[escaping winds are more likely
  for low mass galaxies;][]{cote12}.  Therefore, we have used $R_{\rm
  vir}$ to distinguish CGM from IGM absorbers (see additional support
for this statement in Section 3.2 below) and to place all CGM clouds
into a context close to scale-free with respect to galaxy mass.

We have investigated two different prescriptions for virial mass and
radius as a function of galaxy luminosity for this
paper. \citet{prochaska11a} used an expression for the virial radius
of a galaxy  based on its luminosity $L/L^*$:

\begin{equation}
R_{\rm vir} = 250\,(L/L^*)^{0.2}~{\rm kpc}.                           
\end{equation}

\noindent Compared to the other estimator of virial mass and $R_{\rm
  vir}$ used herein,  Equation (1) yields larger dynamical masses and
virial  radii. Adopting a ``halo matching'' scheme whereby an observed
galaxy luminosity function  \citep [e.g.,][] {marzke94,
  montero-dorta09} is matched with the theoretical halo mass  function
of \citet{sheth99}, somewhat smaller values of total halo mass and
$R_{\rm vir}$ are obtained. Comparisons between various estimators of
halo mass and virial radius are shown  in Figure 1, where Equation (1)
is shown as a blue line and the observed Tully-Fisher  relationship of
\citet{meyer08} is shown in red. In black and yellow lines are results
from the  halo matching schemes for the \citet{marzke94} $B$-band
luminosity function and the SDSS $r$-band luminosity function of
\citet{montero-dorta09}, respectively.  In both cases a faint-end
slope of $\alpha=-1.25$ is assumed to include LSB galaxies.  A fourth
physical size scaling of $L^{0.4}$ has been advocated by
\citet*{chen01} based only on a minimization of the spread in the
observed broad correlation between {\eqw(\lya)} and impact parameter.
While each different scaling finds different fiducial radii for
different luminosity galaxies, some scaling seems appropriate to
compare the CGMs of different luminosity/halo mass galaxies. 

\begin{figure}[!t]
\epsscale{1.15} \centering \plotone{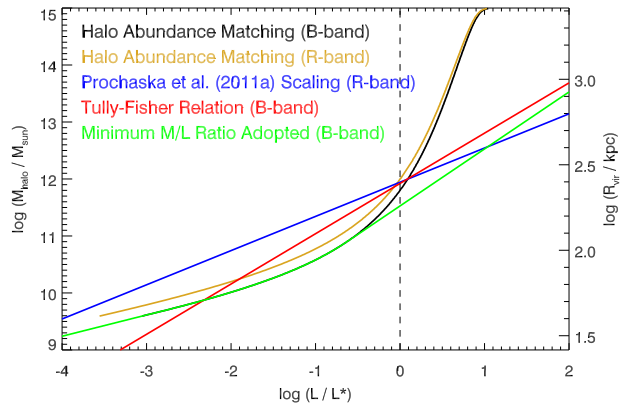}
\caption{Comparison of various galaxy mass models as a function of
  galaxy luminosity alone. The primary mass model used in this paper
  is based on a ``halo-matching'' scheme \citep{trenti10} in which a
  simulated halo abundance is matched to the $B$-band luminosity
  function of \citet{marzke94}. At $L > L^*$ the halo-matching returns
  enormous masses because individual halos encompass many galaxies.
  The green line labeled ``Minimum $M/L$ Ratio'' departs from the halo
  matching above $L=0.2\,L^*$, where the number of galaxies per halo
  begins to rapidly increase \citep{moster10}. We adopt this hybrid
  prescription for virial mass and radius, although the scaling based
  on Equation (1) \citep{prochaska11a} is used on occasion to check
  the sensitivity of the results to these assumed prescriptions.
\label{fig:halomatch}}
\end{figure}

For the halo matching schemes, $L > L^*$ galaxies are matched with
much larger halo masses than either of the scaling relations shown in
Figure 1 because these massive halos now encompass entire small groups
of  galaxies. Halos with associated stellar luminosities $L \geq
0.2\,L^*$ contain an increasing number of galaxies (Moster
et~al. 2010). Due to this multiplicity effect and in order to make a
more accurate  association between galaxy luminosity and mass we use
the halo-matching approach below $0.2\,L^*$ and assume a constant
mass-to-light ratio (50 in solar units) above that point (green curve
in Figure 1). $M_{\rm halo}/L_{\rm gal} = 50~M_{\odot}/L_{\odot}$ is
also the minimum value of this quantity predicted by halo matching
\citep{moster10,moster12}. Below $0.2\,L^*$ the halo mass-to-stellar
light ratio of smaller  galaxies rises significantly, as has been
noted in several studies \citep [e.g.,] [] {peeples11}. 

In this paper we adopt the halo masses and virial radii defined by the
green line in Figure 1,  a ``halo matching'' with the CfA $B$-band
galaxy luminosity function of \citet{marzke94} with  $\alpha =
-1.25$. Steeper faint-end slopes to the luminosity function bring the
results of the halo matching technique into close agreement with
Equation (1) near $L^*$ but fall below  these scalings (i.e.,
Tully-Fisher and Prochaska et~al. 2011a) at $L<0.2\,L^*$. Two
differences compared to Equation (1) are noted: (a) the values of the
halo mass and $R_{\rm vir}$ at $L^*$ are $\sim20$\% smaller than
Equation (1) and (b) the difference between these two models increases
below $L^*$ down to $\sim0.01\,L^*$.     For the present work this
means that Equation (1) and the halo-matching result represent likely
extremes in estimating halo masses and virial radii from total galaxy
luminosities. While we have used  both prescriptions to analyze the
STIS serendipitous absorber-galaxy sample, in the discussion below we
will present results based on the halo-matching technique, commenting
on any differences which occur if Equation (1) is used.
  
Because the escape velocity is calculated using a \citet{salucci07}
mass model in which the total halo mass determines a core radius, the
escape velocities in Tables 2--5 also depend on the adopted model from
Figure 1. While an NFW profile is somewhat ``cuspier'' than the
\citet{salucci07} model we have assumed, this difference has little
effect because we do not probe either the targeted  or serendipitous
absorber host galaxies at small enough impact parameters that the
cusp/core difference is noticeable.  And, because the galaxy density
profile is truncated at the virial radius so that the galaxy mass
doesn't exceed the halo mass,  at $\rho >$ R$_{vir}$ there is a
Keplerian fall-off in $v_{\rm esc}$. Very few absorbers are affected
by this assumed mass truncation.
 
For the galaxy sample we find that our three luminosity bins
(super-$L^*$, sub-$L^*$ and dwarfs) have median luminosities of (2.0,
0.45, 0.03)$\,L^*$ and,  from the halo matching, median halo masses
and virial radii of ($10^{11.8}$, $10^{11.2}$,
$10^{10.3}$)~$M_{\odot}$ and (230, 140, 70)~kpc, respectively. For
comparison, based on equation (1),  the median virial radius values
for the three luminosity bins are (285, 215, 125)~kpc. The Milky Way
has an estimated $B$-band luminosity midway between the median
luminosities for the super-$L^*$ and sub-$L^*$ samples; i.e., for our
Galaxy the two prescriptions yield $R_{\rm vir} = 230$~kpc and 170~kpc
from Equation (1) and from the hybrid halo-matching formalism,
respectively. In general, these values should be taken as {\it
  indicative} in defining the  extent of the CGM.

\subsection{CGM vs. IGM Absorbers}
   
In Figure 2 we show the cumulative distribution function (CDF) of
projected nearest-neighbor distances for the sample of $L>L^*$
galaxies within 1 Mpc of \lya\ absorbers. While there are many more
super-$L^*$ galaxies in the full database (203) than are shown in this
plot, the others are missing for good reasons.  Thirty-four galaxies
are $>1$~Mpc from any absorber, an even larger number  of galaxies
(83) are not included due to having another super-$L^*$ galaxy closer,
and 12 are not in regions surveyed at least to $L^*$
completeness. None of these are plotted in Figure 2 leaving a total of
74 galaxies as ``hits'' ($> L^*$ galaxies with an absorber within 1
Mpc) and 34 as ``misses'' ($> L^*$ galaxies with no absorber within 1
Mpc).  The nearest galaxy distance CDFs for the sub-$L^*$ and dwarf
galaxies are similar in shape to Figure 2  when scaled down in
distance due to their smaller virial radii (see discussion below) but
contain fewer absorber/galaxy pairs in each sample.

\begin{figure}[!t]
\epsscale{1.15} \centering \plotone{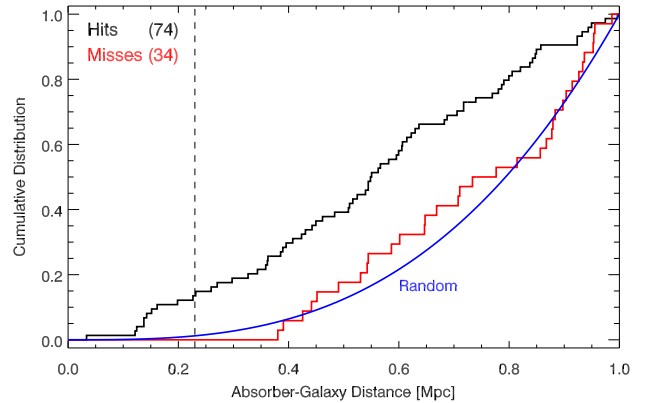}
\caption{Cumulative distribution function of absorber-galaxy distance
  for the super-$L^*$ galaxy sample with $\leq1$~Mpc absorber-galaxy
  separation. The ``hits'' are absorbers $\leq1$~Mpc from luminous
  galaxies while the misses are luminous galaxies $\leq1$~Mpc from the
  sight line with no absorber present at $|\Delta v| \leq
  400$~\kms. The blue line shows the expectation for a random
  placement of galaxies relative to the sight line. Sample numbers are
  shown at upper left.
\label{fig:cdf}}
\end{figure}

\noindent The inferences we draw from Figure 2 are: 

\begin{enumerate}
\item Most local \lya\ absorbers ($>80$\%) are projected significantly
  further from the nearest bright galaxy than the estimated virial
  radius of that galaxy: 200--250 kpc for $L^*$ and 230--285 kpc for
  $2\,L^*$, the median luminosity for the sample.  Most absorbers are
  found at 2--5 virial radii (median {\it projected} distance
  $\sim600$~kpc)  from the nearest bright galaxy, about twice as far
  as the distance between bright galaxies in our sample (Stocke
  et~al. 2006). We classify these absorbers as IGM, not CGM, because
  they cannot be assigned unambiguously to a single bright galaxy (see
  further discussion below). 

\item Bright galaxies are more centrally concentrated around absorbers
  than a random distribution (probability of this distribution
  occurring by chance is 1 part in $10^{11}$ using the K-S test).  On
  the other hand, the ``misses'' have a CDF consistent with a random
  placement relative to the absorber.  The other luminosity bins
  exhibit the same behavior at lesser but still significant
  levels. The central concentration of the ``hits'' in Figure 2
  suggests to us that while the relationship between IGM absorbers and
  galaxies is not close, there is a statistical connection; i.e., both
  galaxies and absorbers trace the same large-scale dark matter
  distribution (see \citealp{dave99} and \citealp{penton02}).
 
\item Only $\sim15$\% of local \lya\ absorbers in Figure 2 are
  projected close enough to bright galaxies to be potentially within
  the estimated virial radius. Since many other galaxies ($83+34$ in
  the full sample) are  farther away (despite being in well-surveyed
  galaxy regions) the number of CGM absorbers is an even smaller
  percentage of the total. Of course, projection effects can only
  diminish the number of CGM absorbers further still. Unfortunately,
  this means that starting with an extremely large sample of galaxies
  around QSO sight lines, only a small fraction ($\sim5$\%) are close
  enough to target sight lines to sample the CGM.

\item Given the ``hits'' and ``misses'' in Figure 2, the covering
  factor of warm gas as traced by  \lya\ absorption several virial
  radii from bright galaxies is $\sim70$\%, consistent with previous
  results \citep[e.g.,][] {penton02, stocke06}. 
This result is also consistent with a recent analysis by
\citet{prochaska11a} showing that only a fraction of local
\lya\ absorbers at $\log{N_{\rm H\,I}} \geq 13.5$ can be accounted for
as very extended (300 kpc radius), fully-covered galaxy CGMs.
\end{enumerate}

\begin{figure}[!t]
\epsscale{1.15} \centering \plotone{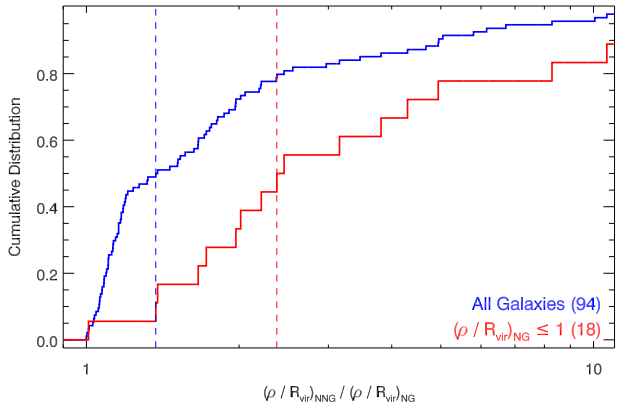}
\caption{Ratio of next-nearest galaxy distance ($\rho_{\rm NNG}$) to
  nearest galaxy distance ($\rho_{\rm NG}$) in  units of virial
  radii. The blue curve shows the CDF of this ratio for all galaxies
  $\leq1$~Mpc from the  sight line while the red curve shows only
  those galaxies with CGM absorbers   at $\rho \leq R_{\rm vir}$. The
  median of the blue CDF (blue dashed vertical line) shows that for
  all galaxy-absorber pairs the next nearest galaxy is typically only
  1.4 times farther away from the absorber than the nearest.  The red
  dashed vertical line shows that for CGM absorbers the next nearest
  galaxy is 2.4 times farther away, making the identification of a CGM
  absorber with a specific galaxy more unique (see Section 4.1 for
  further discussion). Sample numbers are shown in parentheses at
  lower right. 
\label{fig:cgmhist}}
\vspace{1ex}
\end{figure}

Using this same super-$L^*$ sample we can construct the CDF for the
ratio of next-nearest galaxy distance to nearest galaxy distance for
each absorber. This new distribution (Figure 3) has a median ratio of
1.4, meaning that for a typical absorber the next nearest galaxy is
only 40\% farther away in projection from the absorber, making it
problematical to assign the absorber to any one bright galaxy in many
cases. Further, many of these same absorbers also have lower
luminosity galaxies in close proximity, which are not accounted for in
this ratio. Taking this result together with the statistics of Figure
2, we conclude that, while most local \lya\ absorbers are associated
with galaxies in general, a typical \lya\ absorbing cloud cannot be
associated unambiguously with an individual galaxy; i.e., the
association is rather with galaxy filaments or groups. This is the
same conclusion reached by \citet{morris93} and \citet{penton02,
  penton04} and also theoretically by \citet{dave99}.
\citet{penton02} found that $\sim20$\% of all low-$z$ \lya\ absorbers
are located in galaxy voids $>3$~Mpc from any known galaxy; the
remaining $\sim80$\% are located in galaxy filaments although not so
close to any one galaxy to be considered in its CGM. The
\citet{dave99} numerical simulations of \lya\ absorbers exhibit a
similar loose correlation between \lya\ equivalent width and nearest
galaxy distance, a correlation originally discovered observationally
by \citet{lanzetta95}. On the basis of their simulations,
\citet{dave99} concluded that this observed correlation is consistent
with a filamentary origin for most local \lya\ clouds, in agreement
with our assessment here. 

However, if we restrict our scrutiny to just the potential CGM
absorbers, the CDF shown in Figure 3 has a median ratio of $\sim2.4$;
i.e., the typical CGM absorber has a next nearest neighbor galaxy of
comparable size $\sim2.4$ times farther away. For the small subset of
\lya\ absorbers that we term the CGM, there is little ambiguity
concerning the galaxy associated with most absorbers (but see Section
4.1).
 
Figure 4 shows that the CDFs of nearest neighbor distances for all
three luminosity bins exhibit similar behavior once the modest
dependence on galaxy mass is removed by scaling the impact parameter
to the virial radius of each galaxy. In Figure 4 there is a slight
tendency for super-$L^*$ galaxies to have a larger percentage of CGM
versus IGM absorbers near them although this difference is not
robust. If the Equation (1) scaling is used, the difference in the
fraction of CGM versus IGM absorbers is not statistically different
between the three luminosity classes, although the number of true
dwarf CGM absorbers is quite small: 2 of 13 by the halo matching
formalism and 5 of 22 by Equation (1).

\begin{figure}[!t]
\epsscale{1.15} \centering \plotone{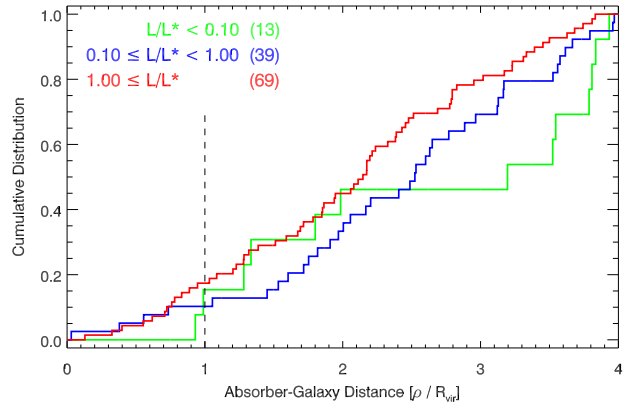}
\caption{Cumulative distribution functions of absorber-galaxy distance
  for the three different luminosity bins used in this paper. While
  there is no strong indication of a difference between these three
  distributions, the statistics for the dwarfs are modest. Sample
  numbers are shown in parentheses at upper left.
\label{fig:cdf_lum}}
\end{figure}

However, a large difference in the column density of CGM vs. IGM
absorbers is evident in Figure 5 with few CGM absorbers present in
this sample at $\log{N_{\rm H\,I}} < 14.5$. On the other hand, 50\% of
all absorbers with $\log{N_{\rm H\,I}} > 14.5$ are CGM, again
consistent with Lanzetta et~al. (1995) and subsequent work by
H.-W. Chen (e.g., Chen \& Mulchaey 2009). Among other things, this
result means that even COS snapshot spectra (${\rm SNR} < 5$ per
resolution element) will be sensitive enough to detect CGM absorbers
given a sufficient pathlength per spectrum.

\begin{figure}[!t]
\epsscale{1.15} \centering \plotone{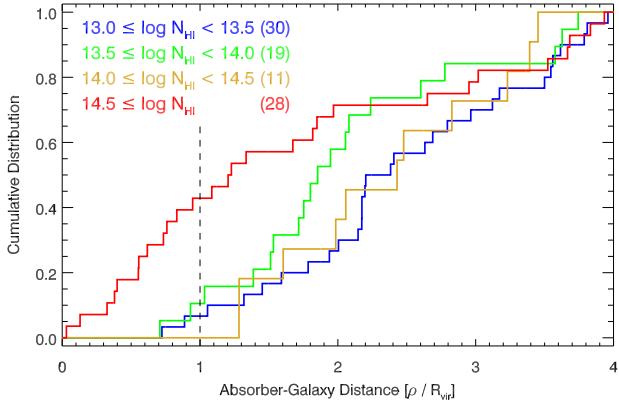}
\caption{Cumulative distribution functions of absorber-galaxy distance
  for different absorber hydrogen column densities including galaxies
  of all luminosities.  While there are a few low column density
  absorbers inside the virial radii of galaxies, most of the CGM
  absorbers, and all CGM absorbers  at $\rho \leq 0.5\,R_{\rm vir}$,
  have $\log{N_{\rm H\,I}} \geq 14.5$. Nearly half of all high-column
  density absorbers are CGM absorbers.  Sample numbers are shown in
  parentheses.
\label{fig:cdf_logN}}
\end{figure}

While we take these statistical results as ample evidence to classify
an absorber as CGM if it is closer to a galaxy than the virial radius,
projection effects can only decrease the number of true CGM
absorbers. Given $d\mathcal{N}/dz = 50$ per unit redshift for
\lya\ absorbers  with $N_{\rm H\,I} \geq 10^{13.0}$, we expect 3--5
projected IGM absorbers in our combined targeted + serendipitous  CGM
absorber sample (see Section 4.4). Those absorbers with $N_{\rm H\,I}
< 10^{14.5}$ are most likely to be projected IGM absorbers
misclassified as CGM.  An example of this distinction is the pair of
absorbers detected at very low-$z$ in the COS spectra of 1ES~1028+511
and 1SAX~J1032.3+5051  presented in Table 2 (see Paper 2 for a full
discussion).  An absorber in the 1ES~1028+511 sight line is detected
27~\kms\ and half of a virial radius away  from an $M_r \approx -14.0$
post-starburst dwarf galaxy.  But absorption near this velocity is not
detected in the other QSO sight line 33~kpc away from the first sight
line and at $\sim R_{\rm vir}$ from the dwarf. However, a second
absorber at $|\Delta v| = 79$~\kms\ is detected at comparable
equivalent width in both sight lines and so is at least 33~kpc in
extent. We identify the higher column density absorber as CGM and the
latter (detected twice) as a single projected IGM absorber, associated
with a large-scale gaseous filament in this region. Therefore, while
the CGM sample must include some projected IGM absorbers, it is likely
that most of these are at $N_{\rm H\,I} \leq 10^{14}$.

\subsection{The Covering Factor of the CGM and the Filamentary IGM in Warm Gas}

In Section 2.1 we found that at least one \lya\ absorber is detected
for each galaxy CGM targeted by our COS spectra regardless of galaxy
luminosity, implying a very high covering factor of CGM gas out to
approximately the virial radius. Low and high ionization metal-line
absorptions were usually but not always found associated with these
CGM \lya\ absorbers (10 out of 17 absorbers; see Table 2).  Similarly,
\citet{tumlinson11} found a very high covering factor of
\OVI-absorbing gas in the CGM of a large sample of low-$z$, $L>L^*$
late-type galaxies. However, to avoid any possibility of selection
bias  in either our targeted sample or in the SDSS galaxy sample of
\citet{tumlinson11} influencing the covering factor,  we employ the
CGM serendipitous sample to determine covering factors. Evidence for
some bias in the selection of the COS Targeted Sample is shown in
Figure 6,   a plot of absorber location relative to the proximate
galaxy's major axis. Several of the QSO targets were chosen to probe
gas along the minor axis of a starburst or dwarf starburst galaxy. The
subset of serendipitous absorbers found near galaxies with good
determinations of disk position angle on the sky shows no obvious
orientation preference with respect to the nearby galactic disk.

\begin{figure}[!t]
\epsscale{1.15} \centering \plotone{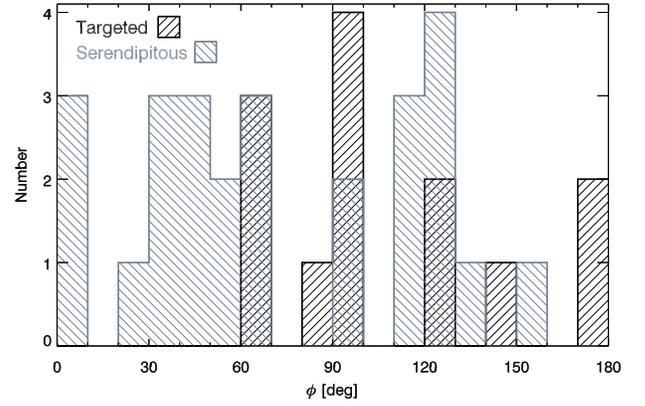}
\caption{Position angle distribution for targeted (black outlined
  histogram shaded with back-slashes)  and serendipitous (grey
  outlined histogram shaded with forward slashes) CGM absorbers, where
  such data are available through NED. The position angle is measured
  from the galaxy to the absorber relative to the galaxy's major
  axis. Thus, $\phi=0\degr$ and $180\degr$ are along the major axis
  and $\phi = 90\degr$ is along the minor axis. QSO/galaxy pairs are
  plotted only when a well-defined disk is present, for which position
  angles  are quoted in NED or SDSS. Because several of the
  COS-Targeted QSO/galaxy pairs were selected to be along the minor
  axis, the serendipitous sample is more randomly chosen with respect
  to galaxy orientation. 
\label{fig:phi}}
\end{figure}

Figure 7 shows the covering factor of \lya\ absorbing gas as a
function of projected absorber/galaxy distance in units of virial
radii for the three luminosity bins used in our serendipitous
sample. These values were computed using the ratio of ``hits'' to
[``hits'' + ``misses''] where ``hits'' are absorber/galaxy matches
within $\pm400$~\kms\ in radial velocity and within the bin of scaled
impact parameter on the abscissa. The full list of 172 ``hits'' comes
from DS08, while the 147 ``Misses'' are galaxies within each impact
parameter bin with no associated \lya\ absorber (i.e, at $|cz_{\rm
  abs} - cz_{\rm gal}| = |\Delta v| \leq 400$~\kms and $\log{N_{\rm
    H\,I}} \geq 13.0$). Because Galactic absorption lines can obscure
the presence of \lya\ absorption, we have excluded portions of the
STIS spectra around the strong Galactic absorption lines:
\SII\ 1250.6~\AA, 1253.8~\AA\ and 1259.5~\AA;  \SiII\ 1260.4~\AA;
\OI\ 1302.2~\AA; \SiII\ 1304.4~\AA;  \CII\ 1334.5~\AA; \CII$^*$
1335.7~\AA; and \SiIV\ 1393.8\AA\ and 1402.8~\AA.  While other strong
Galactic absorption occurs longward of 1450~\AA, any \lya\ obscured by
those transitions would be at a higher redshift than the $z\leq0.2$
limit imposed by this survey.  Only 10 of 157 ``misses'' are excluded
from the sample on this basis.  Errors in the Figure 7 histogram are
computed from sample size Poisson statistics \citep{gehrels86} and are
shown as color-shaded regions in each bin. 

Given the sizes of the Poisson errors, all three galaxy luminosity
bins possess similar CGM area covering factors as a function of impact
parameter once scaled by $R_{\rm vir}$.  Within the virial radius the
super-$L^*$ and sub-$L^*$ samples possess  covering factors ($C$)
consistent with $C=1$ for the inner half virial radius and $C=0.75$
for $0.5 \leq (\rho/R_{\rm vir}) \leq 1.0$. Because the galaxy surveys
available to us become incomplete for $L < 0.1\,L^*$ at $cz >
5000$~\kms, the covering factor  statistics for the dwarf sample are
poor. So while the covering factor around dwarfs may be smaller  than
for more luminous galaxies, the available statistics do not support
this conclusion at high  confidence; more data are needed. The
detection of \lya\ absorption around all five dwarfs in the COS
Targeted Sample supports a high CGM covering factor of
\lya\ absorption for dwarfs within the virial radius.

\begin{figure}[!t]
\epsscale{1.15} \centering \plotone{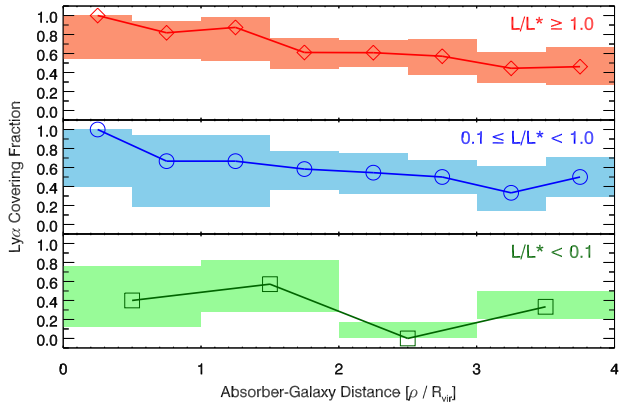}
\caption{The covering fraction ($C$) of \lya-absorbing gas for
  galaxies of various luminosities, where $C = $``hits''/[``hits'' +
    ``misses''].  The shaded regions are the Poisson errors for each
  bin. The super-$L^*$ and sub-$L^*$ samples have statistically
  indistinguishable covering factor distributions. While the dwarfs
  appear  to have somewhat lower covering factors, the statistics are
  too meager for that sample to be certain.
\label{fig:covering}}
\end{figure}

For our first attempt at modeling the ``warm'' ($T\sim10^4$~K) CGM
gas, based on Figure 7 we will assume that super-$L^*$ and sub-$L^*$
galaxies have the same high values of covering factors mentioned above
and dwarfs have $C=0.50$ inside the virial radius. 


\subsection{The Extent of Hotter Gas around Galaxies}

Since many \OVI\ absorbers are found well beyond the virial radii of
galaxies, to maximum impact parameters of $\sim3.5\,R_{\rm vir}$
\citep[$\sim800$~kpc for super-$L^*$ galaxies  and $\sim450$~kpc for
  sub-$L^*$ galaxies;][]{stocke06}, this metal-bearing gas must be
patchy. Figure 7 shows that only $\sim60$\% of the sight lines that
pass through regions at 1--$4\,R_{\rm vir}$ have detectable
\lya\ absorption. From Figure 1 in \citet {stocke06} only $\sim50$\%
of these \lya\ absorbers have detectable \OVI\ at these  distances for
a total \OVI\ covering factor of $\sim30$\%. Once impact parameters
are scaled to virial radii, there is little difference between the
covering factors we find for the super-$L^*$ and sub-$L^*$ samples. As
with the \lya\ absorption, we are less certain about the dwarfs since
the sample  of \OVI\ absorbers associated with dwarfs is very
small. Neither the previous \citet{stocke06} study, the recent
\citet{prochaska11a} study nor the current accounting use
\OVI\ absorber samples which do {\it not} have associated
\lya\ absorption. While indications are that the number of such
systems are small (DS08), the nature of these \OVI-only absorbers and
their relationship to the much more common \lya\ $+$ \OVI\ systems is
not yet clear \citep[although see][and Section 5.1 herein for a
  description of one important \OVI-only system]{savage10}.

If we assume a near unity covering factor in \OVI-absorbing gas around
star-forming galaxies of all luminosities {\it inside} $R_{\rm vir}$
(consistent with the new \citealp{tumlinson11} pointed survey)  and a
30\% covering factor from 1--$3.5\,R_{\rm vir}$ found here, then a
total $d\mathcal{N}/dz \approx 20$ per unit redshift interval for
\OVI\ absorbers at $\log{N_{\rm O\,VI}} \gtrsim 13.2$
\citep[DS08;][]{tilton12}  can be obtained only if all three galaxy
luminosity bins contribute significantly to the total cross-section. 
\citet{tumfang05} found a similar result. 
  
This result differs from the assessment  of \citet{prochaska11a} who
claim that a $\sim100$\% covering factor around sub-$L^*$ galaxies out
to 300 kpc (1.5 to 2 virial radii) can account for all \OVI\ absorbers
in the low-$z$ universe. The difference between these two conclusions
appears to be the assumed value of the covering factor. While we find
no evidence for covering factors near unity like \citet{prochaska11a}
assumed, the numbers of absorbers in these samples are still
modest. The \citet{stocke06} sample of \OVI\ absorbers in regions
surveyed for galaxies to at least $0.1\,L^*$ contains only 17
\lya\ absorber data points (9 \OVI\ detections and 8
non-detections). Using the current galaxy database, which is only
slightly enlarged from \citet{stocke06}, we have increased the
\OVI\ sample near sub-$L^*$ galaxies to 9 detections and 10
non-detections with no statistically significant change in covering
factor from our previous result. A maximum spread of metals around
galaxies of $\sim3.5\,R_{\rm vir}$ is also confirmed, except that we
find one new \OVI\ absorber at $\sim5\,R_{\rm vir}$ (see Section 4.2).
Deeper ($r\sim22$--23) galaxy survey work close to each absorber sight
line is required to reconcile the somewhat differing conclusions
concerning \OVI\ absorption.

\section{The Merged Sample of Targeted plus Serendipitous CGM Absorbers}

\subsection{Defining the Serendipitous CGM Absorbers}

In Table 3 we list the basic properties of the close, serendipitous
QSO/galaxy pairs obtained from the STIS/\fuse\ QSO sample of DS08 and
the combined galaxy redshift database. This Table includes all pairs
in our complete samples (defined in Section 3.1) projected closer than
$1.5\,R_{\rm vir}$ and includes the following information by column:
(1) Target of the STIS/\fuse\ spectroscopy;  (2) Nearest foreground
Galaxy probed by this sight line; (3) and (4) The recession velocities
of the absorber and galaxy. While the STIS absorber velocities are
accurate to a few~\kms, the galaxy velocities have poorer and variable
accuracies depending upon the source of the redshift (see Section
3.1); (5) the impact parameter of the QSO absorber/galaxy pair in kpc;
(6) the galaxy luminosity in $L^*$ units; (7) the completeness limit
galaxy luminosity in $L^*$ units for the galaxy redshift survey in and
around that sight line at the absorber distance; (8) the range of
impact parameter values in units of the virial radius as defined by
the ``halo matching'' scheme (larger number) described  in Section 3.1
and by Equation (1) (smaller number); (9) the range of absorber/galaxy
relative velocities in units of the escape velocity ($v_{\rm esc}$)
obtained assuming the range of virial radii used for the values in
column (8).  Projection effects require that the values in columns (8)
and (9) are lower limits on the three dimensional galaxy-absorber
distance and relative velocity in all cases.  The galaxy chosen in
each case as nearest is that object which is the smallest number of
virial radii away from the absorber (see discussion below). The final
column lists associated metal absorption species detected. 

\begin{turnpage}
\begin{deluxetable*}{llccccccccl}

\tablecolumns{11}
\tablewidth{0pt}

\tablecaption{Serendipitous CGM Absorber Sample
\label{tab:serendipitous}}

\tablehead{ \colhead{Target} & \colhead{Galaxy} & \colhead{$cz_{\rm abs}$} & \colhead{$cz_{\rm gal}$} & \colhead{$\rho$} & \colhead{$L_{\rm gal}$} & \colhead{$L_{\rm lim}$} & \colhead{$\log{N_{\rm H\,I}}$} & \colhead{$\rho/R_{\rm vir}$} & \colhead{$|\Delta v|/v_{\rm esc}$} & \colhead{Associated Metals} \\ & & \colhead{(\kms)} & \colhead{(\kms)} & \colhead{(kpc)} & \colhead{($L*$)} & \colhead{($L*$)} }

\startdata
3C~273         & SDSS~J122815.96+014944.1  &  ~1015 &  ~~911 & ~69 &  0.008 &   0.002  & $14.23\pm0.02$ & 0.73--1.26 &   1.4--3.1   & O\,{\sc vi}? \\
3C~273         & SDSS~J122950.57+020153.7  &  ~1585 &  ~1775 & ~80 &  0.015 &   0.004  & $15.38\pm0.34$ & 0.74--1.32 &   2.2--5.2   & C\,{\sc ii}, Si\,{\sc ii}/{\sc iii} \\
H~1821+643     & J182202.6+642139          &  36324 &  36436 & 156 &  2.0   &   1.2    & $14.17\pm0.08$ & 0.54--0.68 &  0.43--0.59  & \\
               &                           &  36415 &        &     &        &          & $13.81\pm0.16$ &            & 0.080--0.11  & \\
               &                           &  36632 &        &     &        &          & $13.13\pm0.03$ &            &  0.75--1.0   & \\
Mrk~335        & J000529.1+201336          &  ~1957 &  ~1960 & ~96 &  0.047 &   0.006  & $13.99\pm0.10$ & 0.71--1.30 & 0.027--0.066 & \\
               &                           &  ~2275 &        &     &        &          & $13.24\pm0.06$ &            &   2.8--7.0   & \\
PG~0953+414    & SDSS~J095638.90+411646.1  &  42667 &  42759 & 434 &  6.7   &   2.8    & $13.68\pm0.14$ & 1.18--1.26 &  0.39--0.43  & O\,{\sc vi}, C\,{\sc iii} \\
               &                           &  42756 &        &     &        &          & $13.35\pm0.03$ &            & 0.013--0.014 & O\,{\sc vi}, C\,{\sc iii} \\
PG~1116+215    & SDSS~J111905.51+211733.0  &  17698 &  17993 & 131 &  0.12  &   0.11   & $13.53\pm0.02$ & 0.80--1.42 &   2.3--5.5   & O\,{\sc vi} \\
               &                           &  17774 &        &     &        &          &      13.0:     &            &   1.7--4.1   & O\,{\sc vi}, N\,{\sc v}, C\,{\sc iv}, Si\,{\sc iii} \\
               &                           &  18226 &        &     &        &          & $13.21\pm0.04$ &            &   1.8--4.3   & \\
PG~1116+215    & SDSS~J111906.68+211828.7  &  41521 &  41428 & 138 &  2.9   &   0.57   & $16.35\pm0.10$ & 0.45--0.53 &  0.31--0.39  & O\,{\sc vi}, C\,{\sc ii}/{\sc iii}, Si\,{\sc ii}/{\sc iii}/{\sc iv}\\
PG~1211+143    & IC~3061                   &  ~2130 &  ~2317 & 108 &  0.19  &   0.008  & $13.42\pm0.03$ & 0.60--1.03 &   1.2--2.6   & O\,{\sc vi}\\
PG~1211+143    & SDSS~J121409.55+140420.9  &  15302 &  15309 & 136 &  1.9   &   0.081  & $15.67\pm0.35$ & 0.48--0.60 & 0.026--0.036 & O\,{\sc vi}?, N\,{\sc v}, C\,{\sc iii}/{\sc iv}, Si\,{\sc ii}/{\sc iii}/{\sc iv} \\
               &                           &  15425 &        &     &        &          & $14.13\pm0.11$ &            &  0.43--0.59  & \\
               &                           &  15605 &        &     &        &          & $13.58\pm0.04$ &            &   1.1--1.5   & \\
PG~1211+143    & SDSS~J121413.94+140330.4  &  19329 &  19334 & ~71 &  0.16  &   0.13   & $15.17\pm0.10$ & 0.41--0.72 & 0.028--0.062 & O\,{\sc vi}, C\,{\sc iii}/{\sc iv}, Si\,{\sc iii}\\
               &                           &  19467 &        &     &        &          & $13.82\pm0.05$ &            &  0.75--1.6   & O\,{\sc vi}, C\,{\sc iii}/{\sc iv} \\
PG~1216+069    & SDSS~J121930.86+064334.4  &  24125 &  24116 & 500 &  6.4   &   0.19   & $13.87\pm0.28$ & 1.38--1.48 & 0.042--0.047 & \\
PG~1216+069    & SDSS~J121923.43+063819.7  &  37091 &  37204 & ~92 &  0.77  &   0.46   & $14.75\pm0.04$ & 0.39--0.55 &  0.46--0.75  & O\,{\sc vi}?, C\,{\sc iii}, Si\,{\sc ii} \\
PG~1259+593    & SDSS~J130207.44+584153.8  &  ~~686 &  ~~662 & ~58 &  0.019 & $<0.001$ & $13.83\pm0.24$ & 0.51--0.92 &  0.23--0.53  & O\,{\sc vi} \\
PG~1259+593    & SDSS~J130101.05+590007.1  &  13808 &  13862 & 136 &  1.0   &   0.29   & $15.51\pm0.28$ & 0.54--0.75 &  0.24--0.37  & O\,{\sc vi}, C\,{\sc iii}/{\sc iv}, Si\,{\sc iii} \\
               &                           &  13940 &        &     &        &          & $14.75\pm0.38$ &            &  0.34--0.54  & O\,{\sc vi}, N\,{\sc v}, C\,{\sc iii}/{\sc iv}, Si\,{\sc iii} \\
PHL~1811       & SDSS~J215456.65-091808.6  &  15430 &  15453 & 266 &  3.0   &   0.09   & $13.79\pm0.02$ & 0.85--1.01 & 0.098--0.13  & \\
PHL~1811       & 2MASS~J21545996--0922249  &  24222 &  24223 & ~35 &  2.9   &   0.21   & $18.00\pm0.50$ & 0.11--0.14 & 0.002--0.003 & C\,{\sc ii}/{\sc iii}/{\sc iv}, Si\,{\sc ii}/{\sc iv}\\
PHL~1811       & J215506.5--092326         &  39661 &  39758 & 226 &  2.4   &   0.57   & $14.67\pm0.19$ & 0.76--0.93 &  0.41--0.55  & O\,{\sc vi}, C\,{\sc iii}\\
PHL~1811       & J215454.9--092331         &  52926 &  52793 & 351 &  6.1   &   1.0    & $14.87\pm0.03$ & 0.98--1.05 &  0.53--0.59  & O\,{\sc vi}, Si\,{\sc iii}/{\sc iv} \\
PKS~0405--123  & 2MASX~J04080654--1212494  &  24394 &  23990 & 414 &  6.0   &   0.19   & $13.76\pm0.02$ & 1.16--1.25 &   1.7--2.0   & \\
PKS~0405--123  & 2MASX~J04075411--1214493  &  28950 &  28989 & 375 &  5.5   &   0.27   & $14.64\pm0.12$ & 1.07--1.17 &  0.17--0.19  & O\,{\sc vi} \\
               &                           &  29127 &        &     &        &          & $13.32\pm0.09$ &            &  0.58--0.67  & \\
PKS~0405--123  & J040743.9--121209         &  45378 &  45718 & 195 &  3.5   &   0.69   & $13.17\pm0.07$ & 0.61--0.71 &   1.2--1.5   & C\,{\sc ii}?, Si\,{\sc ii}? \\
               &                           &  45624 &        &     &        &          & $13.46\pm0.03$ &            &  0.34--0.42  & \\
PKS~0405--123  & J040751.2--121137         &  50105 &  50065 & 116 &  8.8   &   0.82   & $16.45\pm0.07$ & 0.30--0.31 & 0.094--0.098 & O\,{\sc vi}, N\,{\sc v}, C\,{\sc ii}/{\sc iii}, Si\,{\sc ii}/{\sc iii}/{\sc iv}\\
PKS~1302--102  & NGC~4939                  &  ~3447 &  ~3110 & 295 &  6.4   &   0.39   & $13.31\pm0.07$ & 0.81--0.87 &   1.2--1.3   & \\
PKS~1302--102  & 2MASX~J13052026--1036311  &  12567 &  12759 & 225 &  3.4   &   0.058  & $13.01\pm0.12$ & 0.71--0.82 &  0.74--0.92  & \\
               &                           &  12665 &        &     &        &          & $14.83\pm0.17$ &            &  0.36--0.45  & O\,{\sc vi}, C\,{\sc iv}, Si\,{\sc ii}/{\sc iii}/{\sc iv} \\
PKS~1302--102  & 2MASX~J13052094--1034521  &  28176 &  28304 & 350 &  4.2   &   0.29   & $14.95\pm0.06$ & 1.05--1.20 &  0.57--0.69  & \\
               &                           &  28435 &        &     &        &          & $17.10\pm0.40$ &            &  0.58--0.71  & O\,{\sc vi}?, C\,{\sc ii}/{\sc iii}, Si\,{\sc ii}/{\sc iii}/{\sc iv}\\
PKS~2155--304  & 2MASX~J21584077--3019271  &  16965 &  17005 & 421 &  6.3   &   0.097  & $14.48\pm0.28$ & 1.17--1.25 &  0.17--0.19  & \\
               &                           &  17109 &        &     &        &          & $14.04\pm0.01$ &            &  0.45--0.50  & O\,{\sc vi} \\
Q~1230+011     & SDSS~J123047.60+011518.6  &  23399 &  23585 & ~54 &  0.47  &   0.18   & $15.06\pm0.40$ & 0.25--0.38 &  0.72--1.3   & O\,{\sc vi}?, C\,{\sc iii}/{\sc iv}, Si\,{\sc iii}/{\sc iv}? 
\enddata

\tablerefs{Associated metal line information is taken from \citet{danforth08}.}

\end{deluxetable*}
\end{turnpage}

Notice that there are very few pairs at $\leq0.5\,R_{\rm vir}$ but
many at $0.5\,R_{\rm vir} \leq \rho \leq 1.5\,R_{\rm vir}$ which makes
this serendipitous sample a good complement to the COS targeted sample
described in Section 2. Although the targeted sample contains very few
luminous galaxies, this sample contains quite a few (17 at $L\geq
L^*$), again making the sum total sample quite diverse with respect to
galaxy luminosity and impact parameter. These absorbers are located
relatively isotropically around their nearest associated galaxy and at
$N_{\rm H\,I}> 10^{14.5}$ as shown in Figure 6. 

We have included pairs  projected out to 1.5 virial radii to ensure
that we do not miss any potential absorber associated with an
individual galaxy's CGM because the dividing line between CGM and IGM
absorbers is somewhat arbitrary.  To address this concern, we have
used the scaling of Equation (1) to identify an alternative CGM sample
including all absorber/galaxy pairs separated at $\leq1.5\,R_{\rm
  vir}$. The sample presented in Table 3 and this new alternative CGM
sample are largely the same, but because Equation (1) yields virial
radii $\sim20$\% larger than the halo matching formalism for most
galaxy luminosities, Table 4 lists 18 potential CGM absorber/galaxy
pairs in this alternative sample not listed  in Table 3 (i.e., the
alternative sample includes all absorbers in Tables 3 and 4).  The
columns in Table 4 supply the same information as in Table 3. All but
two (the first two entries) have  impact parameters $ \rho >R_{\rm
  vir}$ by the Equation (1) scaling and only 3/18 have unambiguous
metal-line detections. Using these additional two $\rho < R_{\rm vir}$
absorber/galaxy pairs  in the analyses below does not alter the
conclusions we derive by using the sample defined by ``halo
matching''.  

\begin{turnpage}
\begin{deluxetable*}{llccccccccl}

\tablecolumns{11}
\tablewidth{0pt}

\tablecaption{Alternate CGM Absorber Sample
\label{tab:extended}}

\tablehead{ \colhead{Target} & \colhead{Galaxy} & \colhead{$cz_{\rm abs}$} & \colhead{$cz_{\rm gal}$} & \colhead{$\rho$} & \colhead{$L_{\rm gal}$} & \colhead{$L_{\rm lim}$} & \colhead{$\log{N_{\rm H\,I}}$} & \colhead{$\rho/R_{\rm vir}$} & \colhead{$|\Delta v|/v_{\rm esc}$} & \colhead{Associated Metals} \\ & & \colhead{(\kms)} & \colhead{(\kms)} & \colhead{(kpc)} & \colhead{($L*$)} & \colhead{($L*$)} }

\startdata
3C~351        & SDSS~J170615.84+604218.8  &  ~3465                  &  ~3581 & 172 &  0.24  &   0.020  & $13.52\pm0.05$ & 0.92--1.52 &  0.85--1.8   & \\
              &                           &  ~3597                  &        &     &        &          & $13.66\pm0.03$ &            &  0.12--0.25  & \\
Mrk~876       & NGC~6140                  &  ~~932                  &  ~~910 & 180 &  0.23  &   0.018  & $14.46\pm0.13$ & 0.97--1.62 &  0.17--0.36  & O\,{\sc vi}?, Si\,{\sc iii}\\
PHL~1811      & SDSS~J215517.30-091752.0  &  22032                  &  21951 & 497 &  4.7   &   0.17   & $14.37\pm0.41$ & 1.46--1.63 &  0.41--0.49  & Si\,{\sc iii}/{\sc iv} \\
PHL~1811      & J215447.5--092254         &  23310                  &  23278 & 307 &  0.85  &   0.20   & $14.94\pm0.08$ & 1.27--1.78 &  0.21--0.36  & C\,{\sc ii}/{\sc iii}/{\sc iv}, Si\,{\sc iii} \\
              &                           &  23632                  &        &     &        &          & $14.76\pm0.15$ &            &   2.4--4.0   & \\
PHL~1811      & J215450.8--092235         &  23694                  &  23623 & 235 &  0.30  &   0.20   & $14.65\pm0.11$ & 1.20--1.93 &  0.57--1.2   & \\
PKS~0312--770 & J031201.7--765517         &  17824                  &  17792 & 237 &  0.27  &   0.14   & $13.53\pm0.03$ & 1.23--2.02 &  0.27--0.56  & \\
PKS~0312--770 & J031158.5--764855         &  35466                  &  35732 & 378 &  1.6   &   0.56   & $14.17\pm0.04$ & 1.38--1.78 &   1.6--2.4   & \\
              &                           &  35813                  &        &     &        &          & $13.79\pm0.02$ &            &  0.50--0.74  & \\
PKS~2155--304 & ESO~466--032              &  ~4989                  &  ~5126 & 320 &  1.5   &   0.012  & $13.43\pm0.02$ & 1.18--1.54 &  0.79--1.2   & \\
              &                           &  ~5098                  &        &     &        &          & $13.56\pm0.02$ &            &  0.16--0.24  & \\
              &                           &  ~5166                  &        &     &        &          & $13.21\pm0.03$ &            &  0.23--0.34  & \\
PKS~2155--304 & J215846.5--301738         &  20330                  &  20226 & 330 &  0.61  &   0.14   & $13.07\pm0.03$ & 1.46--2.14 &  0.80--1.4   & \\
PKS~2155--304 & J215845.1--301637         &  31633                  &  31887 & 399 &  2.0   &   0.33   & $14.12\pm0.18$ & 1.39--1.75 &   1.5--2.1   & \\
              &                           &  31736                  &        &     &        &          & $13.47\pm0.03$ &            &  0.89--1.3   & \\
Q~1230+011    & SDSS~J123103.89+014034.4  &  ~1489\tablenotemark{a} &  ~1136 & 119 &  0.022 &   0.003  & $13.62\pm0.05$ & 1.02--1.85 &   4.4--11    & \\
Ton~28        & SDSS~J100618.16+285641.9  &  ~1067                  &  ~1362 & 166 &  0.083 &   0.003  & $14.02\pm0.06$ & 1.09--1.97 &   2.9--7.1   & 
\enddata

\tablerefs{Associated metal line information is taken from \citet{danforth08}.}
\tablenotetext{a}{This absorber is located near the galaxy group CGCG~014--054, of which the listed galaxy is the closest member.}

\end{deluxetable*}
\end{turnpage}

How unique are the host galaxy identifications given in Tables 3 and
4? There are two aspects to this question. First, an identification
would be ambiguous if a different definition of virial radius caused a
different galaxy to be identified as the host galaxy.  The entries in
Table 4 show that this is not the case because there is no individual
absorber which has different galaxies listed as associated  in Tables
3 and 4. But a second aspect of this issue is not addressed by
comparing Tables 3 and 4; viz., are there other galaxies comparably
close to the identified host galaxy but only slightly further away?
Particularly since galaxies are almost always found in groups or
clusters, other comparably close galaxies could be present in some
cases.   While it was shown in Figure 3 that the next-nearest galaxy
to a CGM absorber is typically 2.4 times farther away, there are a few
next nearest neighbors substantially closer. Table 5 addresses this
other aspect of the uniqueness issue by listing all those next nearest
galaxies potentially associated  with absorbers by being $<1.5$ virial
radii away. In some cases this ``next nearest'' galaxy is actually
slightly closer physically to the absorber but farther away in number
of virial radii; e.g., the Table 5 entry for the absorbers in the
PKS~0405--123 sight line is a much less luminous galaxy than the one
listed for these absorbers  in Table 3. 

Two groups of absorbers have multiple entries in Table 5 indicating a
small group of galaxies close to those absorbers: PG~1211+143 at
$cz=15,302$, 15,425 and 15,695~\kms\ \citep{tumlinson05} and
PG~1259+593 at $cz=13,808$ and 13,940~\kms\ \citep{richter04}. In both
of these cases the \lya\ absorption is strong and complex and the
\fuse-detected \OVI\ absorption is significantly broader than
predicted by the \HI\ and lower ionization metal lines. Also in both
cases the nearest galaxy we have listed in Table 3 is classified by
the SDSS as an early-type system based on the color discriminator
advocated by \citet{strateva01},  which uses an SDSS color cut at
$(u-r) = 2.22$. The presence of late-type galaxies comparably close to
the sight line means that even the two early-type galaxies potentially
associated with CGM absorbers are not unambiguous associations. The
only other early-type galaxy association in Table 3 is in the
PG~1116+215 sight line at $cz=41,521$~\kms, a $3\,L^*$ Sa type galaxy
138~kpc from the sight line.  However, the galaxy survey in this
direction is complete only to $0.6\,L^*$ at that distance and there
are several  late-type galaxies closer to the QSO but which lack
redshifts at this writing. In a recent paper from the ``COS/Halos''
Team, \citet{thom12} report the detection of \lya\ absorption in 12 of
16 early-type galaxies, classified as such by their low specific
star-formation rate \citep{thom12, werk12}.  However, just as with the
examples above, \citet{thom12} admit that their fields have not been
uniformly surveyed for galaxies at lower luminosities. While the
evidence for warm gas absorption around early-type galaxies is
strongly suggested by their observations, we contend that the source
and fate of that gas is still uncertain. Deeper galaxy spectroscopy
and further investigation of the absorption associated with small
groups of galaxies is necessary and important to resolve this
question.  

From this analysis we conclude that: (1) the unique identification
between an absorber and its host galaxy is reasonably robust; (2)
there is no compelling evidence for warm CGM clouds in early-type
galaxies, consistent with earlier work by \citet{chen10} and
\citet{tumlinson11} but at variance with \citet{thom12}; and (3) small
groups of galaxies contain complex absorbers in which  the
\OVI\ absorption appears broader than its associated \lya\ and lower
ionization metal lines would predict. In the targeted sample the
galaxies NGC~2611 (PG~0832+251 sightline) and NGC~3511
(PMN~J1103--2329 sightline) have other nearby, lower-luminosity
galaxies at  comparable impact parameters (Paper 2). The absorbers at
the redshifts of these galaxies have multiple velocity components, and
the one target that has (admittedly poor) {\it FUSE} spectroscopy
contains strong O~VI absorption.   Complex, multi-phase gas may be a
hallmark of spiral-rich groups of galaxies; e.g., see also a detailed
discussion of the $cz=50,105$~\kms\ absorber in the PKS~0405--123
sight line (Table 3)  in \citet{prochaska04}, \citet{savage10} and
Section 5.1.

\begin{turnpage}
\begin{deluxetable*}{llccccccccc}

\tablecolumns{11}
\tablewidth{0pt}

\tablecaption{Absorbers Possibly Associated with Galaxy Groups
\label{tab:multiples}}

\tablehead{ \colhead{Target} & \colhead{Galaxy} & \colhead{$cz_{\rm abs}$} & \colhead{$cz_{\rm gal}$} & \colhead{$\rho$} & \colhead{$L_{\rm gal}$} & \colhead{$L_{\rm lim}$} & \colhead{$\rho/R_{\rm vir}$} & \colhead{$|\Delta v|/v_{\rm esc}$} & \colhead{$\chi_{\rm ph}$\tablenotemark{a}} & \colhead{$\chi_{\rm vir}$\tablenotemark{b}} \\ & & \colhead{(\kms)} & \colhead{(\kms)} & \colhead{(kpc)} & \colhead{($L*$)} & \colhead{($L*$)} }

\startdata
3C~273         & SDSS~J123103.89+014034.4  &  ~1015                  &  ~1105 & 168 &  0.022 &   0.002  & 1.44--2.61 &   1.3--3.2   & 2.43 & $2.04\pm0.80$ \\
3C~273         & NGC~4420                  &  ~1585                  &  ~1693 & 288 &  0.59  &   0.004  & 1.28--1.89 &  0.78--1.4   & 3.60 & $1.54\pm0.52$ \\
3C~351         & NGC~6292                  &  ~3465                  &  ~3411 & 301 &  0.70  &   0.020  & 1.29--1.86 &  0.38--0.65  & 1.75 & $1.29\pm0.39$ \\
               &                           &  ~3597                  &        &     &        &          &            &   1.3--2.3   &      &               \\
PG~1116+215    & SDSS~J111905.34+211537.7  &  17698                  &  17697 & 256 &  1.1   &   0.11   & 1.01--1.37 & 0.006--0.009 & 1.95 & $1.07\pm0.34$ \\
               &                           &  17774                  &        &     &        &          &            &  0.44--0.69  &      &               \\
PG~1211+143    & NGC~4189                  &  ~2130                  &  ~2115 & 333 &  0.91  &   0.002  & 1.36--1.89 &  0.10--0.17  & 3.33 & $2.15\pm0.66$ \\
PG~1211+143    & SDSS~J121407.36+140924.8  &  15302                  &  15290 & 398 &  5.3   &   0.081  & 1.14--1.26 & 0.053--0.061 & 2.93 & $2.22\pm0.27$ \\
               &                           &  15425                  &        &     &        &          &            &  0.59--0.69  &      &               \\
               &                           &  15605                  &        &     &        &          &            &   1.4--1.6   &      &               \\
PG~1211+143    & SDSS~J121406.93+140437.9  &  15302                  &  15586 & 180 &  0.19  &   0.081  & 1.00--1.72 &   2.3--5.1   & 1.32 & $2.52\pm0.72$ \\
               &                           &  15425                  &        &     &        &          &            &   1.3--2.9   &      &               \\
               &                           &  15605                  &        &     &        &          &            &  0.15--0.34  &      &               \\
PG~1211+143    & SDSS~J121419.88+140509.8  &  19329                  &  19259 & 149 &  1.2   &   0.13   & 0.58--0.77 &  0.30--0.46  & 2.10 & $1.19\pm0.37$ \\
               &                           &  19467                  &        &     &        &          &            &  0.90--1.4   &      &               \\
PG~1259+593    & SDSS~J130033.95+585857.2  &  13808                  &  13794 & 320 &  3.8   &   0.29   & 0.98--1.13 & 0.061--0.075 & 2.35 & $1.64\pm0.29$ \\
               &                           &  13940                  &        &     &        &          &            &  0.64--0.79  &      &               \\
PG~1259+593    & SDSS~J130100.56+585804.7  &  13808                  &  13854 & 235 &  1.2   &   0.29   & 0.91--1.22 &  0.24--0.38  & 1.73 & $1.65\pm0.36$ \\
               &                           &  13940                  &        &     &        &          &            &  0.46--0.71  &      &               \\
PG~1259+593    & SDSS~J130022.13+590127.2  &  13808                  &  13926 & 358 &  1.3   &   0.29   & 1.36--1.81 &  0.75--1.2   & 2.63 & $2.46\pm0.53$ \\
               &                           &  13940                  &        &     &        &          &            & 0.089--0.14  &      &               \\
PHL~1811       & J215516.5--092408         &  22032                  &  22112 & 341 &  0.67  &   0.18   & 1.48--2.14 &  0.61--1.06  & 0.69 & $1.17\pm0.22$ \\
PHL~1811       & J215450.8--092235         &  23310                  &  23623 & 235 &  0.30  &   0.20   & 1.20--1.93 &   2.5--5.1   & 0.77 & $1.03\pm0.29$ \\
               &                           &  23632                  &        &     &        &          &            & 0.072--0.15  &      &               \\
PHL~1811       & 2MASX~J21545868--0923057  &  24222                  &  24103 & ~89 &  3.8   &   0.21   & 0.27--0.31 &  0.32--0.38  & 2.54 & $2.80\pm0.41$ \\
PKS~0405--123  & J040758.0--121225         &  28950                  &  28916 & 264 &  0.67  &   0.27   & 1.14--1.65 &  0.23--0.40  & 0.70 & $1.25\pm0.23$ \\
               &                           &  29127                  &        &     &        &          &            &   1.4--2.5   &      &               \\
PKS~1302--102  & J130525.6--103923         &  12567                  &  12579 & 313 &  0.65  &   0.058  & 1.37--1.99 & 0.088--0.16  & 1.39 & $2.20\pm0.43$ \\
               &                           &  12665                  &        &     &        &          &            &  0.63--1.1   &      &               \\
Q~1230+011     & NGC~4517                  &  ~1489\tablenotemark{c} &  ~1128 & 345 &  1.0   & $<0.001$ & 1.38--1.90 &   2.4--3.9   & 2.90 & $1.14\pm0.38$
\enddata

\tablerefs{Associated metal line information is taken from \citet{danforth08}.}
\tablenotetext{a}{The ratio of the impact parameter (in kpc) of the tabulated galaxy to that of the nearest galaxy (see Tables~\ref{tab:serendipitous} and \ref{tab:extended}): $\chi_{\rm ph} \equiv \rho/\rho_{\rm NG}$.}
\tablenotetext{b}{The ratio of the impact parameter (in virial radii) of the tabulated galaxy to that of the nearest galaxy (see Tables~\ref{tab:serendipitous} and \ref{tab:extended}): $\chi_{\rm vir} \equiv (\rho/R_{\rm vir}) / (\rho/R_{\rm vir})_{\rm NG}$.}
\tablenotetext{c}{This absorber is located near the galaxy group CGCG~014--054, of which the listed galaxy is the second closest member.}

\end{deluxetable*}
\end{turnpage}

\subsection{Characterizing the Merged CGM Absorber Sample}

Figure 8 combines the targeted and serendipitous absorber samples to
provide the most numerous warm CGM cloud sample outside the Milky
Way. In this plot of the basic observables, impact parameter and
absorber/galaxy radial velocity difference, colored symbols mark
metal-bearing absorbers, while empty symbols are absorbers having no
detected metal lines. Red filled symbols indicate the presence of low-
and high-ionization metal lines like \CII, \SiIII\ and/or \CIV\  found
within the \hst/STIS or \hst/COS bandpasses. The blue filled symbols
indicate those absorbers which contain only \OVI\ 1032,
1038~\AA\ and/or \CIII\ 977~\AA\ without the metal ions found in the
\hst\ bandpass; i.e. these absorbers have only  \fuse\ bandpass metal
absorption detected. Absorbers containing both \hst\ bandpass metal
ion absorption and also \OVI\ are colored red.  We make this
distinction because the COS targeted absorbers do not have
\fuse\ spectra to search for associated \OVI.  Absorbers for which
available data are inconclusive as to the presence of metals in the
\hst\ bandpass (poor SNR at the redshifted wavelengths of \SiII,
\SiIII, \SiIV, \CII\  and/or \CIV, or the wavelengths of these
absorptions are not covered by available spectroscopy) are represented
by symbols with question marks.  Therefore, the dividing line between
``metal-bearing'' and ``\lya-only'' absorbers is poorly defined in
this sample  due to the modest SNR of the STIS spectra, the
availability of \fuse\ spectra, the absorber $N_{\rm H\,I}$ value and
the physical cloud conditions. In both parts of Figure 8 the squares
indicate COS-targeted absorbers and the circles are the STIS
serendipitous absorbers.  The size of the symbols indicate the
luminosity of the host galaxies with the largest symbols being $L >
L^*$, the intermediate size symbols being sub-$L^*$ and the smallest
symbols are absorbers associated with dwarfs.  This plot is limited to
absorbers at $\rho \leq 1.5\,R_{\rm vir}$ using the halo matching
definition of the virial radius  (i.e., absorbers in Tables 2 and 3
only).

\begin{figure}[!t]
\epsscale{2.30} \centering \plottwo{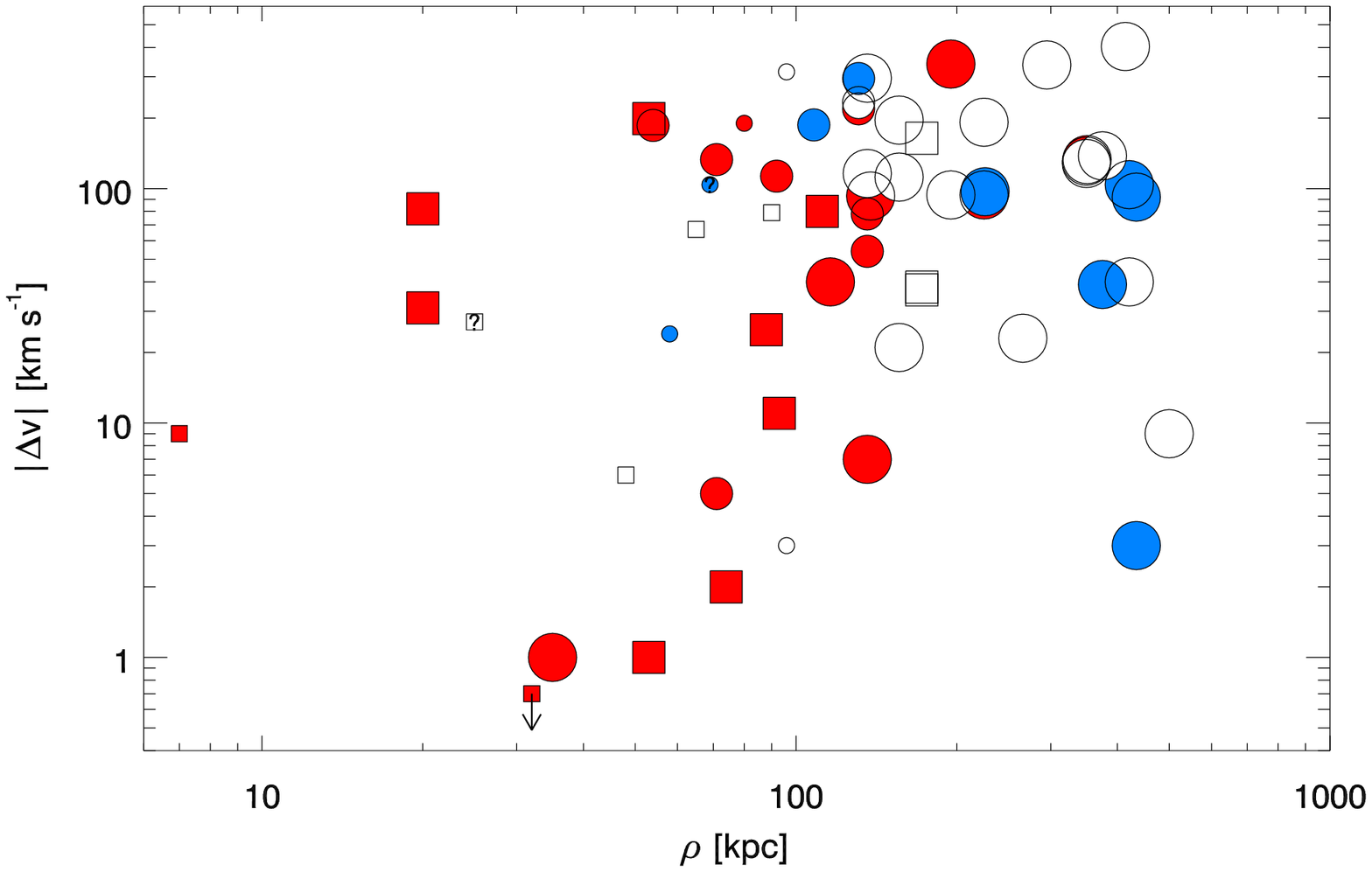}{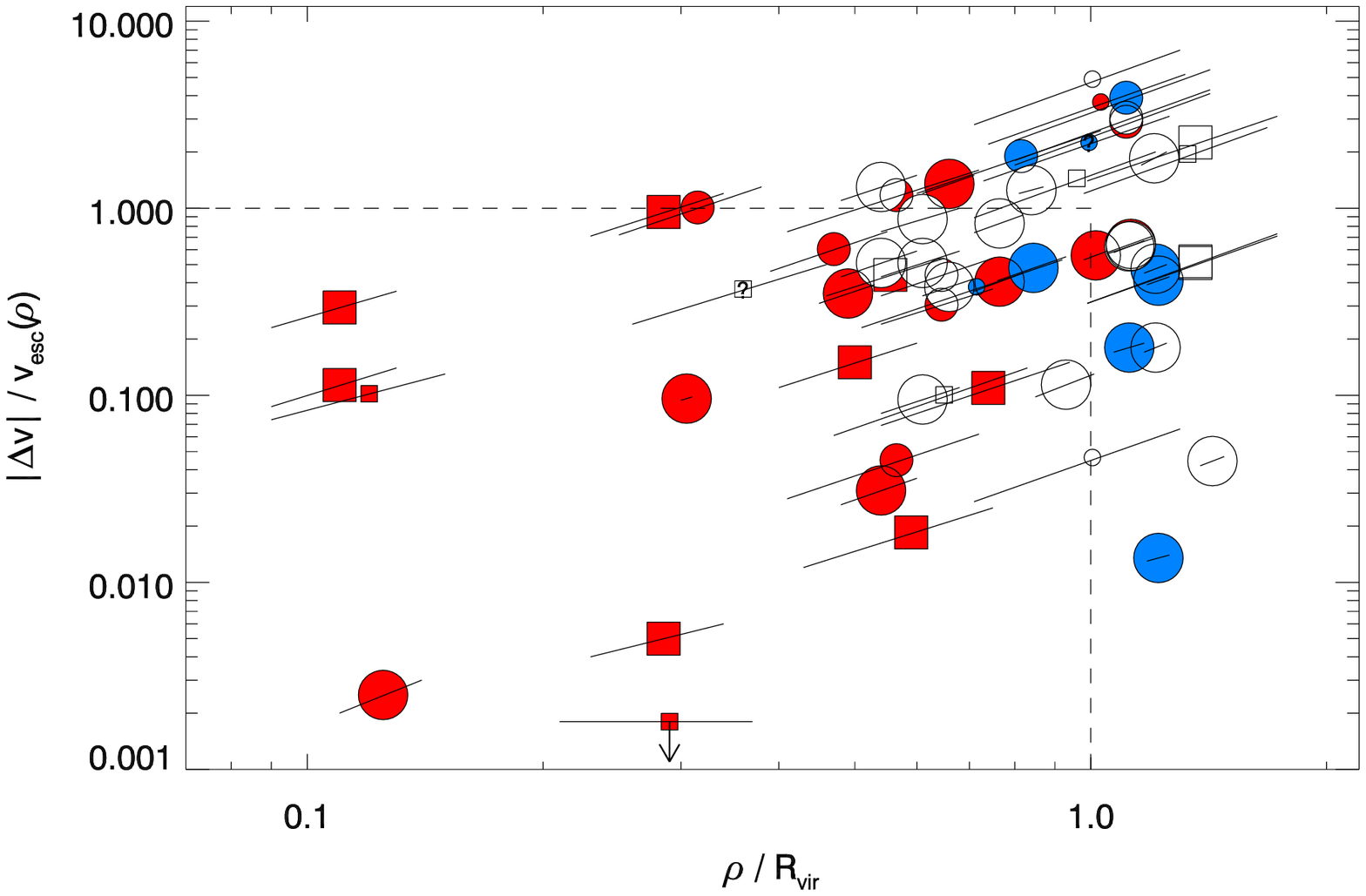}
\caption{Distribution of CGM absorbers in impact parameter and
  absorber-galaxy radial velocity difference. The squares come  from
  the COS-targeted sample and the circles are from the
  STIS-serendipitous sample. Filled symbols are metal-bearing
  absorbers, open symbols are \lya-only absorbers. Red colored symbols
  have metal lines detected in the \hst/STIS and \hst/COS bandpasses
  while blue circles are absorbers with \OVI\ or \OVI\ + \CIII\ metal
  absorption only detected in the \fuse\ bandpass.  Absorbers with
  both \OVI\ and \hst\ bandpass metal lines  are coded red.  Symbol
  size encodes the host galaxy luminosity bin; largest symbols are
  from the super-$L^*$ sample; intermediate size from the sub-$L^*$
  sample and dwarfs are the smallest symbols. The diagonal lines
  indicate the change in location in the plot by adopting a different
  virial radius prescription with the lower-left end using the
  Equation (1) scaling and upper-right end from the halo-matching
  prescription. 
\label{fig:bv1}}
\end{figure}

Figure 9 shows a variation on the now-standard plot of impact
parameter vs \lya\ equivalent width first described by
\citet{lanzetta95} and further investigated by \citet{chen01} and
\citet {chen09}. On this compressed scale the loose correlation
between hydrogen line strength and projected distance from the nearest
galaxy is not so obvious. Inside $0.5\,R_{\rm vir}$ there is little
decline in \eqw\ with distance perhaps because at these small
galaxy-absorber separations the three-dimensional distances are
dominated by projection effects. Outside $0.5\,R_{\rm vir}$ there is
also no obvious correlation but a much larger spread in $N_{\rm
  H\,I}$, although the range of impact parameters is small. More
obvious is the dichotomy of ``metal-bearing'' and ``\lya-only''
absorbers; no absorber at $\eqw \geq 500$~m\AA\ lacks metals in the
\hst\ band.  This may be a column density effect or an ionization
effect or both. 

For example, the three \lya-only absorbers in the COS target
HE~0435--5304 (overlapping open squares at $\eqw \approx 200$~m\AA)
along the minor axis of ESO~157--49 are constrained geometrically to
be outflowing gas \citep{stocke10, keeney12} and so very likely
contain metals at or somewhat below the metallicity of their host
galaxy. These three minor axis clouds may differ from the two
major-axis, metal-bearing clouds only by having lower $N_{\rm
  H\,I}$. Therefore, {\it using the current STIS and \fuse\ spectra},
we cannot constrain the absence of metal absorption lines in these and
other \lya-only absorbers to much better than a few tenths solar
metallicity.  This means that the current metal-bearing/\lya-only
distinction cannot be used as a significant discriminator for the
origin of the \lya-only clouds. Higher SNR spectra are required to
make this distinction for CGM clouds and so to determine the plausible
origin of the \lya-only CGM clouds.

\begin{figure}[!t]
\epsscale{1.15} \centering \plotone{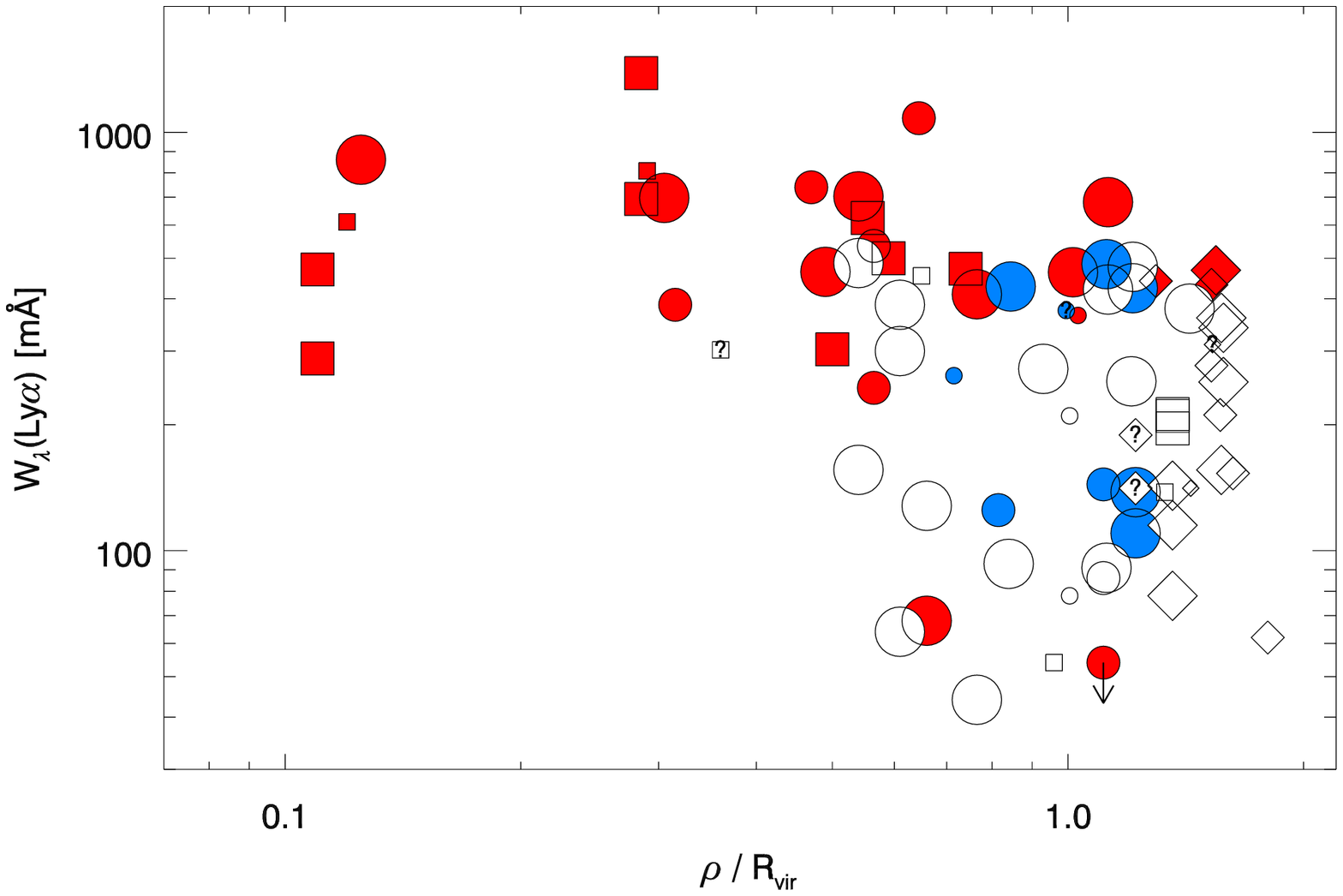}
\caption{Plot of \lya\ equivalent width versus normalized impact
  parameter in units of the virial radius. As in Figure 8 the
  absorbers with metal ions detected in the \hst\ band (e.g., \SiII,
  \SiIII\ and \SiIV) are red-filled symbols, absorbers in blue contain
  metals (\OVI\ or \CIII) detected only in \fuse\  spectra, and
  unfilled symbols have \lya-only detections. Symbols with question
  marks have unknown metal content.  As in Figure 8 the size of the
  symbol indicates the luminosity of the nearby galaxy with squares
  indicating COS targeted absorbers and circles indicating STIS
  serendipitous absorbers. Diamonds denote absorbers described in
  Table 4 and Figure 10 below.  Because of their systematically lower
  $N_{\rm H\,I}$ we cannot rule out the possibility that many of the
  \lya-only clouds have similar metallicities as the metal-bearing
  clouds.
\label{fig:lanzetta}}
\end{figure}


In a plot of the physical quantities, impact parameter and radial
velocity difference (Figure 8a), we find that, regardless of
associated galaxy luminosity, the CGM absorbers enriched with
\hst-bandpass metals (red filled symbols), whether targeted or
serendipitous, are found $\leq150$~kpc in projection from their
nearest galaxy neighbor. This maximum extent of lower-ionization (than
\OVI) metals is the same as found for low-$z$ luminous galaxies using
\CIV\ as a marker for metals \citep{chen01}. \citet{tumlinson11} found
a similar metal enrichment region in high column density \OVI\ around
late-type galaxies but they did not observe QSO sight lines beyond 150
kpc from bright galaxies. The observed radial velocity difference
between the absorbers and their nearest neighbor galaxies for the
metal-bearing absorbers in Figure 8a is also similar to the
distribution found by \citet [][see their Figure 2] {tumlinson11} in
their \OVI\ absorber survey. At larger impact parameter ($\geq
150$~kpc) and/or velocity difference ($|\Delta v| > 250$~\kms), the
fraction of metal-bearing CGM absorbers dramatically decreases, likely
a combination of increased ionization state and decreasing hydrogen
column density. 

Figure 8b plots the same absorber data as in Figure 8a but scaling the
projected impact parameter by the virial radius and the radial
velocity difference by the escape velocity from that impact
parameter. These two coordinates are now partially correlated and each
axis plots a projected value divided by a three-dimensional
(full-space) value. Thus, both coordinate values are strict lower
limits. The diagonal lines show the spread in values associated with
altering the virial radius definition; i.e., the upper-right end of
the line uses the ``halo-matching'' definition of $R_{\rm vir}$ and
the lower-left end uses Equation (1). Because the escape velocity is
computed at the minimum absorber/galaxy distance, the escape velocity
for each cloud could be somewhat less than as plotted. We expect that
many of the absorbers with normalized velocity differences in Figure
8b close to unity have sufficient 3D velocity to escape while all the
absorbers with $|\Delta v| > v_{\rm esc}$ have sufficient velocity to
escape {\it if they are outflowing from the nearest galaxy}. By
scaling the abscissa and ordinate by quantities related to galaxy
mass, there is no indication of segregation by galaxy mass; i.e.,
there is no strong evidence for different physical conditions (e.g.,
sizes or masses) in warm CGM clouds around big and little galaxies.
However, this sample is still modest in size and this conclusion
requires further work to be confirmed.

Dividing Figure 8b into three regions along the $x$-axis, we find an
12:1 ratio of ``metal-bearing'' (red and blue symbols) to \lya-only
CGM absorbers (open symbols) at $\leq0.5\,R_{\rm vir}$, a 14:11 ratio
at 0.5--$1\,R_{\rm vir}$  and a 9:12 ratio beyond $R_{\rm vir}$. Using
the alternate definition of virial  radius from Equation (1) only
changes the statistics outside $R_{\rm vir}$, where three
metal-bearing absorbers and 15 \lya-only absorbers are added, for a
total metal-bearing to \lya-only ratio of 12:27.  While there appears
to be a clear transition of physical cloud conditions that occurs  in
the 0.5--$1\,R_{\rm vir}$ regime, it is not clear what that transition
means in terms of cloud origins, metallicity and physical
structure. Higher SNR spectra certainly will assist in determining the
nature and point of origin of the lower column density \lya-only CGM
clouds. 

If we rather arbitrarily divide this sample vertically at $|\Delta v|
= 0.5\,v_{\rm esc}$,  60\% of the metal-bearing CGM absorbers are
below that line. This suggests that these metal-enriched CGM clouds
are bound to their associated galaxies although we cannot tell if
these clouds are infalling or outflowing (see Section 4.4). The
fraction of metal-bearing clouds at high relative velocities ($|\Delta
v| > 0.5\,v_{\rm esc}$) increases with $\rho$ to the point where 6 of
9 ``metal-bearing'' clouds at $\rho > R_{\rm vir}$ have  $|\Delta v| >
0.5\,v_{\rm esc}$ (Figure 8b).  The changing cloud demographic with
$\rho/R_{\rm vir}$ reinforces the use of the virial radius as an {\it
  approximate} indicator for the boundary of galactic halos.

\begin{figure}[!t]
\epsscale{2.30} \centering \plottwo{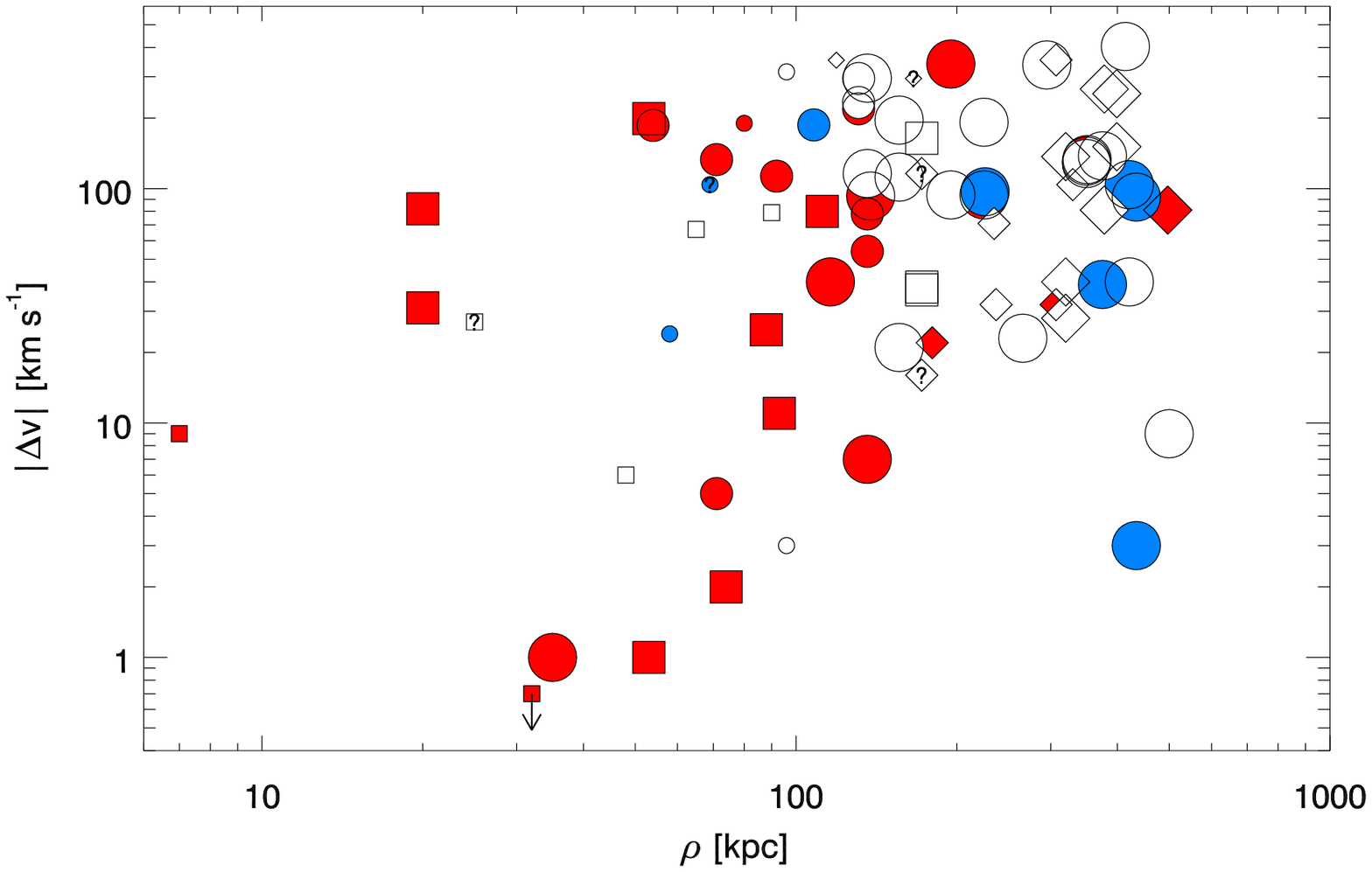}{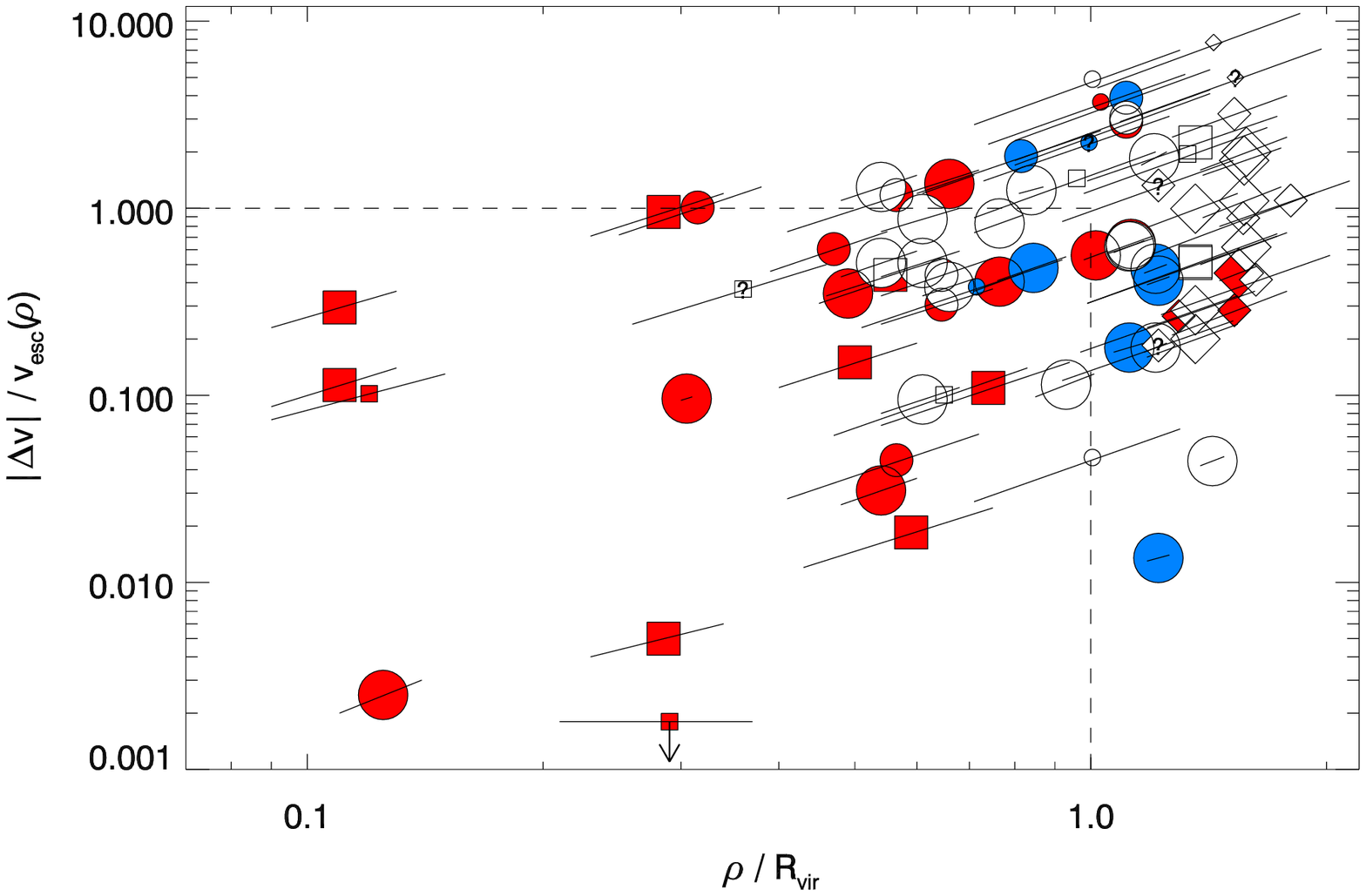}
\caption{Distribution of CGM absorbers in impact parameter scaled by
  virial radius and absorber-galaxy radial velocity difference scaled
  by escape velocity.  The symbols here are the same as in Figure 8
  except that the diamonds are absorbers from the alternate CGM sample
  listed in Table~4. 
\label{fig:bv2}}
\end{figure}

In Figure 10 the alternate CGM absorber sample (Table 4) is added as
diamond symbols to show that using the alternate definition of $R_{\rm
  vir}$ does not change the distribution substantially. The color and
size codings in Figure 10 are the same as for Figure 8.  The alternate
sample absorbers are exclusively at $\rho \gtrsim R_{\rm vir}$ and
possess $|\Delta v|$ indicative of unbound (if  outflowing) gas in
almost all cases. As shown explicitly in Figures 2, 4 and 5, there are
many more absorbers beyond the virial radii of  galaxies on the scale
of galaxy groups and large-scale filaments \citep{penton00b,
  penton04}. Otherwise, the alternate sample offers no novel insights
or trends not present in the primary sample shown in Figure 8. The
sample shown in Figure 8 (and listed in Tables 2 and 3) is  the
primary CGM cloud sample used for analyses in this paper.

\begin{figure}[!t]
\epsscale{2.30} \centering \plottwo{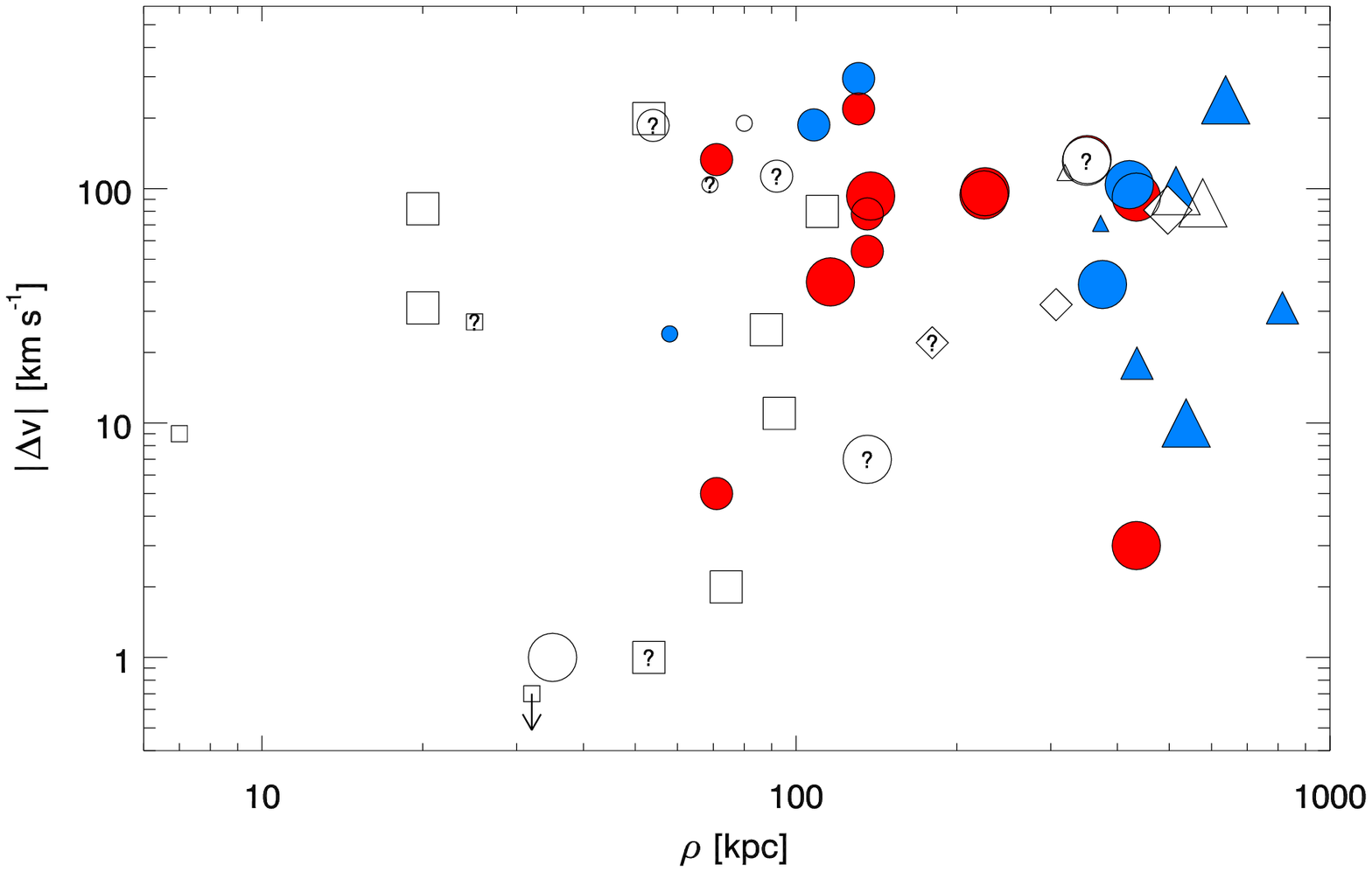}{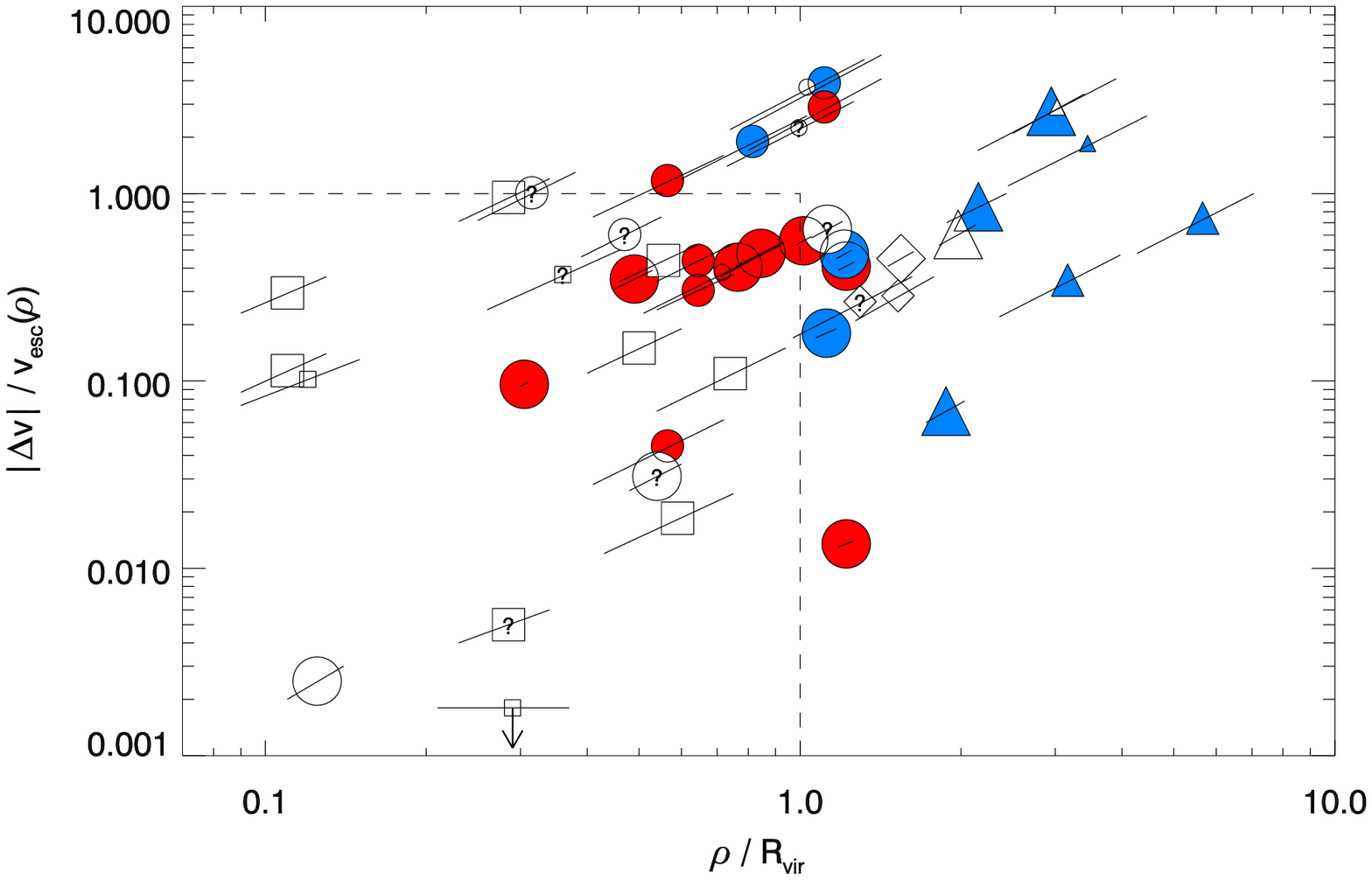}
\caption{Distribution of absorbers in impact parameter scaled by
  virial radii and absorber-galaxy radial velocity difference scaled
  by escape velocity for all metal-bearing absorbers in regions
  surveyed for galaxies completely down to $0.15\,L^*$.  The filled
  symbols all have \OVI\ detections; the open symbols lack
  \OVI\ detections, mostly due to the absence of \fuse\ spectroscopy.
  The blue-filled symbols represent absorbers with \OVI-only while the
  red-filled symbols are absorbers containing both \OVI\ and lower
  ionization absorption.  Squares, circles, and diamonds are defined
  as in Figures 9 and 10, and the triangles show absorbers that are
  unique to this figure.
\label{fig:bv_metals}}
\end{figure}

Figure 11 presents a plot similar to Figure 8 for all CGM metal-line
absorbers plus all other metal-line absorbers in well-surveyed galaxy
regions beyond the virial radius. Detections  beyond the CGM are
included only if the galaxy survey is complete to  below $0.15\,L^*$
at the absorber location. We chose this limit rather than $0.10\,L^*$
because the sample size is a factor of two larger than if we use the
slightly lower luminosity completeness limit. The $0.15\,L^*$
luminosity limit is  still low enough to provide confidence that
almost all sub-$L^*$ galaxies are included in the analysis. The blue
symbols are \OVI-only absorbers; the red symbols are absorbers with
\OVI\ and lower ion absorption and the open symbols have lower
ionization metal lines but no detected \OVI.  As in Figure 8, the size
of the symbol encodes the luminosity of the nearest galaxy while the
symbol shape encodes the sample from which the absorber was taken
(squares = targeted sample in Table 2; circles = serendipitous sample
in Table 3, diamonds = alternate serendipitous sample in Table 4; and
triangles are additional \OVI\ absorbers in regions surveyed
completely to $L \geq 0.15\,L^*$). While lower-ionization metal
absorbers generally do not occur in  regions much beyond $R_{\rm vir}$
from the nearest galaxy, \OVI-only absorbers (i.e., \lya\ +
\OVI\ absorption only) extend much further out ($\sim3.5\,R_{\rm
  vir}$)  as reported previously \citep{stocke06}. One newly
catalogued \OVI-only absorber in the PG~1259+593 sight line near the
starburst galaxy Mrk~232 is $5\,R_{\rm vir}$ (788~kpc) away in
projection, somewhat greater than previously found for any sub-$L^*$
galaxy.   In no case do \OVI\ absorbers extend much farther away from
their nearest associated galaxy than a distance comparable to  the
size of a small, spiral-rich group of galaxies.  Whether due to the
larger ionization parameters of diffuse absorbers far from galaxies
($\log{U} \geq -1.5$) for photoionized clouds or much hotter
temperatures ($T \geq 10^5$~K) for collisionally-ionized clouds,
\OVI\ absorption appears to be the best probe for the extent of metal
transport away from galaxies into the IGM at low-$z$.

There are three \OVI\ absorbers whose nearest galaxy is a dwarf,
suggesting that some metal-line absorbers are associated with very
small galaxies. In general, the data shown in Figure 11 support the
conclusions made in Section 3.4 that extended, patchy (30\% area
covering factor) metal-enriched regions exist around galaxies of all
luminosity classes. The existence of several \OVI\ absorbers at $\rho
> R_{\rm vir}$ with large $|\Delta v|/v_{\rm esc}$ supports the
interpretation  that at least some outflowing absorbers escape into
the IGM. Because the ultimate extent of metals into the IGM is very
sensitive to the amount of feedback \citep [e.g.,][]{chen01}, deep
galaxy surveys in regions of \OVI\ absorbers discovered by \hst/COS at
$z\geq0.12$ are a high priority for future ground-based spectroscopic
surveys of galaxies near targeted sight lines (e.g., Keeney et~al., in
prep). On the longer-term, an O~VI survey at $z<0.1$ using new far-UV
spectroscopic missions are required to measure more accurately the
spread of metals around galaxies into the IGM.     

It is tempting to compare this result with similar metal transport
studies conducted at  high-SNR ($\sim30$--50) using high-$z$ QSO
spectra obtained with Keck/HIRES or VLT/UVES. But all low-$z$ metal
transport studies using STIS (and even COS given limited
\hst\ observing time) are severely SNR limited compared to high-$z$
studies. The \OVI\ absorber extent described in \citet{stocke06},
\citet {prochaska11a}, and the current paper all use absorber samples
with $\log{N_{\rm O\,VI}} \geq 13.2$ (DS08). Much weaker \CIV\ and
\OVI\ lines are detectable at high-$z$ and a much greater extent of
metals through the IGM is inferred \citep*[e.g.,][] {aguirre02},
although not well-tested owing to the lack of very deep
($L\geq0.1\,L^*$) galaxy survey work at $z\sim2$. ${\rm SNR} \sim
30$--50 \hst/COS spectra are required to conduct similarly sensitive
measurments at $z\sim0$, whether with individual absorbers or using
pixel-coaddition techniques \citep{cowie98}. 

\begin{deluxetable*}{lccccccc}

\tablecolumns{8}
\tablewidth{0pt}

\tablecaption{Indicative Photoionization Models of Warm CGM Clouds
\label{tab:photomodel}}

\tablehead{\colhead{Sight Line} & \colhead{$cz_{\rm abs}$} & \colhead{$\log{N_{\rm H\,I}}$} & \colhead{$\log{U}$} & \colhead{$\log{(Z/Z_{\Sun})}$} & \colhead{$\log{n_{\rm H}}$\tablenotemark{a}} & \colhead{$D_{\rm cl}$\tablenotemark{b}} & \colhead{$\log{(M_{\rm cl}/M_{\Sun})}$\tablenotemark{c}} \\ & \colhead{(\kms)} & & & & & \colhead{(kpc)} }

\startdata
HE~0439--5254    & ~1662 & $15.21\pm0.44$\tablenotemark{d} & $-2.4^{+0.3}_{-0.2}$ &       $+0.1^{+0.9}_{-0.4}$ & $-3.7$ & 1.1  & 3.7 \\
PG~0832+251      & ~5227 & $18.48\pm0.17$\tablenotemark{d} & $-3.5^{+0.1}_{-0.2}$ &       $-0.5\pm0.2$         & $-2.6$ & 16   & 8.3 \\
PG~0832+251      & ~5425 & $16.39\pm0.91$\tablenotemark{d} & $-2.4^{+0.4}_{-0.5}$ &       $-0.9^{+0.7}_{-0.5}$ & $-3.7$ & 28   & 7.9 \\
PMN~J1103--2329  & ~1194 & $15.94\pm0.47$\tablenotemark{d} & $-2.2^{+0.4}_{-0.5}$ &       $-0.8^{+0.5}_{-0.4}$ & $-3.9$ & 31   & 7.8 \\
RX~J0439.6--5311 & ~1671 & $15.41\pm0.42$\tablenotemark{d} & $-2.6^{+0.4}_{-0.2}$ &       $-0.3^{+0.6}_{-0.5}$ & $-3.5$ & 1.2  & 4.0 \\
SBS~1108+560     & ~~665 & $17.38\pm0.63$\tablenotemark{d} & $-3.1\pm0.4$         & $\phm{-}0.0^{+1.0}_{-0.5}$ & $-3.0$ & 7.4  & 6.8 \\
SBS~1108+560     & ~~778 & $15.44\pm0.42$\tablenotemark{d} & $-2.3\pm0.3$         &       $-0.5\pm0.3$         & $-3.8$ & 5.7  & 5.7 \\
SBS~1122+594     & ~1204 & $15.92\pm0.42$\tablenotemark{d} & $-2.5\pm0.4$         &       $-0.2\pm0.3$         & $-3.5$ & 3.8  & 5.5 \\
VII~Zw~244       & ~~712 & $15.81\pm0.26$\tablenotemark{d} & $-2.8^{+0.1}_{-0.2}$ &       $-0.2^{+0.1}_{-0.2}$ & $-3.2$ & 0.70 & 3.5 \\
\tableline
3C~273           & ~1585 & $15.85\pm0.09$\tablenotemark{e} & $-3.2^{+0.2}_{-0.1}$ &       $-0.9\pm0.2$         & $-2.9$ & 0.17 & 2.1 \\
PG~1116+215      & 41521 & $16.35\pm0.10$\tablenotemark{f} & $-3.3\pm0.1$         &       $-0.3^{+0.1}_{-0.2}$ & $-2.8$ & 0.28 & 2.8 \\
PG~1211+143      & 15302 & $15.67\pm0.35$                  & $-2.9^{+0.5}_{-0.3}$ &       $-0.5^{+0.3}_{-0.4}$ & $-3.2$ & 0.52 & 3.2 \\
PG~1211+143      & 19329 & $15.17\pm0.10$\tablenotemark{g} & $-2.4^{+0.1}_{-0.2}$ &       $-0.9\pm0.1$         & $-3.7$ & 2.2  & 4.5 \\
PG~1211+143      & 19467 & $13.82\pm0.05$                  & $-2.1\pm0.1$         &       $-0.2^{+0.2}_{-0.1}$ & $-3.9$ & 0.14 & 0.9 \\
PG~1259+593      & 13808 & $15.51\pm0.28$                  & $-2.2^{+0.3}_{-0.9}$ &       $-1.1^{+0.9}_{-0.3}$ & $-3.9$ & 14   & 6.7 \\
PG~1259+593      & 13940 & $14.75\pm0.38$                  & $-1.7^{+0.3}_{-1.3}$ &       $-0.6^{+0.8}_{-0.5}$ & $-4.4$ & 21   & 6.9 \\
PHL~1811         & 22032 & $14.88\pm0.09$\tablenotemark{h} & $-2.7^{+0.3}_{-0.2}$ &       $-0.3^{+0.2}_{-0.3}$ & $-3.3$ & 0.14 & 1.4 \\
PHL~1811         & 23310 & $14.94\pm0.08$                  & $-2.7\pm0.2$         &       $-0.2\pm0.2$         & $-3.3$ & 0.15 & 1.5 \\
PHL~1811         & 24222 & $18.00\pm0.50$\tablenotemark{f} & $-3.5^{+0.3}_{-0.9}$ &       $-0.7^{+0.8}_{-1.4}$ & $-2.5$ & 4.1  & 6.5 \\
PHL~1811         & 52926 & $14.87\pm0.03$                  & $-2.6\pm0.5$         &       $-0.5^{+0.4}_{-0.5}$ & $-3.5$ & 0.33 & 2.3 \\
PKS~0405--123    & 50105 & $16.45\pm0.07$\tablenotemark{f} & $-3.0\pm0.1$         &       $+0.1\pm0.2$         & $-3.1$ & 1.2  & 4.4 \\
PKS~1302--102    & 12665 & $14.83\pm0.17$                  & $-2.8\pm0.1$         &       $+0.2\pm0.2$         & $-3.3$ & 0.08 & 0.6 \\
PKS~1302--102    & 28435 & $17.10\pm0.40$\tablenotemark{f} & $-3.1^{+0.5}_{-0.3}$ &       $-1.7^{+0.6}_{-0.4}$ & $-3.0$ & 6.0  & 6.6 \\
Q~1230+011       & 23399 & $15.06\pm0.40$                  & $-2.2^{+0.4}_{-0.7}$ &       $-0.2\pm0.4$         & $-3.9$ & 3.0  & 4.8
\enddata

\tablecomments{Column densities are given in units of ${\rm cm}^{-2}$ and densities in ${\rm cm}^{-3}$.}
\tablenotetext{a}{Total hydrogen column density, $\log{n_{\rm H}} = -6.074 - \log{U}$.}
\tablenotetext{b}{Cloud thickness along the line of sight.}
\tablenotetext{c}{Cloud mass, assuming spherical clouds with diameter $D_{\rm cl}$, uniform density $n_{\rm H}$, and purely hydrogen+helium composition.}
\tablenotetext{d}{We have constrained the H\,{\sc i} column densities of the targeted absorbers by assuming that the H\,{\sc i} and metal lines reside in a single photoionized phase.  See \citet{keeney12,keeney13} for details.}
\tablenotetext{e}{We have adopted the column densities of \citet{tripp02} to model this absorber rather than those of \citet{danforth08}.}
\tablenotetext{f}{These column densities have been modified from the DS08 values by including information on the Lyman limit decrement for high column density absorbers. The column density for the PKS~0405--123 absorber is from \citet{prochaska04}.}
\tablenotetext{g}{The DS08 column density for this absorber ($\log{N_{\rm H\,I}} = 15.73\pm0.32$) has been revised using a new curve-of-growth analysis that favors $b=33\pm4$~\kms\ and $\log{N_{\rm H\,I}} = 15.17\pm0.10$.}
\tablenotetext{h}{DS08 lists two absorbers at $cz=21,995$ and 22,050~\kms, but further scrutiny reveals no evidence of multiple velocity components.}

\end{deluxetable*}

\subsection{Synopsis of Photoionization Modeling of Warm CGM Clouds}

In general, where metals are associated with these CGM absorbers, the
higher ions associated with warm, photoionized gas (\SiIV\ and \CIV)
predominate over lower ions (\SiII\ and \CII) in both the targeted and
the serendipitous samples. No strong \MgII\ ($\eqw \geq 0.5$~\AA)
absorption was found for any of the  few targeted absorbers observed
with the G285M grating and only one targeted and three serendipitous
absorbers have stronger low ions (i.e., \SiII\ $>$ \SiIV). The $cz =
5,225$~\kms\ absorber in the PG~0832+251 sight line is one of the few
absorbers dominated by lower ions and is a Lyman limit system at
$\log{N_{\rm H\,I}} \approx  18.4$ (but \MgII\ was not observed for
this, previously unknown, Lyman limit system).  A dust lane across the
disk of the nearby low-level starburst galaxy NGC~2611 and the sign of
the absorber/galaxy velocity difference require that this absorber is
outflowing from NGC~2611 \citep [] [Paper 2] {stocke10} but its low
relative radial velocity ($|\Delta v|/v_{\rm esc} \approx 0$; see
Table 2) suggests that it will not escape into the IGM. A companion
absorber at $cz = 5437$~\kms\ has a much larger $|\Delta v|/v_{\rm
  esc} = 1.2$ and is infalling based on the sign of its velocity
difference from NGC~2611. This second component has a much higher
ionization spectrum, more typical of the remainder of the sample. 

There are a few other Lyman-limit systems (LLS) we have modeled based
on direct detections of flux decrements in \fuse\ spectra: PG~1116+215
at $cz = 41,521$~\kms\ \citep{tripp98}, PHL~1811 at $cz =
24,222$~\kms\ \citep{jenkins03}, PKS~0405--123 at $cz =
50,105$~\kms\ \citep{prochaska04} and PKS~1302--102 at $cz =
28,435$~\kms\ (see Table 6). These LLSs are all associated with
$L\geq0.3\,L^*$ galaxies  consistent with many earlier studies of
strong \MgII/LLS absorbers \citep[e.g.,][]{steidel95, churchill00,
  kacprzak11, chen10}. 

Another of the very few low ionization absorbers is at 1585~\kms\ in
the 3C~273 sight line  in the serendipitous sample (see Table
3). Because this absorber already had been studied in detail by
\citet{sembach01}, \citet {tripp02}, and \citet {stocke04}, we modeled
this absorber as a check on our CLOUDY analysis process.  Standard
photoionization modeling by \citet{tripp02} found a surprisingly high
hydrogen density of $n_{\rm H}=10^{-2.8}~{\rm cm}^{-3}$, low
temperature ($\sim10^4$~K) and thus quite a small line-of-sight size
of 70~pc.  The metallicity of 6\% solar and a super-solar Si/C ratio
suggesting recent Type II supernova  enrichment are consistent with
the absorber originating in a nearby ($\rho \approx 70$~kpc)
post-starburst dwarf galaxy  of similar metallicity
\citep{stocke04}. The large absorber/galaxy radial velocity
difference, large impact parameter (see Figure 8b;  top right small
red filled circle) and small host galaxy mass require that this
absorber will escape into the IGM. While the equivalent width
measurements for the various low ions are slightly different in DS08
as compared to those used by \citet{tripp02}, we recover very similar
cloud parameters (see Table 6) including small cloud size and low
metallicity.  This absorber is one of the few (three) metal-bearing
absorbers with no \OVI\ detection. There are a few other targeted and
serendipitous absorbers  which have modeled line-of-sight sizes
$\sim100$~parsecs but all of these have higher ionization parameter
and \OVI\ detected.
   
We have attempted photoionization modeling only for those 15
serendipitous absorbers in Tables 3 and 4 plus nine targeted absorbers
in Table 2 with multiple metal ion detections. We required detections
of two or more  ionization states of the same element in order for a
satisfactory model to be calculated; i.e.,  (\CII, \CIII, \CIV) in the
serendipitous sample where \fuse\ spectra are available or (\SiII,
\SiIII, \SiIV) where only \hst/STIS or \hst/COS spectra exist like in
the targeted sample. Column densities of \OVI\ are not used in this
modeling both because this high ion likely traces
collisionally-ionized gas and because its redshifted wavelength has
been observed for only one of the  targeted absorbers, PG~0832+251 at
$cz = 5225$~ \kms. While the targeted CGM absorbers have only
\lya\ with which to determine a hydrogen column density, the
serendipitous absorbers all have higher Lyman line detections from
\fuse\ and thus much more accurate $N_{\rm H\,I}$ values from
curve-of-growth analysis (DS08). While the nine targeted absorbers
that meet the metal-line criteria have poorly determined $N_{\rm
  H\,I}$ values (this is particularly true of the M~108 aborbers
because the COS spectrum of  SBS~1108+560 contains a higher-redshift
Lyman limit system that reduces the SNR at \lya\ significantly), we
have used other constraints to reduce the  errors on the hydrogen
column density considerably. Specifically, the range of plausible
$N_{\rm H\,I}$ values in Table 6 requires consistency with a
single-phase photo-ionization model constrained by the observed
metal-line ratios, an absorber size $<$ the impact parameter, and an
absorber metallicity $\lesssim$ the galaxy metallicity (which sets a
lower bound on $N_{\rm H\,I}$).  
The procedure which uses these constraints to create viable
photo-ioization models is shown in detail in \citet{keeney12}  for the
two metal-bearing absorbers in Table 6 in the HE~0439-5254 and RX
J0439.6-5311 sightlines. 
 
These cloud models are derived from single homogeneous CLOUDY
\citep{ferland98} calculations   assuming only photo-ionization by an
external radiation field as specified in \citet{haardt12}.  Even the
most proximate absorbers in the targeted sample are farther away from
their associated galaxies than their ``proximity distance'' where
additional ionization from hot stars begins to contribute
significantly if the escape fraction is as high as $\sim5$\%
\citep{giroux97}. All of the models were produced using the same,
standard procedure explained in detail on a case-by-case basis in
Paper 2. Here, by way of a synopsis, we list the results of these
models in Table 6, which contains the following information by column:
(1) the sight line target name; (2) the absorber heliocentric
recession velocity (\kms); (3) the log of the neutral hydrogen column
density ($\log{N_{\rm H\,I}}$) in cm$^{-2}$; (4) the log of the
ionization parameter ($\log{U}$); (5) the log of the mean metallicity
as determined from CLOUDY modeling [$\log{(Z/Z_{\Sun})}$] assuming
solar abundance amounts and ratios from \citet{grevesse10};  (6) the
log of the total hydrogen density ($\log{n_{\rm H}}$) in cm$^{-3}$;
(7) a characteristic cloud diameter ($D_{\rm cl}$) as determined from
a modeled line-of-sight cloud depth in kpc; and (8) the  logarithm of
the estimated total cloud mass ($\log{M_{\rm cl}}$) in solar
masses. As a check on our CLOUDY modeling, we have used our procedure
to reproduce the physical conditions derived by others for some of the
same absorbers in Table 6 including 3C~273 at
$cz=1585$~\kms\ \citep{tripp02} and PKS~0405-123 at
$cz=50,105$~\kms\ \citep{prochaska04}. The sizes and masses of our
models differ from others by factors of 2--3 and 4--10 respectively,
even in cases where we have used the same measured column
densities. These differences underscore that these models are
indicative, not precise.

\begin{figure}[!t]
\epsscale{1.15} \centering \plotone{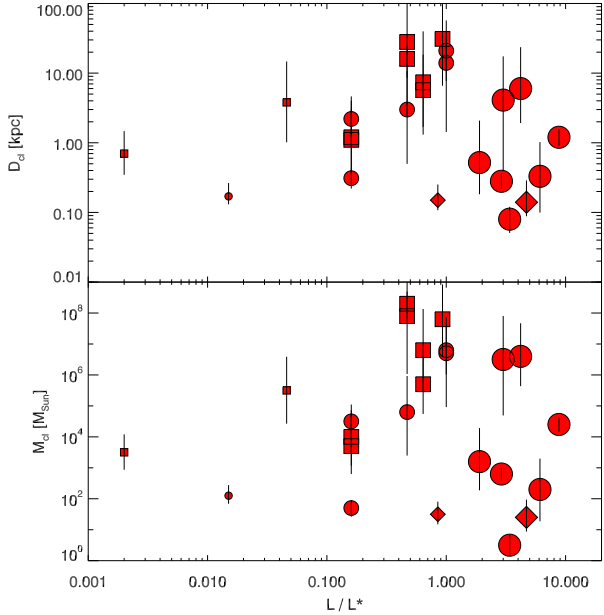}
\caption{The distribution of modeled CGM cloud diameters (top) and
  cloud masses (bottom) as a function of host galaxy luminosity. The
  symbology is the same as for Figure 10 and all symbols are filled
  (indicating metal-bearing clouds)  because a requiremement for our
  applying CLOUDY modeling to absorbers in our sample is that at least
  two different ionization states of the same element (i.e., either C
  or Si) must be present.  
\label{fig:clouds_lum}}
\end{figure}

Before using the model results of Table 6 to assess physical
conditions in the CGM of late-type galaxies, the next set of figures
explores whether these derived quantities are dependent on either
galaxy luminosity or impact parameter. Despite concerns that the
virial radius and escape velocity scalings may mask trends between CGM
cloud parameters and galaxy luminosity, Figure 12 shows that neither
cloud diameter (top) nor mass (bottom) are dependent upon galaxy
luminosity (although data below $0.1\,L^*$ are quite sparse). The
physical conditions derived from the CLOUDY modeling of these clouds
(temperature, density and pressure) also show no dependence on host
galaxy luminosity. Warm CGM clouds are similar regardless of the host
galaxy nearby, excepting that no LLSs have been detected around
dwarfs, consistent with earlier results \citep*{steidel95, mclin98}.
However, there are weak trends in cloud diameter and mass with scaled
impact parameter that are similar to the \lya\ equivalent width
dependence seen in  Figure 9 with smaller, less massive clouds at
larger scaled impact parameters (Figure 13): $D_{\rm cl}=
(0.19\pm0.04~{\rm kpc})(\rho/R_{\rm vir})^{-1.7\pm0.2}$ and $M_{\rm
  cl}= (36\pm2~M_{\odot}) (\rho/R_{\rm vir})^{-6.2\pm0.5}$ (quoted
errors are statistical only).  While these best-fit power-laws decline
with radius, the scatter is substantial and the fits quantified above
are poor (reduced $\chi^2 \gg 1$). Large, massive clouds exist at all
radii.  However, Ly$\alpha$-only clouds are found almost exclusively
at $\rho \approx R_{vir}$.

Despite the generally declining cloud size and mass with impact
parameter, the pressure within CGM clouds estimated from mean cloud
densities (Table 6; column 6) and temperatures  (not shown but always
near $\log T \approx 4.0$--4.3 K) does not  appear to decline very
steeply with scaled impact parameter regardless of galaxy luminosity
(Figure 14).  The best-fit power-law to the data in Figure 14 is:
$(P/k) = (12\pm2~{\rm cm^{-3}\,K})(\rho/R_{\rm vir})^{-0.3\pm0.2}$,
and has a reduced $\chi^2$ value of 3.3. Clearly any trend in these
data is minimal.  The surprising absence of a clear trend in cloud
pressure with impact parameter will be addressed in Section
5.1. Neither cloud diameter, mass nor pressure vary significantly with
absorber-galaxy  relative velocity for this sample of 24 modeled CGM
absorbers.

\begin{figure}[!t]
\epsscale{1.15} \centering \plotone{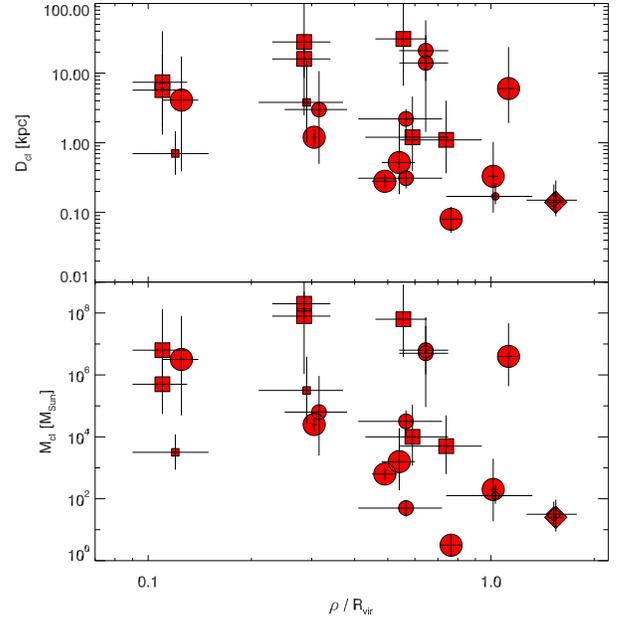}
\caption{The distribution of modeled CGM cloud diameters (top)  and
  cloud masses (bottom) as a function of scaled impact parameter. The
  symbology is the same as for Figure 10 and all symbols are filled
  (i.e., ``metal-bearing'') as required for the CLOUDY modeling.  The
  declining CGM cloud sizes and masses as a function of impact
  parameter is similar to the trend of  \lya\ equivalent width with
  radius seen in Figure 9. While a power-law has been fit to these
  data (see text for best-fit  parameters), these fits are poor
  (reduced $\chi^2 \gg 1$).
\label{fig:clouds_radius}}
\end{figure}

\subsection{Limited CGM Cloud Origin Information}


The following information which bears on the origins of the CGM clouds
in this sample is available: (1) absorber metallicity for 24 modeled
clouds (Table 6);  (2) infall/outflow determination made using the
three-dimensional orientation of the galaxy (possible for only four
galaxies and eight absorbers in this sample); (3) the position angle
on the sky relative to the galaxy orientation \citep [i.e., absorber
  projected closer to the major  or minor axis;] [] {bouche12} as
shown in Figure 6 (39 absorbers have these data); and (4) absorber
velocity relative to its host galaxy as a  fraction of escape velocity
(Figure 8b, which plots 58 absorbers from Tables 2 and 3). The last
two sets of data are suggestive in some cases but not conclusive as to
cloud infall/outflow kinematics and origin.

\subsubsection{Absorber Metallicity vs. Host Galaxy Metallicity}

While absorber and host galaxy metallicities are important in
determining whether the absorbing gas originated in the nearby galaxy,
this information is not entirely  definitive because modeled
metallicity values can have large associated errors ($\pm$ 0.2--0.4
dex; see Table 6 and Paper 2). Also, a dilution of metal-enriched gas
by more pristine gas in the galaxy's vicinity can decrease absorber
metallicity by an unknown amount so that an absorber which originates
in the host galaxy can have $Z_{abs}<Z_{gal}$. Recently
\citet{lehner13} found a strong bifurcation of luminous galaxy CGM
absorber metallicities at $Z_{abs}$ = 0.1--0.2 solar. These authors
conclude that the more metal-rich absorbers likely originate in the
nearby large galaxy and may be outflowing or recycling gas while the
lower metallicity gas is infall. We adopt a similar  criteria to make
our tentative assessments of this sample of absorbers. Of the modeled
absorber metallicities for the 24 cases in Table 6,  many values are
broadly consistent with the metallicities of their host
galaxies. Whether these absorbers are infalling or outflowing is not
specifically known but  nine of them have $|\Delta v| \ll v_{\rm esc}$
so we identify these as likely recycling gas. Six of these are at
$\rho \leq 0.5\,R_{\rm vir}$.  For most other absorbers in Table 6 the
modeled absorber metallicities  are not sufficiently accurate or
sufficiently different from their host galaxy metallicities to provide
secure cloud origin determinations. 

However, there are a few absorbers for which we can be more
definite. Three absorbers have modeled metallicities consistent with
gas originating in the  nearby host galaxy (i.e., $Z_{abs}\approx
Z_{gal}$) but with $|\Delta v| > v_{\rm esc}$.  We identify these
three absorbers as likely outflowing gas. One of these is the 3C~273
absorber at  $cz = 1585$~\kms, which has a metallicity of a few
percent solar \citep [] [and Table 6] {tripp02}. Its host galaxy is a
nearby, post-starburst dwarf, which also has a few percent solar
metallicity  suggesting that the absorber is outflowing and its radial
velocity ($\Delta v = 5.2\,v_{\rm esc}$) requires it to be unbound
\citep{stocke04}.
 
Four other absorbers have metallicities of $\leq20$\% solar with
super-$L^*$ associated galaxies \citep{lehner13}, much less than
expected if the gas originated in their host galaxies even with
dilution by pristine gas. We identify these absorbers as likely
infalling gas. One targeted sample absorber associated with the
starburst galaxy NGC~2611 (Table 2) has $\sim$10\% solar metallicity
and is constrained to be infalling gas by the geometrical argument
made in the next sub-section.   A low-metallicity ($\sim$15\% solar)
absorber in the targeted PMN~J1103--2329 sightline is associated with
an $0.9\,L^*$ spiral and is also constrained geometrically to be
infalling.  A low-metallicity serendipitous absorber ($\sim$3\% solar)
is identified with a $2\,L^*$ galaxy in the PKS~1302-102 sightline
($cz=28,435$~\kms\ absorber).  The fourth low-metallicity
absorber/galaxy pair is somewhat more ambiguous. The PG~1211+143
absorbers at $cz = 19,329$~\kms\  and 19,467~\kms\ have metallicities
of $\sim15$\% solar and solar. In this case we interpret the higher
metallicity absorber as  originating, not in the nearest $0.16\,L^*$
galaxy (Table 3), but in a $1.2\,L^*$ galaxy (Table 5) only slightly
further  away at $0.77\,R_{\rm vir}$ in projection. The lower
metallicity absorber  could either originate in the lower luminosity
galaxy or be falling into either of the two nearby galaxies,
originating elsewhere. Given that this system is  most simply
described as a large galaxy with a small companion in its halo, the
least ambiguous interpretation of these two absorbers is entirely with
respect to the $1.2\,L^*$ spiral \citep{tumlinson05} with the solar
metallicity absorber being outflow and the sub-solar absorber being
infall.

\begin{figure}[!t]
\epsscale{1.15} \centering \plotone{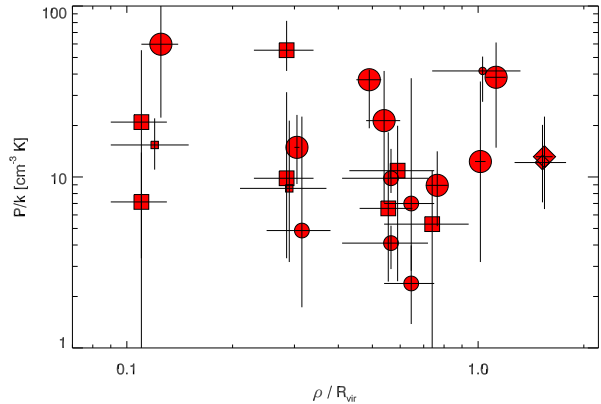}
\caption{The modeled mean pressure ($P/k$ in cm$^{-3}$\,K) inside warm
  CGM clouds as a function of scaled impact parameter.  Surprisingly,
  there is no indication of a strong trend as might be expected if
  these clouds are in pressure equilibrium with a hot ($T \approx
  10^6$~K) gaseous halo, whose density is declining rapidly with
  radius around the host galaxy.  While a power-law has been fit to
  these data (see text for best-fit parameters), this fit is poor
  (reduced $\chi^2 \approx 3.3$).
\label{fig:pressure}}
\end{figure}

\subsubsection{Three-Dimensional Orientation of Host Galaxy}

Although the sign of the absorber/galaxy radial velocity difference is
known, whether the absorber is infalling or outflowing is not known
unless the three dimensional orientation of the galaxy can be
determined \citep{stocke10}. This is the primary reason why we have
chosen to plot the absolute value of the galaxy-absorber velocity
difference in Figures 8 and 10 . Three dimensional orientation
determinations are possible only for those few galaxies with both
well-defined disks inclined at an intermediate angle  relative to us
on the sky and also some global internal extinction pattern (i.e., one
side of the galaxy disk is more obscured than the other from our
perspective, and therefore more distant from us).  Even possessing
this information, an infall/outflow determination is possible only for
absorbers within about $45\degr$ of the galaxy's minor axis, for which
we {\it assume} a total motion perpendicular to the galaxy's
disk. Therefore, we have solid determinations  for only eight
absorbers near four galaxies in the targeted list:
HE~0435--5304/ESO~157--49 \citep [three components with \lya-only
  detected;][]{keeney12},  SBS~1108+560/M108, PNM J1103-2329/NGC~3511
and PG~0832+251/NGC~2611 (see Paper 2 for detailed descriptions).  The
geometry and sign of $\Delta v$ require that the two lower redshift
HE~0435--5304 absorbers are outflowing while the third is infalling.
The lower redshift, higher metallicity SBS~1108+560 absorber is found
to be outflowing, the other infalling. PG~0832+251 possesses one
outflowing  [the $\log{(Z/Z_{\Sun})} = -0.5$ LLS] and one infalling
[$\log{(Z/Z_{\Sun})} \sim -0.9$] absorber (Table 6). The
PNM~J1103-2329 absorber associated with NGC~3511 is infalling and has
low metallicity [$\log{(Z/Z_{\Sun})} \sim -0.8$].  No absorber/galaxy
pairs in the serendipitous list have geometrically well-determined
directions for the gas flow.

But the case of the three \lya-only absorbers located along the minor
axis just beyond the virial radius of ESO~157--49 is a cautionary tale
for this analysis. The geometry of the host galaxy requires that two
of these are outflowing absorbers but no metals are detected in the
COS spectra \citep{keeney12}.  However, the absorbers' lower $N_{\rm
  H\,I}$ values makes the absence of metal absorptions reasonable if
these clouds have similar metallicity and physical conditions to the
absorbers  found in the other two sight lines around ESO~157--49
\citep{keeney12}. Therefore, the assignment of these two absorbers
specifically, and the other  ``\lya-only'' clouds in general, as very
low metallicity infalling gas is not justified by the limited
sensitivity of the current data. Given the substantial fraction of CGM
absorbers at $\rho > 0.5\,R_{\rm vir}$ which are ``\lya-only'' clouds,
a confident accounting of low-metallicity gas   infall onto galaxies
requires higher SNR spectra than in-hand currently.

\subsubsection{Absorber Location Relative to Host Galaxy's Minor Axis}

Another piece of information that is available for many absorbers is
absorber location relative to the host galaxy's major axis (see Figure
6).  This determination is possible for 26 serendipitous CGM absorbers
and 14 (all but three) targeted absorbers based on galaxy sky
orientations from NED.  Most of these values come from SDSS and have
been checked by us for plausibility.  Several host galaxies simply
lack such information while others have no well-defined major axis
(e.g., a dwarf Magellanic spiral) and some galaxies are face-on.
Using a sample of very low-$z$ \MgII\ absorbers, \citet{bouche12}
found most absorbers concentrated around either the host galaxy's
major or minor axis.  They identified the major axis absorbers with
infalling gas and the minor axis absorbers as outflowing gas. But this
assignment is far from certain \citep[e.g., the 3C~232 absorber near
  NGC~3067 and one absorber each in the three sightlines mentioned in
  the last sub-section are located along the minor axis but
  constrained by geometry to be infalling;][Section
  2.1]{stocke10}. Therefore, until this dichotomy can be tested
further, we must view it as interesting but unreliable. This is
especially true for the serendipitous sample absorbers, which are
rather isotropically configured around their host galaxies. The
absence of a bi-modal, major axis/minor axis concentration in absorber
position angle persists even for absorbers at absorber/galaxy
separations of $\leq100$~kpc.  Therefore, we make no cloud origin
determinations based on these data.

\subsubsection{Absorber Velocity Relative to Host Galaxy}

In Section 4.2, Figure 8b was divided into plausible regions in both
coordinates which we  interpret further here using the additional
information described above:

\begin{enumerate}
\item Inside $0.5\,R_{\rm vir}$ almost all absorbers (12 of 13)
  contain metals. The one exception is a dwarf associated with a COS
  pointing. Although the data quality of this spectrum is good (${\rm
    SNR}\sim20$ in G130M and $\sim12$ in G160M), the metallicity of
  gas  associated with this dwarf could also be quite low, plausibly
  below 0.1 solar which would not be detectable in the spectrum given
  the observed $N_{\rm H\,I}$ value. 
Six of these 13 absorbers have radial velocity differences, $|\Delta
v| \leq 0.1 v_{\rm esc}$, so low that  we conclude that these six are
bound clouds, despite not knowing whether they are outgoing or
infalling at the present time. These six have $Z_{abs}\approx
Z_{gal}$. Geometry  constrains two of these 13 absorbers to be outflowing
and one to be infalling (see Section 4.4.2). Two other absorbers (at
$cz=5425$~\kms\ in the PG~0832+251 sightline and at $cz=1194$~\kms\ in
the PMN~J1103--2329 sightline), which are constrained to be infalling
by geometry, have large $|\Delta v|/ v_{\rm esc} = 1.2$ and 0.54,
respectively.  Both of these absorbers have low metallicities. The
$|\Delta v| > v_{\rm esc}$ for the PG~0832+251 absorber strongly
suggests  that this is infalling gas originating outside NGC~2611; the
case of the PMN~J1103-2329 absorber is less certain. The remaining
metal-bearing clouds at small impact parameter have intermediate
$|\Delta v|/ v_{\rm esc} =0.2$--0.8 values (see Figure 8b) and so are
less securely classified. 

\item At $0.5\,R_{\rm vir} < \rho < R_{\rm vir}$, 14 absorbers have
  metals detected at $\log{(Z/Z_{\Sun})} \gtrsim -1.0$.  Four of these
  have $\log{(Z/Z_{\Sun})}$ close to Solar metallicity and so we
  identify these four as originating inside their $L \geq L^*$ host
  galaxies. Three of these four have  $|\Delta v|/ v_{\rm esc} < 0.1$
  and are identified as  likely bound recycling gas clouds.   The
  remainder have intermediate values of $|\Delta v|/ v_{\rm esc}$ and
  metallicity (or uncertain metallicity) so that their kinematic
  origin and fate are unclear. The 11 \lya-only clouds in this region
  have unknown metallicities and so unknown origin and fate. 

\item Outside the virial radius to our arbitrary limit of $1.5\,R_{\rm
  vir}$ there are nine metal-enriched absorbers, four of which have
  relative velocities much greater than $v_{\rm esc}$, while four
  others  have a velocity indicative of probable escape. Of the three
  absorbers that  have $|\Delta v| \gg v_{\rm esc}$, two have $Z_{abs}
  \approx Z_{gal}$ and are  identified as outflowing, unbound gas
  (including the 3C~273 absorber discussed in Section 4.4.1); the
  third absorber has a very low  [$\log{(Z/Z_{\Sun})}\sim-1.7$]
  metallicity so we identify it as infalling (see Section 4.4.1). 

The \lya-only absorbers in this region have varying relative
velocities and so their origins and fates are unknown, excepting for
the three in the HE~0435--5254 sight line along the minor axis of
ESO~157--49, discussed above. 

Given the substantial sky area surveyed around these galaxies for
absorbers, and the $d\mathcal{N}/dz = 50$ per unit redshift  for
$\log{N_{\rm H\,I}} \geq 13.0$ \citep{penton04}, a few (three to five)
of the absorbers at  large impact parameters could be projected
systems; i.e., the absorbers are actually several Mpc away radially
despite having redshifts within $\pm400$~\kms\ of the galaxy.
\end{enumerate}

In summary, based upon using a combination of the factors listed
above, we have not yet obtained secure origins and fates for CGM
clouds in a majority of cases. However, probable assignments can be
made in some cases.  Nine absorbers at $\rho < R_{\rm vir}$ (6 at
$\rho < 0.5 R_{\rm vir}$) have metallicities broadly consistent with
their host galaxies and very low  absorber-galaxy velocity differences
(radial velocity difference $\leq10$\% of $v_{\rm esc}$). The latter
population is very likely bound, recycling ``galactic fountain'' gas
although we cannot distinguish which clouds are outgoing and which are
infalling at the present time. Five CGM absorbers with metallicities
comparable to their host galaxy metallicities have sufficient
velocities to escape and so we identify them as unbound outflow; three
of these are constrained by geometry (sub-section 4.4.2) to be
outflowing gas.  Importantly, three absorbers with $Z_{abs} \ll
Z_{gal}$ are identified as infalling. One of these three is
constrained geometrically to be infalling gas.  The bulk of the
absorbers, particularly the \lya-only clouds, have unknown origins at
the present time. While current data do not provide unambiguous origin
and fate for the majority of CGM absorbers, our analysis suggests that
this is  a realizable goal. Future, higher-SNR UV spectra with COS can
determine cloud metallicities at $<10$\% solar and ground-based
spectroscopy of \HII\ regions in the host galaxies will provide more
accurate absorber/galaxy metallicity differences to help determine the
origin of these clouds.



In the next Section, we use the photoionization modeling of enriched
clouds  with multiple ion metal detections described in Section 4.3
to determine warm CGM cloud properties in bulk. Based on the analysis
in this Section, a substantial fraction ($\sim$ one-third) of these
absorbers is identified  as ``galactic fountain'' clouds based on
their metallicities and kinematics.

\begin{deluxetable*}{lccccccc}

\tablecaption{Model Parameters for Warm CGM Clouds
\label{tab:cloudmodel}}

\tablehead{ \colhead{Sub-sample} & \colhead{$\langle L \rangle$} & \colhead{$\langle R_{\rm vir} \rangle$} & \colhead{$\langle F \rangle$\tablenotemark{a}} & \colhead{$\langle N_{\rm cl}\rangle$\tablenotemark{b}} & \colhead{$\langle \log{M_{\rm cl} \rangle}$\tablenotemark{c}} & \colhead{$\langle \log{M_*} \rangle$} &  \colhead{$\langle \log{M_{\rm vir}} \rangle$} \\ & \colhead{($L^*$)} & \colhead{(kpc)} }

\startdata
Super-$L^*$ & 2.0~ & 230--290 &   3--6\%   & ~3000--4500  & 10.0--10.4 & 10.4--10.7 & 11.8--12.1 \\
Sub-$L^*$   & 0.45 & 140--210 &  ~5--10\%  &  1000--1500  &  9.5--9.9  & ~9.7--10.3 & 11.2--11.7 \\
Dwarfs      & 0.03 & ~70--120 & 0.5--1\%~~ &   150--250   &  7.8--8.3  &  8.8--9.6  & 10.3--11.0
\enddata

\tablecomments{All masses are given in units of $M_{\Sun}$. Quantities listed as a range are calculated using virial radii determined from halo abundance matching and from Equation (1), respectively.  Estimated errors in warm cloud masses are $\pm0.2$~dex for super-$L^*$ and sub-$L^*$ galaxies, and $\pm0.5$~dex for dwarfs.}
\tablenotetext{a}{Volume filling factor of warm CGM clouds in the inner half ($R \leq 0.5\,R_{\rm vir}$) of the CGM; the outer half ($0.5\,R_{\rm vir} < R \leq R_{\rm vir}$) has $\langle F \rangle$ values $\sim8$ times smaller.}
\tablenotetext{b}{Total number of warm CGM clouds per galaxy with diameters $>1$~kpc. All super-$L^*$ and sub-$L^*$ galaxies have $\gtrsim10,000$ tiny clouds $\leq1$~kpc in size.}
\tablenotetext{c}{Total mass per galaxy in warm CGM clouds.}

\end{deluxetable*}

\subsection{CGM Physical Characteristics Derived from the Combined Samples}

As shown in Table 6, the warm, photoionized CGM gas clouds typically
have 10\% Solar to Solar metallicities, ionization parameters of
$\log{U} = -2.2$ to $-3.0$, cloud sizes (characteristic diameters) of
0.1--10 kpc, total  densities of $\sim10^{-3}$ to $10^{-4}~{\rm
  cm}^{-3}$, temperatures (not shown) of 16,000--24,000~K, and masses
of  $\sim 10$--$10^8~M_{\odot}$. Many of the absorbers in Table 6 have
locations projected within the inner half virial radius  of the
associated galaxy and relative velocities indicative of being bound,
recycling gas. There are six Lyman-limit and partial Lyman-limit
systems in Table 6 that have considerably higher densities ($n_{\rm H}
\geq 10^{-3.0}~{\rm cm}^{-3}$) than the other absorbers and are among
the more massive CGM clouds. The cloud parameters found in Table 6 are
similar to those found for highly-ionized high velocity clouds around
the Milky Way \citep{shull09, lehner11, putman12} but most of  these
clouds have lower ionization parameters, lower mean densities, and
larger sizes and masses. Among the CGM clouds modeled using CLOUDY,
three absorbers appear to be unbound, outflowing gas (see
Section~4.4). There is no obvious distinction between the models for
unbound vs bound clouds in this sample, although four of the unbound
absorbers (3C~273 at $cz = 1585$~\kms\ and PKS~0405--123 at $cz =
45,378$~\kms\ in Table 3 and PHL~1811 at $cz = 22,032$~\kms\ (see
discussion of this absorber below) and PHL~1811 at $cz =
23,310$~\kms\ in Table 4) have no detected \OVI\ ($\log{N_{\rm O\,VI}}
\leq 13.2$).  Only one other metal-bearing serendipitous absorber has
undetected \OVI\ (the LLS in PHL~1811 at $cz = 24,222$~\kms).

The discussion in Section 4.3 and Figures~12--14 show that, except for
the  marginally-significant decline in CGM cloud size and mass with
impact parameter,  there is little distinction between CGM clouds as a
function of galaxy luminosity (for galaxies with $L\geq 0.1\,L^*$),
radial location or relative velocity once the virial radius scaling is
applied. The dwarf sample has lower covering factor and cloud masses
but these conclusions are based upon smaller numbers of
examples. Since the clouds we have modeled using CLOUDY are a modest
fraction of the full ensemble of CGM clouds, we must assume something
about the  clouds we have not modeled (or cannot model), those
containing few or no metal absorptions. In order to proceed we will
assume that  all other CGM clouds are physically similar to the models
in Table 6. Almost all of the CGM clouds at $\rho \leq 0.5\,R_{\rm
  vir}$ have  been modeled and the inner half virial radius contains
the more massive clouds statistically, so the most massive portion of
the CGM is well-modeled by the available data.  For the outer half of
the CGM, to be conservative  we assume that the ``\lya-only'' clouds
are scaled-down versions of the clouds modeled in Table 6, smaller and
less massive but with the same $\log{U}$, consistent with the
available data. But, it is possible that some or even many of the
\lya-only CGM clouds at large impact parameter are more highly-ionized
(many are detected in \OVI) with $\log{U} \geq -2.0$, which could mean
that they are larger and more massive. If this is the case our
calculations have  under-estimated the total CGM warm cloud mass
somewhat.  


While we have measured very high CGM cloud covering factors, these
near unity values do not require very large filling factors of CGM
clouds because we are viewing these galaxies from afar, a situation
geometrically quite different from our view inside the Milky Way
looking outward where large observed area covering factors for HVCs
{\it do imply} large volume filling factors \citep{shull09,
  lehner11}. The current situation is analogous to viewing clouds in
our own Earth's atmosphere around the setting Sun. A similar geometry
causes the covering factor of CGM clouds around external galaxies to
be near unity, even if the filling factor of clouds is $\lesssim10$\%
(see below). 

In the present case, we have measured a very high covering factor for
the CGM out to approximately $R_{\rm vir}$. Now we need to estimate
the filling factor for clouds of varying cloud sizes from these
covering factors. To do this we assume that a galaxy halo extends
spherically to its virial radius and is  partially filled with
spherically-shaped warm CGM clouds whose sizes are determined from the
modeled sizes found in Table 6.  Since we are viewing these galaxies
from an arbitrary direction, the assumption of cloud sphericity seems
reasonable; i.e., the line-of-sight cloud diameters found by
photoionization modeling characterize cloud sizes averaged over
viewing angle, which is well-approximated by circular clouds inside a
circular CGM region on the sky.  For the calculation below, a covering
factor of $C\sim100$\% inside $1/2\,R_{\rm vir}$ and $C\sim75$\%
between $0.5\,R_{\rm vir}$ and $R_{\rm vir}$ is assumed, values
consistent with Figure 3  for the super-$L^*$ and sub-$L^*$
samples. For the dwarfs we assume $C = 0.5$.

An additional, important geometrical factor to consider is
``shadowing''.  When one or more clouds lie behind another from our
perspective, the observed covering factor ($C$)  implies a larger
filling factor ($F$) of CGM clouds, augmented by the amount of
``shadowing'' ($S$). $S$ is the mean number of distinct clouds along a
sight-line  within any one galaxy CGM (i.e., at $|\Delta v| \leq
400$~\kms) and can be measured statistically using the percentage of
multiple CGM systems (e.g., in Tables 2, 3 and 4 individual galaxy
entries are accompanied by muliple absorber recession velocities where
$S > 1$). For the combined serendipitous and targeted CGM sample shown
in Tables 2 and 3, $S = 1.4\pm0.2$ with no evidence for a significant
difference  in shadowing between the inner and outer half virial
radii. This means that for a given value of the covering factor, the
number of CGM clouds is 40\% larger by taking shadowing into
account. A modest amount of shadowing requires a modest  filling
factor if cloud sizes are much less than the virial radius. If the
filling factor of CGM clouds were close to unity  (which it is not),
virtually every sight line would include multiple components (i.e., $S
> 2$). There is no evidence for a different value of $S$ for the
dwarfs.
  
Based on these geometrical assumptions, the covering factor $C$ is
given by: 

\begin{equation}
C = \frac{N_{\rm cl} \times \pi r_{\rm cl}^2} {S \times \pi R_{\rm
    CGM}^2},
\end{equation}

\noindent where $N_{\rm cl}$ is the number of CGM warm clouds, $r_{\rm
  cl}$ is the median cloud radius and $R_{\rm CGM}$ is the radius of
the galaxy's CGM, which we take to be the virial radius. In this case
the volume filling factor ($F$) is given by:

\begin{equation}
F = N_{\rm cl} \, \frac{4/3 \pi r_{\rm cl}^3}{4/3 \pi R_{\rm CGM}^3}.
\end{equation}

\noindent Equations (2) and (3) imply that

\begin{equation}
N_{\rm cl}= C \times S \times \frac{R_{\rm CGM}^2}{r_{\rm cl}^2},
\end{equation}

\noindent and 

\begin{equation}
F = C \times S \times \frac{r_{\rm cl}}{R_{\rm CGM}}.
\end{equation}

By this formulation, the filling factor is given by the covering
factor times the shadowing factor diminished by the ratio of cloud
size to CGM size. The surface area and volume of an annular region
(i.e., the outer half of the CGM: $0.5\,R_{\rm vir} \leq \rho \leq
R_{\rm vir}$) are related slightly differently from the circular area
and spherical volume assumed above, requiring small numerical
correction factors of order unity to the above equations. 
 
We combine the physical cloud parameters found in Table 6 with the
covering factors shown in Figure 7 to estimate a mass per galaxy in
warm CGM clouds. That is, we determine the range of cloud masses and
their relative numbers in our sample based on the modeling results in
Table 6 and assume that these model results are representative of the
full cloud population. The observed covering and shadowing factors are
then used to determine cloud numbers (Equation 4) and filling-factors
(Equation 5) for the various sizes of clouds found. While the covering
factor is dominated by a large number of little clouds, the filling
factor (and thus the mass) is dominated by the few bigger clouds,
which are slightly more frequent in the inner half CGM volume. For the
cloud sizes in Table 6,  the overall filling factor is $\sim15$ times
less than the covering factor in the absence of shadowing (modest
shadowing and small cloud sizes require small CGM filling
factors). The numbers of clouds and their masses as a function of
cloud size then result in a total CGM mass in warm gas. We use the
available data in two radial bins, $\rho \leq 0.5\,R_{\rm vir}$ and
$0.5\,R_{\rm vir} < \rho  \leq 1\,R_{\rm vir}$, and two host galaxy
luminosity bins  (super-$L^* +$ sub-$L^*$ and dwarfs). We have
combined the super-$L^*$ and sub-$L^*$ absorber samples because these
galaxies have similar CGM  covering factors (Figure 7) and similar
modeled cloud parameters (Figure 12). On the other hand, we have
treated the inner and outer half CGM regions ($\rho < 0.5R_{\rm vir}$
and $0.5\,R_{\rm vir} \leq \rho < R_{\rm vir}$) separately to account
for the slightly lower covering factor at larger radii (see Figure 7),
the slightly lower average mass of modeled CGM clouds (Figure 13) and
the significantly smaller number of absorbers modeled (Figure 8 and
Table 6), Importantly, almost all of the Ly$\alpha$-only absorbers are
in the outer half virial radius.  Most clouds at $0.5\,R_{\rm vir}
\leq \rho < R_{\rm vir}$ had insufficient metal absorption-line
detections to be modeled which is likely due to these absorbers being
physically smaller with smaller masses since they have smaller $N_{\rm
  H\,I}$. We conclude that, despite the nearly twice larger volume
projected onto the outer half virial radius, three-quarters of the CGM
warm cloud mass is contained within the well-modeled inner half virial
radius.

Table 7 shows the results of the mass estimates based on Equations
(2)--(5) using cloud parameters based on the CLOUDY model results in
Table 6 for the 24 clouds with multiple ion  detections taken as
representative of the CGM warm gas clouds as a whole. Data in Table 7
include by  column: (1) luminosity sub-sample; (2) median luminosity
of the sub-sample in $L^*$ units; (4) median virial radius (in kpc);
(5) the filling factor of warm CGM clouds summed over cloud size
within the inner half virial radius. The outer half has filling
factors which are $\sim8$ times less than the values in this column ;
(6) the total number of CGM clouds with sizes $>1$~kpc; (7)--(9) the
logarithm of the estimated total mass in CGM warm gas, in stars, and
in baryons plus dark matter inside the virial radius (i.e., column (9)
is the ``halo mass''), respectively. All masses are given in solar
masses. All values are referenced to the median luminosity galaxy in
each sub-sample. For quantities in columns (3)--(9) the range of
values given refers to the range of virial radii and halo masses
assuming the two different prescriptions for these quantities
presented in Section 3.1. The smaller values are from the
halo-matching algorithm while the larger values are found by using
Equation (1). 

For all three luminosity bins the estimated ensemble mass in warm
photoionized CGM clouds is substantial.  In the largest galaxies the
CGM warm clouds contain $\sim 10^{10}~M_{\Sun}$ of gas, about half the
amount of baryons as in the stars, gas and dust in the galaxy's
disk. These masses are quite uncertain ($\pm0.2$~dex) since they
depend on the small number of  very massive CGM clouds in this sample
and on their masses as determined by photo-ionization modeling. In
turn this modeling depends on line strengths of \HI\ and metal
absorption lines measured in modest to good SNR spectra from STIS and
COS, respectively. By comparing the column densities of key metal ions
in DS08 and in other publications modeling the same absorbers \citep
[e.g.,] [] {tripp02, tilton12}, these measurements can differ by
10-20\%, affecting the modeling significantly. For example, Table 7
lists a cloud size for the 3C~273 absorber at $cz = 1585$~\kms\ that
is $\sim3$ times larger than found by \citet{tripp02}. However, we see
no systematic variation in the cloud parameters determined from the
COS spectra (which lack higher order Lyman lines for accurate
curve-of-growth \HI\ column densities) and the STIS spectra, which
generally have significantly less SNR but, in conjunction with FUSE
spectra, allow curve-of-growth determinations of hydrogen column
density. The main uncertainty in the estimates in Table 7 is the
modest number of CGM clouds which were successfully modeled by CLOUDY.
This is particularly true for the dwarfs where only three absorbers in
our sample have been modeled. For that sub-sample we estimate a larger
uncertainty for the mass estimate in Table 7 of $\pm$ 0.5 dex.

Because we have modeled these clouds entirely using their lower,
largely photo-ionized ions,  the total warm CGM mass in column (6)
does not explicitly include any ``warm-hot'' gas collisionally-
ionized by shock fronts associated with the motion of these clouds
through a hotter substrate. \OVI\ is detected in all but three
serendipitous metal-bearing clouds (Table 3), as well as in the one
targeted absorber for which a \fuse\ spectrum is available. This
suggests that a ``warm-hot'' interface is associated with virtually
all warm CGM clouds. The recent ``COS-halos'' study
\citep{tumlinson11} found a high covering factor of \OVI\ to at least
$0.5\,R_{\rm vir}$ in a sample of $L>L^*$ star-forming galaxies. The
similar extent and covering factor of absorbers strongly suggests that
these absorbers are related. From the recent \citet{tumlinson11}
analysis, this hotter gas comprises a somewhat smaller, but still
substantial, amount compared to the photo-ionized gas mass we find
here.

The median luminosities for the super-$L^*$ and sub-$L^*$ samples
bracket the total luminosity estimated for the Milky Way. Therefore,
for the statistics, physical models and halo-matching scalings adopted
here, we expect that the Milky Way's CGM should extend to
$\sim200$~kpc and should contain $\sim 2000$ warm CGM clouds $>1$~kpc
in size with a filling factor of order 4\%. This inferred filling
factor means that approximately one in every 25 \SiIII-absorbing HVCs
should be located at distances of 50--150 kpc, rather than in the low
halo \citep[$<10$--20 kpc above the disk;][]{collins09,
  lehner11}. Although much fewer in numbers than the \SiIII\ HVCs
discovered and  inventoried by \citet{shull09} and \citet{collins09},
these more distant CGM clouds are estimated to contain an order of
magnitude more mass ($\sim10^{10}~M_{\odot}$) than the highly-ionized
HVCs. Recently \citet{lehner11} have used background stars to infer
that many highly-ionized HVCs  are close to the disk, $\leq20$~kpc
away. Because the detection rate of these HVCs is $\sim50$\% in
absorption against distant halo stars and the detection rate of
\SiIII\ HVCs against extra-galactic targets is $\sim80$\%
\citep{shull09},  many Milky Way highly-ionized HVCs do not have good
constraints on their distances and so may be much more distant than 20
kpc above the disk. These covering factors leave open the possibility
that the Milky Way has a distant (50--150 kpc), highly-ionized and
much more massive CGM cloud population like we have found around other
galaxies. Indeed, there is no reason to expect our Galaxy to be
different in this respect.

\section{Discussion}

\begin{deluxetable*}{lccccccc}

\tablecolumns{8}
\tablewidth{0pt}

\tablecaption{Spiral Galaxy Baryon Inventory
\label{tab:baryon}}

\tablehead{ \colhead{Sub-sample} & \colhead{$\log{M_{\rm vir}}$} & \colhead{$\log{M_{\rm bar}}$} & \colhead{$M_*/M_{\rm bar}$} & \colhead{$M_{\rm warm}/M_{\rm bar}$} & \colhead{$M_{\rm O\,VI}/M_{\rm bar}$} & \colhead{$M_{\rm coronal}$/$M_{\rm bar}$} & \colhead{$M_{\rm missing}/M_{\rm bar}$} }

\startdata
Super-$L^*$ & 11.8 & 11.0 & 20\% &   10\% &     6\%  &  $\leq10$\%    &  $\gtrsim50$\% \\
Sub-$L^*$   & 11.2 & 10.4 & 20\% &   15\% &    [9\%] & [$\leq10$\%]   &  $\gtrsim50$\% \\ 
Dwarfs      & 10.3 & ~9.5 & 20\% & $<5$\% & [$<1$\%] & [$\leq10$\%]   &  $\gtrsim65$\%
\enddata

\tablecomments{All masses are given in units of $M_{\Sun}$. All values use $M_{\rm vir}$ and $R_{\rm vir}$ from the halo matching technique.  All percentage values are approximate.}

\end{deluxetable*}

\subsection{The Baryon Budget in Spiral-rich Galaxy Groups}

Rich clusters of galaxies are often considered to be fair samples of
the universe due to their enormous size and deep gravitational
potential wells. In this context a related assumption is that clusters
contain the cosmological baryon to dark matter ratio \citep{white93}.
Observationally, both in rich clusters and in smaller groups of
galaxies dominated by massive ellipticals, the intra-cluster and
intra-group diffuse gas emits copious X-rays \citep{sarazin88,
  mulchaey00} and contains most of the baryons \citep{white93}.
However, \citet{mulchaey96} failed to detect thermal X-rays from
spiral-rich groups of galaxies and, based upon the lower velocity
dispersion of spiral-rich groups compared to elliptical-dominated
groups,  they speculated that the only viable method for detecting
such gas was through absorption lines in the spectra of bright
background QSOs. While \OVI\ 1032, 1038~\AA\ absorption was
specifically mentioned by them as likely transitions to conduct such a
search, \HI\ \lya\ remains sensitive to diffuse gas at temperatures
slightly in excess of $10^6$~K, but requires high-SNR spectra to
detect even modest column densities ($\log{N_{\rm H\,I}} \geq 13$),
which imply total hydrogen columns $\log{N_{\rm H}} \geq 19$ due to
the extremely small neutral fractions at $T>10^6$~K.

Using a very high-SNR ($\sim50$) COS spectrum of PKS~0405--123,
\citet{savage10} reported the discovery of the first broad,
symmetrical \OVI\ absorption which matches the \citet{mulchaey96}
predictions quite well and is arguably the first detection of diffuse,
hot gas in spiral groups. In this case, the very hot temperature of
this gas is required both by the line width of the symmetrical
\OVI\ absorption and the absence of detectable \lya. The broad
\OVI\ absorber in PKS~0405--123 lies at $\rho >100$~kpc in projection
from two late-type galaxies  and also has associated warm-gas
absorption at  $cz=50,105$~\kms\ which is included in Tables 3 and 6
\citep[see also][]{prochaska04}.  In another high-SNR COS spectrum, a
broad \lya\ (BLA) absorption was found in HE~0226--4110 by
\citet*{savage11b} blended with strong, much narrower \lya\ associated
with warm, photoionized gas. The BLA appears associated with much
hotter gas including \OVI\ and \ion{Ne}{8} absorptions. A few other
BLAs blended with narrower \lya\ have already been found in an
on-going  search of all high-SNR COS spectra \citep [] [Danforth
  et~al. in prep] {savage11a}.  The relatively strong
\OVI\ absorptions ($\log{N_{\rm O\,VI}} \geq 14.2$) discovered by
\citet{tumlinson11} are not the same absorber type as the broad,
shallow \OVI\ seen by \citet{savage10},  and may actually mask the
presence of the broad, shallow absorption in many cases even in
relatively high-SNR spectra \citep {savage11a}.

The COS UV spectrograph on \hst\ is sensitive enough that a full
accounting of BLAs and broad, shallow \OVI\ absorptions can be
made. This census can confirm the apparent abundance of BLAs
tentatively seen at lower SNR with STIS \citep*{richter04, lehner07,
  danforth10b}.  A large number of BLAs per unit redshift are expected
if these are detections of hot, extended gas in spiral-rich
groups. Because spiral-rich groups far exceed elliptical-dominated
groups and clusters in number density, the detection of a massive
intra-group medium in spiral groups could contribute a significant
number of baryons to the universal budget \citep [$\sim20$\%;] []
{savage10}.   

Here we have found strong evidence for high covering factor warm gas
clouds around galaxies of all luminosities and used their derived
properties from photoionization modeling to infer the mass in these
clouds. We find no strong distinctions between the warm CGM cloud
properties around late-type galaxies of differing luminosities at
$L\geq0.1\,L^*$.  While warm clouds are detected around dwarfs, the
current statistics are poor. We find no strong evidence for warm
clouds around  early-type galaxies at all \citep [but see] []
{thom12}. Using the results of our warm gas CGM inventory in Table 7,
Table 8 shows our best estimates for the percentage of baryons in the
various reservoirs in  spiral galaxies of varying luminosities. This
Table uses the results on the CGM from this survey and results from
other works to characterize the location and physical conditions of
the baryons detected thus far in late-type galaxies. The stellar
baryon fraction is taken to be 20\% for all galaxies based on a
constant mass-to-light ratio of $M/L \approx 1$--2 in solar units but
could be a somewhat smaller fraction in lower luminosity galaxies
\citep{moster12}. Molecular and atomic disk gas and dust usually are a
small fraction of the total and HVCs detected in 21-cm emission are
likewise a negligible amount, although the percentage of baryons in
these reservoirs could be much more significant in dwarfs
\citep{peeples11}. A detailed accounting of the relative number of
baryons in stars and disk gas as a function of galaxy mass is beyond
the scope of this paper; therefore, these collapsed baryon reservoirs
are included with the stars in Table 8 and are assumed to total 20\%
of the baryon inventory \citep{fukugita98, shull12} for all three
luminosity sub-samples.

The 10--15\% baryon fraction estimate for warm CGM clouds in luminous
galaxies is based on the accounting and photo-ionization modeling in
this paper but also assumes that the ``\lya-only'' clouds are
physically similar to the modeled clouds and that their lower $N_{\rm
  H\,I}$ values correspond to smaller sizes and masses  as the modeled
clouds. It is possible that these lower column density \HI\ clouds
could be much more highly ionized (\OVI\ is detected in many of them)
and more massive than the clouds we have modeled using CLOUDY.  If
this is the case, the values in column (5) may somewhat underestimate
the warm CGM cloud mass.  The 6\% estimate of hot,
collisionally-ionized gas in luminous galaxies comes from the recent
\citet{tumlinson11} ``\hst/COS halos survey''. The \OVI\ absorption
found by that group has a similarly  high covering factor and physical
extent around galaxies as the warm clouds we have detected, so it is
natural to suggest a relationship between the two. Indeed, where
\OVI\ could be detected in these clouds it usually was detected (only
five firm non-detections in Tables 3 and 4). Although the
photo-ionized  and the shock-heated gas may overlap considerably,
creating a ``double-counting'' of baryons issue, CGM clouds and their
shocked interfaces  may account for as much as 15--25\% of all massive
spiral galaxy baryons. Since the \citet{tumlinson11} survey observed
only the most massive galaxies, the other listings in Table 8 column
(6) are extrapolations to lower luminosities based on assuming  a
scaling between the warm CGM  clouds and the warm-hot gas seen in
high-column density \OVI\ absorption. These scaled values are shown in
brackets to indicate that they are not based on actual observations
and are quite uncertain. 

Table 8 includes the following information by column: (1) Luminosity
sub-sample; (2) the logarithm of the total virial mass  ($M_{\rm
  vir}$; dark matter + baryons) in solar masses; (3) the logarithm of
the baryonic mass ($M_{\rm bar}$) in solar masses determined by
assuming that a late-type galaxy  contains the 5:1 cosmic ratio  of
dark matter to baryons \citep {larson11}; (4) the baryon fraction in
stars and disk gas and dust \citep*{fukugita98, peeples11} as
discussed above; (5) the baryon fraction in warm CGM clouds; (6) the
baryon fraction in WHIM gas probed by strong \OVI\ \citep{tumlinson11}
assuming no ``double counting'' with the warm, photoionized baryons.
Estimates in brackets are scaled  values assuming the same fraction of
the warm cloud mass as seen in the massive galaxies; (7) the limit on
the percentage of very hot ($ T\gtrsim 10^7$~K) coronal gas around
spirals  set by the failure to detect hot ($kT \sim 300$~eV) gas
around the Milky Way or other nearby galaxies \citep{bregman07,
  anderson10}; but see \citet{anderson11} and the discussion of \citet
{gupta12} below. Since this limit is set  by observations of Milky
Way-sized galaxies, the value for dwarfs is a very uncertain
extrapolation and is listed in brackets for that reason;  (8) the
percentage of ``missing baryons'' assuming that the stars, CGM warm
clouds, CGM warm-hot interface gas traced by \OVI\ and coronal gas
potentially detected in X-rays are the only major reservoirs and also
assuming that spirals and other late-type galaxies are ``closed
boxes'' for which the cosmic ratio  of baryons to dark matter applies
(but see below). That is, column (8) is 100\% minus the sum of columns
(4), (5), (6) and (7). The values in the last column are all shown as
approximate due to the substantial uncertainties for the values in the
preceding columns. By this accounting we confirm that the amount
``missing'' is large.

The listings in Table 8 use the the total virial mass and the virial
radius values calculated using the halo matching scalings shown in
Figure 1. The scalings from Equation (1) yield similar baryon
percentages because all mass amounts scale upwards by a similar amount
under that assumption (see Section 3.1). 

The hot gas ($T \gtrsim 10^6$~K) predicted by \citet {mulchaey96} and
potentially discovered by \citet{savage10, savage11a} using high-SNR
COS spectra  is a candidate for this ``missing baryon'' reservoir. If
this hot gas extends over the full virial radius of a spiral-rich
group (400 kpc radius is assumed here), it could contain as much as
$7\times10^{11}~M_{\odot}$ of baryons \citep{savage10}; this is enough
to account for the ``missing baryons'' in an entire spiral-rich galaxy
group.  Since we have seen that warm CGM clouds likely are infalling
and outgoing through the virial radius of an individual spiral galaxy,
a single galaxy is  not necessarily a closed box and so may not sample
the cosmological ratio of baryons to dark  matter individually. The
largest extent of metal-enriched gas away from galaxies is probed by
H~I $+$ \OVI\ absorbers, which are found up to $\sim$ 800 kpc from
luminous galaxies (see Section 4.2 and Figure 11). Since this is
comparable to the physical size of a spiral-rich group,  it is
possible that metals are not spread beyond the extent of a single
group of galaxies. Also the warm cloud kinematics shown in Figures
8-11 shows that even the CGM clouds with the largest $|\Delta v|$ only
modestly exceed $v_{\rm esc}$. Based on the current census, it is
likely that most CGM clouds are confined to a single galaxy group and
do not escape into the diffuse IGM. Therefore, a case can be made that
a region the size of an entire spiral-rich group like the Local Group
can be considered a ``closed box'' for baryon content and chemical
evolution modeling. If significant amounts of gas and metals escape
beyond the bounds of spiral-rich groups, this process could have
occurred mostly at higher redshifts and earlier cosmic times when the
spiral-group gravitational potential well had not developed fully. 

Using the halo matching scaling in Figure 1 a spiral-rich group with a
total luminosity of a few $L^*$ has a total halo mass of
$10^{12.7}$--$10^{13.0}~M_{\odot}$ and a total baryonic mass 0.8~dex
less than that amount. The reservoir of hot gas suggested by the
\citet{savage10, savage11a} detections amounts to $\sim
7\times10^{11}~M_{\odot}$ of gas or 40--80\% of the baryons predicted
to be present in such a group. Thus, this hot gas reservoir could
account for the remaining  ``missing baryons'' in spiral galaxy
systems. If present, this hot gas would also be the largest baryon
reservoir in such systems,  a factor of two or more larger than all
the stars, gas and dust in the disks of the group galaxies combined. 

Indirect evidence for the existence of a very extensive hot gas
surrounding late-type galaxies is shown in Figure 14, which displays
warm CGM cloud pressure as a function of scaled impact
parameter. There is no obvious trend in this plot despite declining
cloud sizes and masses as a function of scaled impact parameter (see
Figures 12 and 13) as well as declining \lya\ absorption \eqw\  with
impact parameter (Figure 9).  If these warm clouds are in near
pressure equilibrium with a hot diffuse gas then such a flat pressure
profile is unexpected unless either  the scale height of this gas is
much larger than the virial radii of the more luminous galaxies in
this sample (i.e., $\geq200$--300~kpc in radius) and/or the density
profile of this hot gas is unexpectedly flat with radius from these
galaxies \citep*{fang12}. Since the latter hypothesis may require an
unphysically flat mass profile, this speculation, and the cloud
pressure data from the CLOUDY modeling which support it, must be
treated with some caution until confirmed by new observations.
 
Neither circumstance (large scale-height or flat density profile) is
observed for the hot coronal gas detected around the Milky Way
\citep{bregman07} where scale heights of only a few kpc are
inferred. Also very extended X-ray emitting halos have not been
detected  in general around nearby spiral galaxies \citep{bregman07,
  anderson10}. But X-ray imagers are not sensitive to gas with
temperatures near 10$^6$ K, so that very extended gas could be present
and remain largely undetectable to {\sl Chandra} or {\sl XMM-Newton}.
Given the median pressure shown in Figure 14 ($P/k \approx 10~{\rm
  cm^{-3}\,K}$) over a size $\geq300$~kpc, and by assuming pressure
balance between   warm CGM clouds and this putative hot diffuse gas at
T$\sim$ 10$^6$ K, then the total baryonic mass of this gas is $\geq
2\times10^{11}~M_{\odot}$. The density, pressure, temperature and
total hot gas mass inferred from pressure balance with warm CGM clouds
is in close agreement with the very recent, adiabatic model of
\citet{fang12}.  This amount is comparable to the ``missing baryons''
in a spiral galaxy group. 
 
Recently, \citet{gupta12} reanalyzed the X-ray spectroscopy of eight
bright AGN which all probe the Galaxy halo and Local Group
CGM. Although the location, size  and thus mass of the
\OVII\ absorption found in these spectra at $\log{N_{\rm O\,VII}}
\approx 16$ is controversial,  this column density in \OVII\ is close
to the amount predicted by the \OVI\ column density found by
\citet{savage10, savage11a} if $T\sim10^6$~K; i.e.,  this Local Group
detection could be very extended hot gas consistent with the broad,
shallow \OVI\ found in other spiral-rich groups.  \citet{gupta12}
suggest that if the \OVII\ extent is $\geq150$~kpc it would have a
very large mass  $>10^{11}~M_{\odot}$ \citep[see also][]{fang12},
similar to what we calculate based on both the PKS~0405--123 broad
\OVI\ absorber and on the warm CGM cloud pressures. 

However, other non-detections appear inconsistent with this
interpretation. Excepting a possible  {\sl Chandra} detection of
\ion{O}{8} in the PKS~2155--304 sight line  \citep*{fang07}, only one
other plausible detection of \OVII\ has been made with the {\sl
  Chandra} and {\sl XMM-Newton} spectrometers \citep{buote09}.  If
this hot gas is a common feature    of most or all spiral groups, it
should have been detected in X-ray absorption lines along other sight
lines, and has not been, even in co-added {\sl Chandra}  spectra
\citep{yao10}. But current X-ray spectrometers have limited spectral
resolution and poorly-characterized systematic noise \citep{yao12}, so
that non-detections may be due primarily to these issues at the
current time. Better spectral resolution using well-characterized
detectors are necessary to make advances in this field at X-ray
wavelengths. On the other hand,  a sensitive COS census of BLAs and
broad \OVI\ absorbers is possible and should find $d\mathcal{N}_{\rm
  BLA}/dz \approx 10$--20 per unit redshift  if spiral-rich groups
contain significant amounts (and extents) of hot gas. A solid
measurement of the $d\mathcal{N}/dz$ for such systems can help infer
their sizes, confirming the large hot gas masses suggested here. A BLA
search using COS is currently underway using only high-SNR ($>20$:1)
spectra (Danforth et~al., in prep).

\subsection{Input for Galactic Chemical Evolution Models}

Any accurate accounting of the mass infall and outflow rates into and
out of spirals is premature due to the uncertainty in the physical
structure, kinematics, metallicity and thus origin for many CGM clouds
in our sample. While metal-enriched and \lya-only clouds are present
in the current CGM cloud sample, the distinction between these two
types is not well-defined owing to the limited SNR of the UV
spectroscopy used. \lya-only clouds could be enriched at levels
similar to the metal-bearing clouds but with their metal absorption
lines currently undetectable given their generally lower \HI\ column
densities \citep[for examples, see Section 4.4.1
  and][]{keeney12}. However, the most basic  prediction coming from
Galactic chemical evolution models, the necessary accretion of
low-metallicity gas, finds some preliminary support from this study.
Only four good examples of low-metallicity gas in the CGMs of $\gtrsim
L^*$ spirals have been found using current data (see Section
4.4.1). This suggests that higher SNR COS spectroscopy will be able to
characterize many other examples of gas infall by  detecting metal
absorption at levels significantly below the metallicity of the host
galaxy (i.e., at $<10$\% solar for an $L^*$ galaxy).  Since
low-metallicity gas has already been detected in the Milky Way halo
\citep [e.g., Complex C;] []{richter01, collins07, shull11},  external
galaxy CGM studies using high-SNR COS spectroscopy can add a
statistical accounting of the amount of gas accreted,  outflowing and
recycling to generalize the Milky Way results to other spirals.

What is possible now is a first, rough accounting of the amount of
``galactic fountain'' gas which is being recycled in the CGM of these
galaxies.  In Section 4.4 based on the data shown in Figure 8, we
identified 9 CGM absorbers as good candidates for high metallicity
recycling gas, which are $\sim15$\% by number and $\sim$40\% by mass
of our full CGM cloud sample. This fraction is a lower limit on the
recycling gas mass since there are almost certainly other recycling
gas clouds  which have not been identified unambiguously by our
accounting. Assuming that $\gtrsim40$\% of the mass of the CGM is in
recycling gas and using the total warm baryonic CGM cloud mass in
Table 7 for a $2\,L^*$ galaxy  (and ignoring any modest contribution
from shock-excited gas traced by \OVI\ to be conservative), a typical
super-$L^*$  galaxy is recycling $\gtrsim 4\times10^9~M_{\odot}$ of
high-metallicity gas, about half of which is infalling at any one
time.  Based on their locations in Figure 8a, we assume that the
infalling high-metallicity ``galactic fountain clouds''  are falling
ballistically from a total distance of $\sim$100 kpc at a median total
speed of $\sim$30~\kms (i.e.,  correcting the impact parameter and
radial velocity difference to three-dimensional quantities
statistically).  Thus, recycling CGM gas can provide
$\gtrsim0.6~M_{\odot}\,{\rm yr}^{-1}$  of enriched gas accretion onto
the galactic disk.  Scaling this result to a Milky Way size galaxy
predicts $\gtrsim 0.3~M_{\odot}\,{\rm yr}^{-1}$ of infalling,
recycling gas, about one-third or more of the  infall rate estimate
from \SiIII-absorbing HVCs \citep{shull09}. Comparing the estimated
accretion rate we have obtained with the total infalling  gas estimate
based on \SiIII\ HVCs suggests an origin for the remaining
$\lesssim0.7~M_{\odot}\,{\rm yr}^{-1}$ from outside the galaxy.  This
is not at all unexpected given the example of several Milky Way HVCs
like Complex C.  Increasing the sample size of CGM absorbers at $\rho
\leq 0.5\,R_{\rm vir}$ by targeting new, close QSO/galaxy pairs  will
allow a substantial increase in the known population of CGM clouds
which are recycling galactic gas and will improve this crude, first
estimate. 

Because we have found evidence for a larger reservoir of hot,
metal-enriched gas in spiral galaxy groups \citep{savage10}, any gas
expelled from dwarfs and spirals may largely accumulate there.  Due to
the lower escape speeds for lower mass galaxies, much of this hot gas
could come from dwarfs in the group. The lower covering factor  and
much smaller total masses of CGM clouds around dwarfs found in
Sections 3.3 and 5.1 are modest support for this picture. At present
there are only a couple examples of absorbing gas escaping from dwarf
galaxies in this sample, but any accretion of gas onto larger spirals
probably  comes mostly at the expense of nearby, lower luminosity,
lower metallicity galaxies in the same galaxy group. Low-metallicity
absorbers accreting at much higher velocities (open circles at
$\rho/R_{\rm vir}=0.5$--1.0 and $|\Delta v| \approx 0.3$--$1.0\,v_{\rm
  esc}$) may  make up the additional $\lesssim 0.7~M_{\odot}\,{\rm
  yr}^{-1}$ of infalling gas required by the current rate of Milky Way
star formation. This suggests that, while the Milky Way is not a
``closed box'' for galaxy evolutionary models, the Local Group might
be. Since the star formation histories of local dwarfs \citep{mateo98,
  skillman05} as well as the  histories of our Galaxy and M31
\citep{dalcanton12} are now being constructed, it may be possible in
the near future to construct a chemical evolution  model for the
entire Local Group assuming overall mass conservation. 

\section{Summary}

In this study we have used two samples of absorbers near galaxies
found with HST to investigate and characterize the CGM of low-$z$,
late-type galaxies.  The COS GTO Team has observed 11~QSOs projected
near foreground galaxies, detecting warm, photoionized gas around all
ten $L<L^*$ spiral and  irregular galaxies probed by these \hst/COS
spectra. These galaxies include modest starbursts, normal spirals and
dwarfs, and one low surface brightness galaxy. There are no  obvious
distinctions between the absorptions found around any of these
different types. Absorbers range from ``\lya-only'' clouds at
$\log{N_{\rm H\,I}} \approx 13.5$,  where we find no metals detected
in available spectra,  to a Lyman limit system with numerous metal
detections and $\log{N_{\rm H\,I}} \approx 18.5$. \HI\ \lya\ and metal
ions typical  of photo-ionized ``warm'' clouds are detected in many of
these absorbers, but higher ions like \CIV\ and \SiIV\ typically have
larger equivalent widths than lower ions like \CII\ and \SiII,
indicative of ionization parameters for these clouds of $\log{U} =
-2.0$ to $-3.5$. 

In order to increase the sample size of CGM absorbers, particularly at
$L>L^*$, and to investigate a more random selection of galaxy CGMs,
we have gathered a ``serendipitous'' CGM cloud sample of $\sim60$
absorbers using the \hst/STIS sample compiled by DS08 from 35
well-observed sight lines, most of which also have
\fuse\ spectroscopy. The \fuse\ data are important, providing coverage
of the higher Lyman lines (and accurate $N_{\rm H\,I}$ values from the
curve-of-growth technique) and the important \OVI\ ion, sensitive to
diffuse photo-ionized or  collisionally excited gas. A few absorbers
were found to have Lyman limit decrements in their \fuse\ spectra.
The targeted \hst/COS QSOs do not (except for one) have complementary
\fuse\ data. The CGM absorbers discovered in these serendipitous
spectra are similar to the targeted absorbers. While the spectra, line
measurements and detailed absorber modeling and host galaxy properties
are presented elsewhere (Paper 2), in this paper we presented the
analysis of these data which lead to the following conclusions:

\begin{enumerate}

\item  Only $\sim$5\% of all \lya\ absorbers are projected close
  enough to galaxies to probe their CGM; i.e., most \lya\ absorbers
  are IGM not CGM if the virial radius is taken as the rough dividing
  line between these two populations (Section 3.2).

\item  The covering factor of warm gas clouds inside the virial radius
  of late-type galaxies is very high, consistent with unity inside
  $0.5\,R_{\rm vir}$ and $\sim75$\% between $0.5\,R_{\rm vir}$ and
  $R_{\rm vir}$ for luminosities $\geq 0.1\,L^*$.   These high
  covering factors are consistent with the ubiquity of \lya\ absorbers
  found in both the COS targeted survey presented here and other
  recent surveys \citep {prochaska11a, tumlinson11}. While CGM
  detections and statistics are sparse for dwarf galaxies at $L
  <0.1\,L^*$, the covering factors around dwarfs are still
  substantial, $\sim50$\% inside $R_{\rm vir}$ (Section 3.3).

\item  We find no strong evidence for warm CGM clouds around
  early-type galaxies \citep [but see] [] {thom12};  the only three
  candidate early-type galaxies associated with absorbers in this
  sample all have late-type galaxy alternate identifications (Section
  4.1).

\item  While this survey uses a similar galaxy and absorber database
  as \citet{stocke06}, new galaxy survey work  \citep [chiefly new
    SDSS data releases and] [] {prochaska11a} allows a reassessment of
  the extent of \OVI\ absorbers away from galaxies which confirms our
  earlier results. We find that the \OVI\ absorption screen around
  galaxies is patchy with an approximate covering factor $\sim0.3$ at
  $\log{N_{\rm O\,VI}} \geq 13.2$ out to 3.5--4\,$R_{\rm vir}$ around
  galaxies of all luminosities.  While the current data are still
  sparse at low galaxy luminosities, we extrapolate that dwarfs and
  sub-$L^*$ galaxies are the major sources for \OVI\ absorbers
  (Sections 3.4 and 4.2, Figure 11).

\item The merged sample of COS-targeted and STIS-serendipitous
  absorbers allows a detailed characterization of the CGM of late-type
  galaxies  from super-$L^*$ spirals to sub-$L^*$ spirals and
  irregulars and, with limited statistics, to dwarfs at $L<0.1\,L^*$.
  Inside $0.5\,R_{\rm vir}$ almost all absorbers detected are
  ``metal-bearing'' and many have velocities too low to easily escape
  from the host galaxy. Even  without knowing their direction of
  motion we identify nine of these clouds as likely recycling
  ``galactic fountains''. A few absorbers have metallicities $<0.2$
  solar, too low to be easily ascribed to gas originating in their
  nearby ``host'' galaxy and a few absorbers can be identified
  unambiguously as gas  originating in the host galaxy which is
  escaping into the IGM. These galactic wind candidates have
  metallicities comparable to their associated galaxy's metallicity
  and high radial velocities with respect to their associated galaxy
  ($|\Delta v| > v_{\rm esc}$). Because of the limited SNR of the UV
  spectra, some of the COS-targeted absorbers and most of the
  STIS-serendipitous absorbers cannot be so easily classified. A
  complete accounting of the origin and fate of CGM clouds must await
  new absorber samples found in higher-SNR ($>20$:1) COS spectra
  (Section 4.2).
 
\item Photo-ionization (CLOUDY) modeling of those 24 targeted and
  serendipitous absorbers with multiple ionization states of the same
  element (at least  two of \SiII/\SiIII/\SiIV\ or \CII/\CIII/\CIV)
  finds CGM cloud ionization parameters of  $\log{U} = -2.0$ to
  $-3.5$, typical metallicities of 10\%-- 100\% solar values, total
  cloud densities of $n_{\rm H} = 10^{-3}$ to $10^{-4}~{\rm cm}^{-3}$,
  cloud diameters of 0.1--30~kpc and masses of $10$--$10^8~M_{\odot}$
  (see also Paper 2). The small clouds are best-sampled by this survey
  as  they provide the largest covering factor,  while the massive
  clouds provide the greatest filling factor (5--8\% for the most
  massive galaxies) and the most total mass. However, there are only
  ten of these large ($\gtrsim3$~kpc), massive
  ($\gtrsim10^5~M_{\odot}$) clouds in this sample which means that the
  total  CGM mass in warm gas is not very tightly constrained. Also
  $\gtrsim50$\% of the CGM clouds do not have sufficient metal
  absorption to provide  adequate input for modeling. For this study
  we have assumed that these unmodeled clouds (``\lya-only'' clouds or
  clouds  with only marginal metal-line detections) have similar
  physical conditions to the modeled clouds (Section 4.3). 

\item Unlike the geometry created by observing Milky Way HVCs and
  highly-ionized HVCs from a location inside the distribution of these
  clouds,  very large covering factors of external galaxy CGMs do not
  translate into near unity filling factors. In the simple viewing
  geometry assumed herein, the filling factor depends on the covering
  factor, the mean number of clouds detected in any one CGM and the
  ratio of cloud size to the size of the CGM  (see Equations [2]--[5]
  in Section 4.5).  For a Milky Way size galaxy we estimate a CGM
  volume filling factor of $\sim4$\%, which means that only a small
  fraction of highly-ionized \SiIII\ HVCs \citep{collins09, lehner11}
  are much more distant CGM clouds. This is consistent with  Lehner \&
  Howk's recent result, which places many (but not all) \SiIII\ HVCs
  within a few kpc of the galactic disk.  A low filling factor is also
  consistent with our observation that only a few CGM absorbers have
  multiple velocity components (Section 4.5).

\item Based on our analysis, which includes derived filling factors
  and CGM warm cloud parameters, a Milky Way size galaxy has $\sim
  10^{10}~M_{\odot}$ of warm CGM clouds in its ``halo'' (inside its
  virial radius). Placing the warm CGM cloud population studied here
  into the overall context of the baryon inventory of spiral galaxies
  finds that the warm CGM can account for $\sim10$--15\% of the full
  baryon content of  luminous spiral galaxies assuming that these
  systems contain the cosmic ratio of baryons to dark
  matter. Including the amount of  ``warm-hot'' gas traced by
  \OVI\ recently found around luminous spirals by \citet{tumlinson11},
  and assuming minimal ``double counting''  between these two
  reservoirs, means that the CGM contains a total number of baryons
  comparable to those present in galactic disks, including stars, gas
  and dust (see Tables 7 and 8; Sections 4.5 and 5.1).

\item Although our survey contains only a modest number of absorbers
  associated with dwarf galaxies, we tentatively conclude that warm
  CGM clouds account for $\leq$ 5\% of the baryons in dwarf
  galaxies. Comparing this result with the massive warm CGMs we have
  found around more luminous galaxies suggests that  many of the warm
  CGM clouds escape from the dwarfs into the IGM.

\item Adding in a maximum allowable 10\% of very hot
  ($T\approx10^7$~K) coronal gas near spiral galaxies
  \citep{bregman07, anderson10} leaves $\gtrsim 50$\% of spiral galaxy
  baryons unaccounted for. The recent COS discovery of very broad
  \lya\ (BLAs) and broad \OVI-absorbing gas at $T>10^6$~K   by
  \citet*{savage10, savage11a}, and the suggestion based on STIS
  spectroscopy that BLAs may be quite plentiful \citep{richter04,
    lehner07,  danforth10b}, leads to the hypothesis that the Local
  Group and other spiral-rich groups could contain
  $\gtrsim10^{11}~M_{\odot}$ of very hot gas,  analogous to the
  intra-cluster and intra-group medium detected in elliptical
  dominated clusters and groups \citep [see also] [] {fang12,
    gupta12}.  If a spiral intra-group medium of this amount is also
  ubiquitous, these gaseous reservoirs are a significant fraction
  ($\sim20$\%) of the universal baryon inventory.  \citet{mulchaey96}
  had predicted that this spiral group gas was most easily detectable
  as \OVI\ absorption in the spectra of background QSOs. This
  prediction now has tantalizing confirming evidence in a couple of
  cases \citep{savage10, savage11a}. Analysis of high-SNR COS
  spectroscopy currently underway (Danforth et~al. in preparation) can
  confirm this conjecture (Section 5.1).

\item Indirect support for the existence of a hot ($T\sim10^6$~K),
  extensive ($>300$~kpc), and thus massive spiral intra-group medium
  comes  from the analysis of the physical structure of the warm,
  photo-ionized CGM clouds in our sample.  Figure 14 shows that,
  despite having slightly declining cloud sizes and masses with
  increasing impact parameter, CGM clouds show no clear evidence for a
  strongly declining pressure with impact parameter. If these clouds
  are in pressure equilibrium with an external intra-cloud medium,
  then the pressure of that medium must also be nearly constant with
  radius away from the galaxy to a distance comparable to the virial
  radii of these systems ($\sim150$--250~kpc for sub-$L^*$  and
  super-$L^*$ galaxies). The pressures shown in Figure 14 do suggest a
  rather flat density profile with radius which may be unphysical and
  so these results from the CLOUDY modeling of warm CGM clouds and its
  interpretation must be treated with some caution. But see supporting
  evidence in \citet{fang12} and \citet{gupta12} for  this hot gas in
  our own Local Group.  Nevertheless, by applying pressure balance and
  assuming $T \approx 10^6$~K and a gaseous extent of $>300$~kpc in
  radius suggested by Figure 14  finds $> 7 \times 10^{11}~M_{\odot}$
  of gas, sufficient to account for most if not all of the ``missing
  baryons'' in spiral galaxies. An accurate inventory of BLAs and
  broad \OVI\ absorption can independently confirm very large
  cross-sectional absorbing areas for these systems (Section 5.1).

\item Although a robust accounting of the kinematics, metallicity
  (relative to the host galaxy) and thus origin and fate of individual
  CGM clouds is not yet possible, this analysis has provided some
  tentative information on the gas that fuels star formation in
  late-type galaxies.  Specifically, we estimate that low impact
  parameter, low $\Delta v$, high metallicity clouds plausibly
  associated with ``galactic fountains''  account for $\gtrsim
  0.3~M_{\odot}\,{\rm yr}^{-1}$ of gas infall rate in Milky Way size
  galaxies. This amount is $\sim$ one-third of the amount found in the
  close-in HVC population to be  infalling onto the Galactic disk
  \citep{shull09}. While this suggests that the bulk of the accreting
  gas comes from outside the system, we were able to identify only a
  very few (4) plausible examples of infalling, low-metallicity gas
  clouds from this sample at this time.  To identify these infalling,
  low-metallicity clouds definitively will require both higher-SNR COS
  spectroscopy to determine cloud metallicities to levels $<$ 0.1
  solar and also accurate host galaxy metallicities from ground-based
  spectroscopy. Nevertheless, this study offers  tantalizing
  preliminary evidence for the detection of low-metallicity gas
  accretion required by Galactic chemical evolution models
  \citep[Section 5.2]{larson72, chiappini01}.
\end{enumerate}

The present work is a start to the process of characterizing the CGM
of spiral galaxies. A more definitive characterization of the CGM
will require an \hst/COS QSO/galaxy survey using $\sim500$--1000
orbits to explore the parameter space of galaxy luminosity (and thus
mass and virial radius extent), galaxy type, star formation rate and
metallicity.  High SNRs ($>$ 20:1) are essential for accurate warm
cloud models, cloud metallicities (in comparison with ``host'' galaxy
metallicities),  physical extents around galaxies, and importantly to
search for broad, shallow absorption indicative of hot, spiral
intra-group gas. Detailed study of spiral groups with known or
suspected hot gas reservoirs is required to make a definitive
confirmation; i.e., the gas temperature derived from the absorption
line widths should be comparable to the velocity dispersion of the
galaxies in these groups. Several possible BLAs in spiral groups
identified herein can provide a viable test of the presence of a
massive spiral intra-group gas. A detailed census of the warm and hot
CGM in late-type galaxy groups is a necessary piece to construct an
accurate model for galactic structure and evolution.

\acknowledgements

This work was supported by NASA grants NNX08AC146 and NAS5-98043 to
the University of Colorado at Boulder for the \hst/COS project.   BAK
gratefully acknowledges additional support from NSF grant AST1109117.
Further, we gratefully acknowledge M. Trenti for halo abundance
matching expertise and results, and J. Werk, Y. Yao, and J. Bullock
for helpful conversations.   This research has made use of the
NASA/IPAC Extragalactic Database (NED) which is operated by the Jet
Propulsion Laboratory,  California Institute of Technology, under
contract with the National Aeronautics and Space Administration. 

{\it Facilities:} \facility{HST (COS)}, \facility{HST (STIS)},
\facility{FUSE}, \facility{WIYN (HYDRA)}, \facility {APO (DIS,
  SPIcam)}

\end{document}